\documentclass[]{jfm}

\usepackage{graphicx}
\usepackage{dcolumn}
\usepackage{bm}
\usepackage[utf8]{inputenc}
\usepackage{subfig}
\usepackage{graphicx}
\usepackage{ragged2e}
\usepackage{amsmath}
\usepackage{mathtools}
\usepackage{caption}
\usepackage{color}
\usepackage[most]{tcolorbox}
\usepackage{xcolor}
\usepackage{float}
\usepackage{bigints}
\usepackage[export]{adjustbox}
\usepackage{tabularx}
\usepackage{epstopdf, epsfig}
\usepackage[toc,page]{appendix}
\usepackage{nomencl}
\usepackage{mathtools,amssymb,lipsum}
\usepackage{newtxtext}
\usepackage{newtxmath}
\usepackage{natbib}
\usepackage{hyperref}
\usepackage[normalem]{ulem}
\hypersetup{
	colorlinks = true,
	urlcolor   = pink,
	citecolor  = blue,
}

\usepackage{xcolor}

\usepackage{siunitx}
\usepackage{soul}

\newcommand{\RomanNumeralCaps}[1]
\linenumbers

\newcommand{\bx}[1]{\mathbf{x}{#1}}

\usepackage[normalem]{ulem}
\usepackage[textwidth=\dimexpr\textwidth-2cm\relax]{todonotes}
\makeatletter
\@mparswitchfalse%
\makeatother
\normalmarginpar 

\title{Focussing of concentric free-surface waves}
\author{Lohit Kayal\aff{1}
	,Vatsal Sanjay\aff{2}, Nikhil Yewale\aff{1}, Anil Kumar\aff{1} \and Ratul Dasgupta\aff{1}\corresp{\email{dasgupta.ratul@gmail.com}}}

\affiliation{\aff{1}Chemical Engineering, Indian Institute of Technology, Bombay, India 400 076,\aff{2}Physics of Fluids Group, Max Planck Center for Complex Fluid Dynamics, Department of Science and Technology, and J. M. Burgers Centre for Fluid Dynamics, University of Twente,  P. O. Box 217, 7500 AE Enschede, The Netherlands}

\begin{document}
	\maketitle
	\newcommand{\mj}{{\mathrm{J}}}
	\begin{abstract}
        Gravito-capillary waves at free-surfaces are ubiquitous in several natural and industrial processes involving quiescent liquid pools bounded by cylindrical walls. These waves emanate from the relaxation of initial interface distortions, which often take the form of a cavity (depression) centred on the symmetry axis of the container. The surface waves reflect from the container walls leading to a radially inward propagating wave-train converging (focussing) onto the symmetry axis. Under the inviscid approximation and for sufficiently shallow cavities, the relaxation is well-described by the linearised potential-flow equations. Naturally, adding viscosity to such a system introduces viscous dissipation that enervates energy and dampens the oscillations at the symmetry axis.
        However, for viscous liquids and deeper cavities, these equations are qualitatively inaccurate. \textcolor{black}{In this study, we decompose the initial localised interface distortion into several Bessel functions and study their time evolution governing the propagation of concentric gravito-capillary waves on a free-surface. This is carried out for inviscid as well as viscous liquids}. For a sufficiently deep cavity, the inward focussing of waves results in large interfacial oscillations at the axis, necessitating a second-order nonlinear theory. We demonstrate that this theory effectively models the interfacial behavior and highlights the crucial role of nonlinearity near the symmetry axis. \textcolor{black}{This is rationalised via demonstration of the contribution of bound wave components to the interface displacement at the symmetry axis} Contrary to expectations, the addition of slight viscosity further intensifies the oscillations at the symmetry axis \textcolor{black}{although the mechanism of wave-train generation here is quite different compared to bubble bursting where such behaviour is well-known \citep{duchemin2002jet}}. This finding underscores the limitations of the potential flow model and suggests avenues for more accurate modelling of such complex free-surface flows.
	\end{abstract}

	\begin{keywords}
		Surface waves, nonlinear waves, jet formation, wave focussing, Cauchy-Poisson problem, viscous waves
	\end{keywords}

	\section{Introduction to wave focussing}
	Focussing of moderate amplitude, progressive surface waves can often in turn produce unexpectedly large waves. At oceanic scales, spatial wave focussing, where large amplitude waves form persistently in specific regions \citep{torres2022wave,chavarria2018geometrical}, can produce waves powerful enough to damage or capsize ships. A famous example is the Aghulas current region \citep{Britannica} known for giant waves and shipping accidents \citep{mallory1974abnormal,smith1976giant}.
    The role of current generated refractive focussing leading to the birth of such giant waves, specifically in the Agulhas, was anticipated by \cite{peregrine1976interaction} (also see fig. $8$ in \cite{dysthe2008oceanic} and section $2$ in \cite{white1998chance}). Refractive focussing of surface waves \citep{peregrine1986approximate} has also been exploited to design `lenses' i.e., submerged structures in a water basin which focus incoming divergent, circular waves (see fig. $1$a in \cite{stamnes1983nonlinear}), these being motivated from wave generation of power \citep{mciver1985diffraction,murashige1992ideal}.

    In addition to \textit{spatial focussing}, \textit{spatio-temporal focussing} also occurs \citep{dysthe2008oceanic}, where large wave amplitudes manifest at specific locations in space, albeit briefly. Spatio-temporal focussing has obvious relevance not only towards understanding, for example, rogue (freak) waves in the ocean \citep{RogueWave} but also to our current study (next section).
    The physical mechanisms underlying spatio-temporal focussing have been distinguished further into \textit{linear} and \textit{nonlinear} dispersive focussing (section $4.2,4.3$, \cite{dysthe2008oceanic}). Linear dispersive focussing of progressive waves relies on constructive interference exploiting the dispersive nature of surface gravity waves and is particularly simple to understand in the deep water limit. For uni-directional wave packets in deep water, generated from a wave-maker oscillating harmonically at frequency $\Omega$ at one end of a sufficiently long wave flume, the energy propagation velocity (group velocity) of the packet is $c_g = \frac{g}{2\Omega}$ where $g$ is acceleration due to gravity.
    If the wavemaker frequency varies linearly from $\Omega_1$ to $\Omega_2$ ($\Omega_1 > \Omega_2$) following $\frac{d\Omega}{dt} = -\frac{g}{2x_f}$ within the time interval $\left[t_1, t_2\right]$, \citet{longuet1974breaking} showed that the energy of each wave packet emitted during this period will converge at $x = x_f$ simultaneously at $t = t_f$ (see \cite{brown2001experiments}). This focussing of wave energy thus causes a momentary but significant increase in energy density at $x_f$ manifested as a transient, large amplitude wave at that location around time $t_f$. This technique has been discussed in \cite{davis1964testing} and its variants have been employed extensively to generate breaking waves in the laboratory in a predictable manner in two \citep{rapp1990laboratory} and three dimensions \citep{wu2002breaking,johannessen2001laboratory,mcallister2022wave} as well as in other related contexts such as generation of parasitic capillary on large amplitude waves \citep{xu2023parasitic}.

    On the other hand, in non-linear dispersive focussing, the modulational instability \citep{benjamin1967disintegration} of a uniform, finite-amplitude wave-train (Stokes wave) plays a crucial role. This instability can cause the wave-train to split into groups, where focussing within a group can produce a wave significantly larger than the others \citep{zakharov2006freak}. For further details on nonlinear focussing, we refer readers to the review by \cite{onorato2013rogue}.

    \subsection*{Spatio-temporal focussing at gravito-capillary scales}

    Following this brief introduction to large-scale focussing, we now shift our attention to length scales where gravitational and capillary restoring forces are nearly equivalent. Our study aims to achieve an analytical understanding of wave focussing at these shorter scales. Below, we illustrate two examples where such small-scale focussing can be readily observed.

\begin{figure}
    \centering
        \subfloat[]{\includegraphics[scale=0.38]{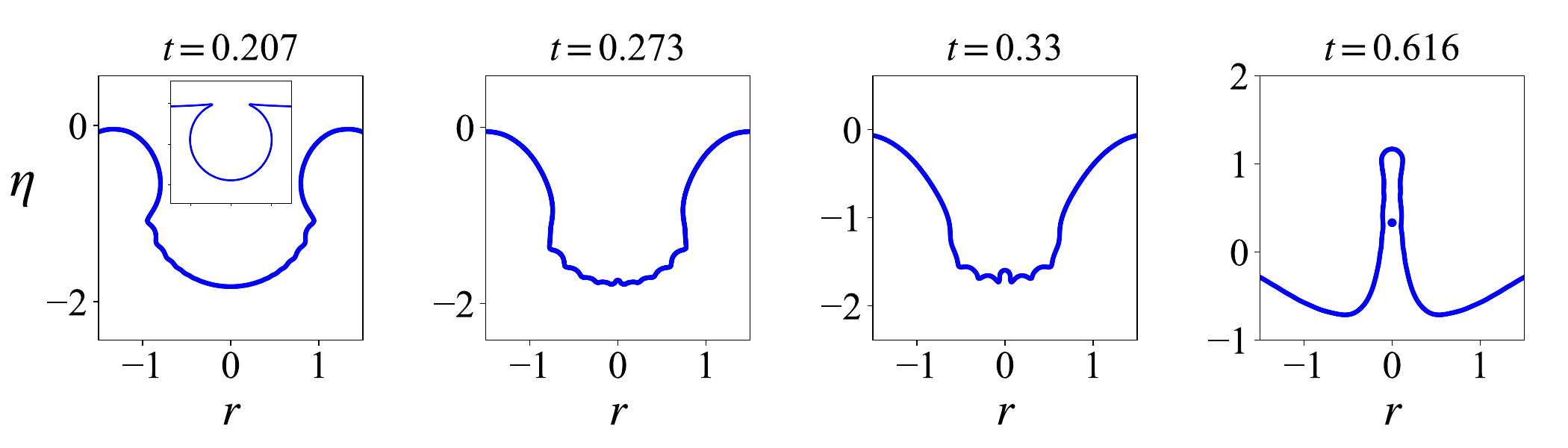}}\\
        \subfloat[]{\includegraphics[scale=0.38]{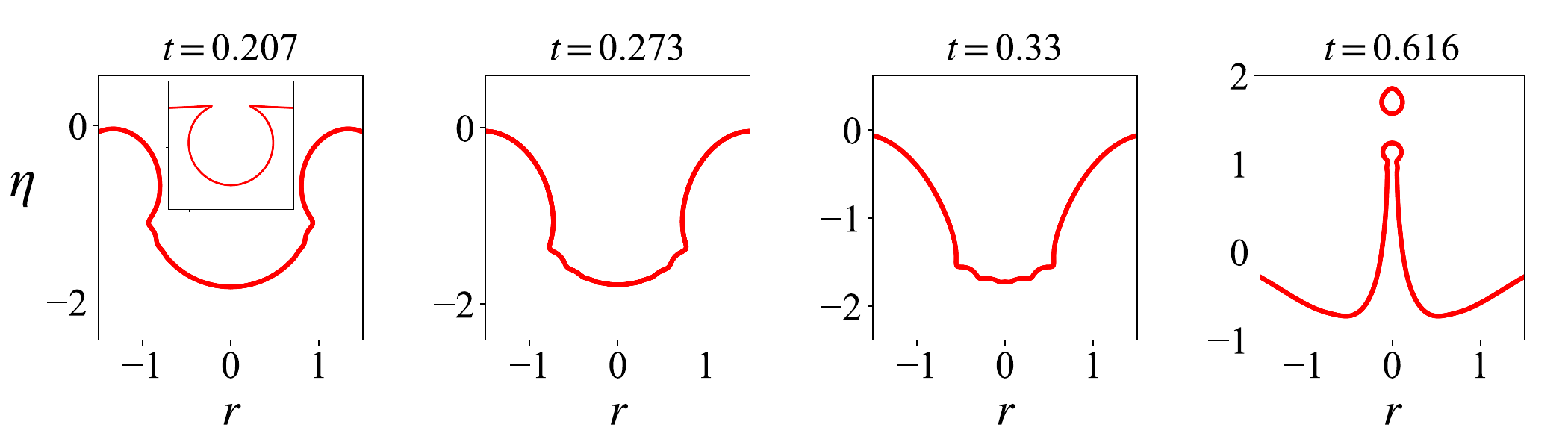}}
        \caption{An example of capillary wave focussing obtained from direct numerical simulations (DNS) conducted using the open-source code Basilisk \cite{popinet-basilisk}. The initial cavity shape (inset in first figure of upper and lower panels) is obtained by solving the Young-Laplace equation with gravity to determine the shape of a static bubble at the free surface (without its cap). In CGS units, initial bubble radius $0.075$, surface tension $T=72$, gravity $g=-981$, density $\rho^{\text{L}}=1.0$ and $\rho^{\text{U}}=0.001$ for upper and lower fluid. Upper panel (blue) simulations are conducted using zero viscosity for both gas (above) and liquid (below). (Red, lower panel) simulations have dynamic viscosity $\mu^{\text{U}}=0.0001$ and $\mu^{\text{L}}=0.01$. Axes are non-dimensionalised using initial bubble radius. Time is non-dimensionalised using the capillary time-scale $t = \frac{\hat{t}}{\sqrt{\frac{\rho \hat{R}_b^3}{T}}}$. For the upper panel $Bo \equiv \frac{\rho^{\text{L}} g\hat{R}_b^2}{T}=0.076$ and Oh$=\frac{\mu^{\text{L}}}{\sqrt{\rho^{\text{L}} T \hat{R}_b}} = 0$.
        For the lower panel Bo$=0.076$ and Oh$=0.0043$.}
        \label{fig1}
    \end{figure}

    \cite{stuhlman1932mechanics} investigated the formation of drops from collapsing bubbles with diameters under $0.12$ cm in water-air interfaces and $0.15$ cm in benzene-air interfaces.
    He hypothesised that these drops emerged from Worthington jets created by the collapse of the bubble cavity. However, contemporary research identifies this as just one of two mechanisms responsible for drop generation \citep{villermaux2022bubbles}.
    The first high-speed ( $\approx 6000$ frames per second) images of jet formation were reported by \cite{macintyre1968bubbles,macintyre1972flow} (see original experiments by \cite{kientzler1954photographic}). Interestingly, these studies demonstrated that the surface ripples are created by the retraction of the circular rim of the relaxing bubble cavity. These ripples travel towards the cavity base before the jet emerges.
    In the words of \cite{macintyre1972flow} (see abstract) ``\textit{..an irrotational solitary capillary ripple precedes the main toroidal rim transporting mass along the surface at about $90\%$ of its phase velocity.
    The convergence of this flow creates opposed jets...}". The seminal work by \cite{duchemin2002jet} of collapsing bubbles (much smaller than their capillary length scale) at a gas-liquid interface was able to resolve this focussing process, via direct numerical simulations (DNS) of the axisymmetric Navier-Stokes equations without gravity. Figure~\ref{fig1} depicts the generation of an axisymmetric, wave-train focussing towards the base of the bubble cavity (also the symmetry axis) for two different Ohnesorge numbers $(Oh)$ and at a fixed Bond number $(Bo)$. The Bond number $Bo\equiv\frac{\rho^Lg\hat{R}_b^2}{T}$ determines the bubble shape, and Ohnesorge number $Oh\equiv\frac{\mu^L}{\sqrt{\rho^LT\hat{R}_b}}$ accounts for the ratio of viscous to capillary forces. Here $\rho^L,\,\mu^L,\,T, \hat{R}_b$ are the lower fluid density, lower fluid viscosity,  coefficient of surface tension and equivalent radius of the bubble respectively. We refer the readers to \citet{deike2022mass, VatsalThesis, gordillo2023theory} for recent advances on study of bubble collapse and jet formation mechanisms.

    Another example of axisymmetric focussing of surface waves was highlighted in the study by \cite{longuet1990analytic}, where several interesting observations were noted. \cite{longuet1990analytic} studied the inverted conical shaped `impact cavities' seen in experiments and simulations \citep{oguz1990bubble} of a liquid droplet falling on a liquid pool. The author compared these cavities to an exact solution to the potential flow equations without surface tension or gravity \citep{longuet1983bubbles}, where the free-surface (gas-liquid interface) took the form of a cone at all time. The apex of this cone (i.e. the impact cavity) is often seen to contain a bulge (see fig. $2$a in \cite{longuet1990analytic}) and the formation of this was attributed to (we quote, section $6$ first paragraph in \cite{longuet1990analytic}) ``\textit{a ripple on the surface of the cone converging towards the axis of symmetry}'', thus highlighting the role of wave focussing once again. \cite{longuet1990analytic} insightfully remarked that this convergence process would be similar to the radially inward propagation of a circular ripple on a water surface. The interface shape could thus be approximated as being due to the linear superposition of an initial, localised wave packet (generated by distorting an initiallly flat surface) whose Fourier-Bessel representation $F(k)$ ($k$ being the wavenumber) slowly varies on a time-scale $\bar{t}$ (i.e. slow compared to the wave packet propagation time-scale $t$). \cite{longuet1990analytic} thus posits that the shape of the perturbed interface $\eta(r,t, \bar{t})$ may be represented as

    \begin{align}
    \eta(r,t,\bar{t}) = \int_{\Delta k} F(k,\bar{t})\mj_{0}(kr)\exp\left(I\sigma(k) t\right)k dk, \label{eqn1}
    \end{align}

    \noindent where $\mj_{0}$ is the Bessel function, $r$ is the radial coordinate and the spectrum of the surface perturbation $F(k,\bar{t})$ evolves slowly on a time-scale $\bar{t}$, $I\equiv \sqrt{-1}$ and $\sigma(k)$ satisfies the dispersion relation for capillary waves (see eqn. $6.2$ in \cite{longuet1990analytic}). Note that if the slow variation of $F(k,\bar{t})$ over $\bar{t}$ is supressed, eqn. \ref{eqn1} represents the solution to the linearised Cauchy-Poisson problem with an initial surface distortion whose Hankel transform is $F(k)$. \cite{longuet1990analytic} however did not report any systematic comparison of available experimental or simulational data \citep{oguz1990bubble} with eqn. \ref{eqn1} although the author anticipated that nonlinearity could become important during the convergence; see last para in page $405$ of \cite{longuet1990analytic}.

    Our current study is partly motivated by the aforementioned observations of \cite{longuet1990analytic} and \cite{duchemin2002jet} and aims at obtaining an analytical description of spatio-temporal wave focussing at these short scales. We seek an initial, \textit{localised} surface distortion which produces a wave-train, and whose radial convergence may be studied analytically, at least in the potential flow limit. We refer the reader to the review by \cite{Eggers2024} where this limit corresponding to Ohnesorge $Oh=0$ is discussed. In the next section we present a localised initial surface distortion which is expressable as a linear superposition of Bessel functions (Fourier-Bessel series). It will be seen that this distortion generates a surface wave-train which focuses towards the symmetry axis of the container. We emphasize that the wave-trains or the solitory ripple seen in \cite{kientzler1954photographic} and  \cite{longuet1990analytic} respectively, have different physical origins compared to the ones we study here. However, following \cite{longuet1990analytic} we intuitively expect there are aspects of their convergence which do not sensitively depend on how these are generated in the first place.

    \textcolor{black}{Of particular relevance to us, is also the interesting study by \cite{fillette2022axisymmetric} who investigated forced capillary-gravity waves in a cylindrical container. These waves were generated via a vertically vibrating ring at the gas-liquid interface. The authors showed that the steady shape of the interface is well represented by the third order, (nonlinear) time-periodic solution due to \cite{mack1962periodic}. The agreement between the analytical model and experimental data, is particularly good around $r=0$, although differences persist away from the symmetry axis (see their fig. $4$b). With increasing forcing amplitude, the authors note an interesting transition from linear to the nonlinear regime followed by a jet ejection regime. We demonstrate in fig.~\ref{fig_flow_wave} (Appendix D) that a similar transition is also seen for our initial condition (see discussion in next paragraph) albeit our study excludes external forcing. Due to the absence of forcing, it becomes feasible to carry out a first principles mathematical analysis of the wave-focussing regime, as has been reported here.}

    \textcolor{black}{While wave-train convergence and jet formation may often be concomitant, as apparent from the bubble collapse simulations in fig. \ref{fig1}, the two phenomena are distinct. Fig. ~$3$ of \cite{deike2018dynamics}, for example, describes experimental investigations of an air bubble bursting at silicon oil-air interface producing a jet, but without any visible signature of a converging wave-train towards the collapsing bubble base. On the other hand, the converging wave-train in the shape oscillations generated due to two coalescing bubbles \citep{zhang2008satellite} (their fig. $12$), lead to rapid interfacial oscillations at the focal point, but no signature of pinchoff or a liquid jet. When a converging wave-train and a liquid jet are both present, the dynamics of the latter can be affected by the former quite non-trivially. The fastest jet in such cases can occur at an ``optimal'' value of liquid viscosity, rather than in the inviscid limit; see experiments and fig. $3$ (panel b) of \cite{ghabache2014physics} in the context of bubble bursting. In view of this rather complex aforementioned behaviour, it becomes desirable to have first principles studies of cavity collapse with and without an accompanying wave-train. The spatially \textit{localised} interface deformation considered in this study (fig. \ref{fig3}), permits access to these phenomena independently, through a tuning parameter. As shown in Appendix D, for small cavity depth (relative to its width), the initial distortion generates a train of radially inward focussed waves (after reflection), which we label as `wave focussing' and whose physics is of interest here. At larger cavity depth, a jet emerges already at short time due to `flow focussing' (see description below fig.~ \ref{fig_flow_wave}). Notably, this jet is formed significantly before wall reflections can generate a radially inward propagating wave-train. The study in \cite{basak2021jetting} investigated such a jet, albeit obtained from a single Bessel function. In contrast, we study here the wave focussing regime where no such jet is generated.}.

\begin{figure}
	\includegraphics[scale=1.0]{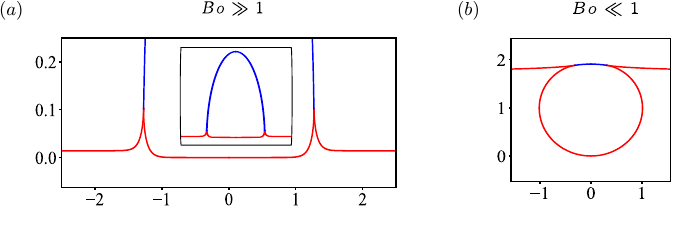}
	\caption{\textcolor{black}{The effect of change of Bond number on the shape of a static bubble. (Right panel) an air bubble corresponding to Bo $=0.076<< 1$ (also see inset in panel (a) of fig. \ref{fig1}). (Left panel of current figure) The bubble shape for Bo $=222 >> 1$. As the Bond number is increased, an increasingly larger fraction of the bubble shifts upwards (compared to the mean interface level at large distance) and its `rim' (see sharp corners on the right panel) distorts into \textit{vertically pointing} kinks seen in the left panel. For Bo$>>1$, the bubble shape is a single-valued function $\eta(r)$ (the red curve on the left panel) and provides the motivation for the initial interface distortion (albeit significantly smoother) in fig. \ref{fig3} and treated analytically in this study. The curves in blue in both panels represent the bubble cap. The inset on the left panel, depicts the entire bubble including its cap while the main figure, highlights the bubble `rim'.}}
	\label{fig2}
\end{figure}

    \textcolor{black}{As further motivation of our current study, we note that the bubble whose collapse is described in fig. \ref{fig1}, is nearly spherical initially as its Bond number is low ($<< 1$). The highly deformed, multi-valued initial shape of such a bubble (inset of first panel in fig.~\ref{fig1}) precludes expressing it as Fourier-Bessel series. In contrast, fig. \ref{fig2} (left panel) depicts the bubble shape in the converse limit of large Bond number. Here the bubble shape appears like a cavity albeit with sharp protrusions. Such an initial shape (with some smoothening of the protrusions) is amenable to expression in a Fourier-Bessel series, whose coefficients may be evaluated in time. The cavity treated in this study, may thus be considered a crude approximation to a bubble at high Bond number. For numerical reasons, we have chosen our initial deformation to be a cavity with smooth humps (see fig. ~\ref{fig3}) in contrast to the bubble shape with kinks in panel (a) of fig. ~\ref{fig2}. We emphasize that for such an initial deformation as studied here, the physical origin of the focussing wave train that appears in our simulations is different from that of \cite{gordillo2019capillary}. Consequently, the focussing of the wave-train is not the same as that of the wave-train in classical bursting of bubbles at low $Bo$. However, qualitative similarities in certain aspects may be expected between the two situations and are studied here (see Appendix E).}

    We develop an inviscid nonlinear theory for the focussing of a concentric wave-train resulting from the aforementioned \textit{a priori} imposed free-surface deformation. This theory developed from first principles here has no fitting parameters and helps delineate those aspects of focussing which may be accounted for by linear theory compared to nonlinear features.
    In a series of earlier theoretical and computational studies from our group \citep{farsoiya2017axisymmetric,basak2021jetting,kayal2022dimples,kayal2023jet}, we have solved the initial-value problem corresponding to delocalised, initial interface distortions in the form of a single \textcolor{black}{Bessel function} $(\mj_{0}(kr))$ at gravity dominated large-scales (\cite{kayal2023jet}), gravito-capillary intermediate scales \citep{farsoiya2017axisymmetric,basak2021jetting} and capillarity dominated small-scales (\cite{kayal2022dimples}) (also see the recent study in \cite{dhote2024standing} for a  delocalised initial perturbation on a sessile bubble). In contrast to these studies \textcolor{black}{where the initial perturbation was spatially delocalised}, we study here a \textcolor{black}{spatially} \textit{localised} initial excitation. Apart from the obvious advantage of easier experimental realisation of this (see \cite{ghabache2014liquid} for experiments at gravity dominated scales), this initial condition has the additional advantage that already at linear order, a radially propagating concentric wave-train is obtained and one can ask how does this converge at the axis of symmetry? In contrast, for the single \textcolor{blue}{Bessel function} initial excitation as in \citep{basak2021jetting,kayal2022dimples,kayal2023jet}, at linear order one obtains only a standing wave and it is necessary to proceed to quadratic order and beyond to generate the focussing wave-train.

    The manuscript is structured as follows: \S~\ref{sec:2} illustrates the time evolution of a relaxing cavity and introduces the analytical equations for wave evolution. \S~\ref{sec:3} compares these analytical results with direct numerical simulations (DNS). Finally, the paper culminates with discussions and outlook in \S~\ref{sec:4}.

	\section{Time evolution of a relaxing cavity}
    \label{sec:2}

    \begin{figure}
    	\centering
    	\includegraphics[scale=0.27]{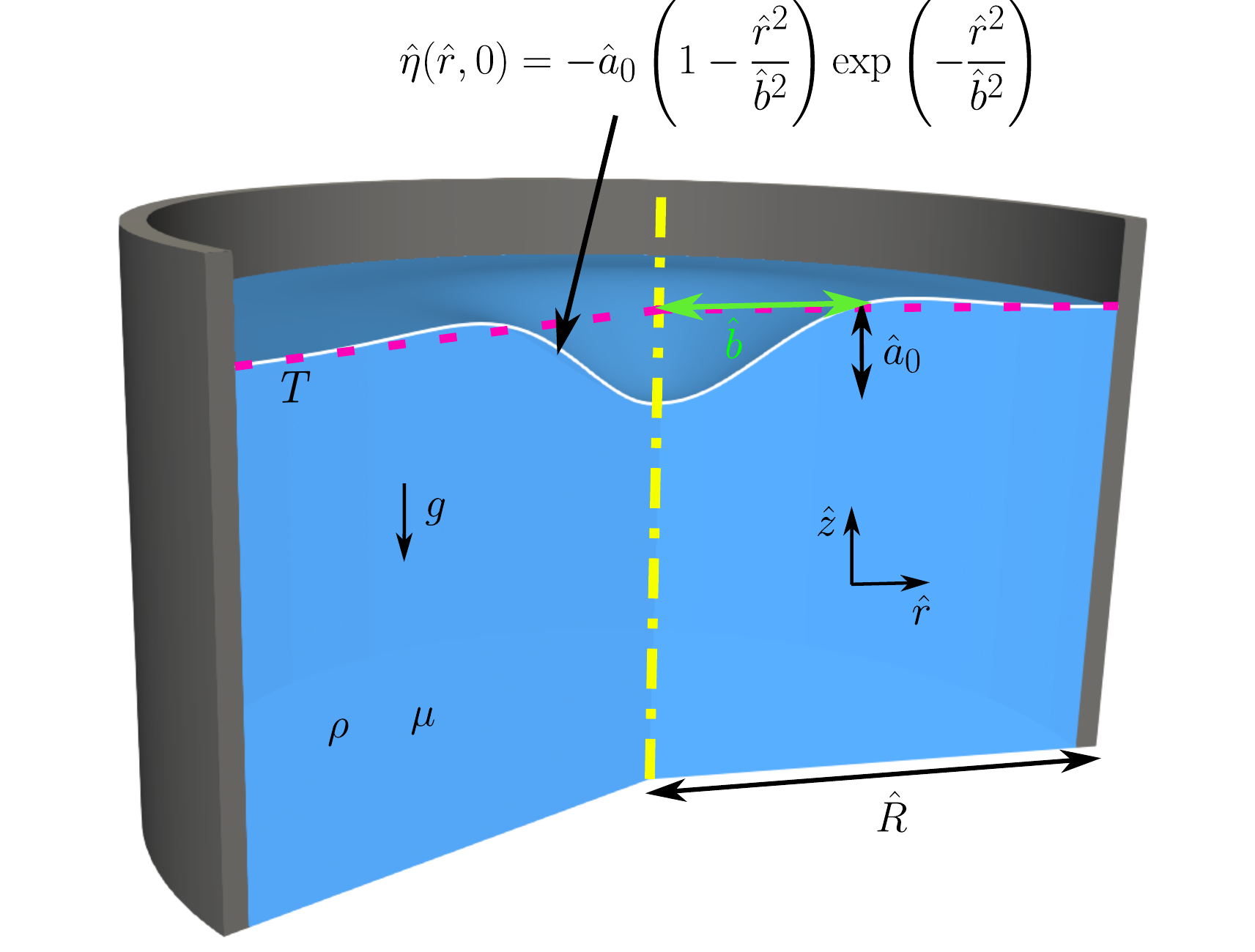}
    	\caption{A (not to scale) cross-sectional representation of the initial interface distortion $\hat{\eta}(\hat{r},0)$ shaped as a cavity of half-width $\hat{b}$ and depth $\hat{a}_0$ in a cylinder of radius $\hat{R}$ filled with liquid (in blue). The functional form chosen for $\hat{\eta}(\hat{r},0)$ was first proposed by \cite{miles1968cauchy} and represents a volume preserving distortion on radially unbounded domain. The red dotted line indicates the unperturbed level of the free-surface of the liquid pool. The gas-liquid surface tension is $T$. Liquid density and viscosity are $\rho$ and $\mu$ respectively, $g$ is gravity. \textcolor{black}{The cavity shape can be considered as a crude representation to the $Bo>>1$ bubble shape in fig. \ref{fig2} with the kinks smoothened drastically. It must be emphasized that our initial condition and the resulting wave-train are significantly different from that of a bursting bubble. However, we intuitively expect that there may be qualitative similarities between the two processes and that it is possible to learn something about one by studying the other, which incidentally has the advantage of analytical tractability.} }
    	\label{fig3}
    \end{figure}

    As shown in fig.~\ref{fig3}, the system consists of a cylindrical container of radius $\hat{R}$ filled with quiescent liquid (indicated in blue). As we do not model the upper fluid in our theory, here onwards the superscript L is dropped from the variables representing fluid properties. For simplicity of analytical calculation, the cylinder is assumed to be infinitely deep and the gas-liquid density ratio is \textcolor{black}{kept} fixed at $0.001$ \textcolor{black}{for DNS only}.
    In our theoretical calculations, we approximate the gas-liquid interface as a free-surface and neglect any motion in the gas phase (\textcolor{black}{although, it is modeled in our DNS}).
    Some of the relevant length scales are the gravito-capillary length $l_c  \equiv \sqrt{T/\rho g} \approx 2.7\,\si{\milli\meter}$ and the visco-capillary length scale $l_\mu \equiv \mu^2/\rho T \approx 0.01\,\si{\micro\meter}$.
    For our chosen \textcolor{black}{half-width of the initial} interface perturbation \textcolor{black}{($\hat{b}=8.0\,\si{\milli\meter}$)}, these length scales justify the inclusion of both capillarity as well as gravity in the theoretical calculation while neglecting viscosity at the leading order.
    However, we stress that viscosity is known to have a non-monotonic effect on wave focussing in a collapsing bubble, as demonstrated by \cite{ghabache2014physics}. Their fig. $3$ shows that the jet velocity during bubble bursting varies non-monotonically with increasing viscosity. Thus, the fastest jets occur not in an inviscid system but at an `optimal' viscosity. In what follows, we employ potential flow equations in our theory and do not treat the boundary layers expected to be generated at the air-water interface and the cylinder walls \citep{mei1973damping}. We will address the inclusion of viscous effects later in the study.

     Before delving into the theoretical formulation, it is instructive to discuss the phenomenology of the problem. Fig.~\ref{fig4}, panels (a)-(i) depict the interface at various time instants as obtained from DNS. These are obtained by solving the inviscid, axisymmetric, and incompressible Euler's equations with surface-tension and gravity in cylindrical coordinates \citep[Basilisk,][]{popinet-basilisk} (script file is available as supplementary material \citep{url}). The images in fig.~\ref{fig4} are obtained by generating the surface of revolution of axisymmetric DNS data. As shown in panel (a), the interface is initially distorted in the shape of an axisymmetric, stationary, and \textit{localised} perturbation. As this cavity relaxes, waves are generated which travel outward reflecting off the wall (between panels (e) and (f)). This produces a wave-train which focusses at the symmetry axis of the container ($r=0$).
     One notes the formation of a small dimple-like structure at the symmetry axis in panel (h). In \S~\ref{sec:3}, we will demonstrate that neither the dimple nor other interface features around the symmetry axis can be explained by the linear theory.

     \subsection{Governing equations: potential flow}

\begin{figure}
    \centering
    \subfloat[t=0]{\includegraphics[scale=0.13]{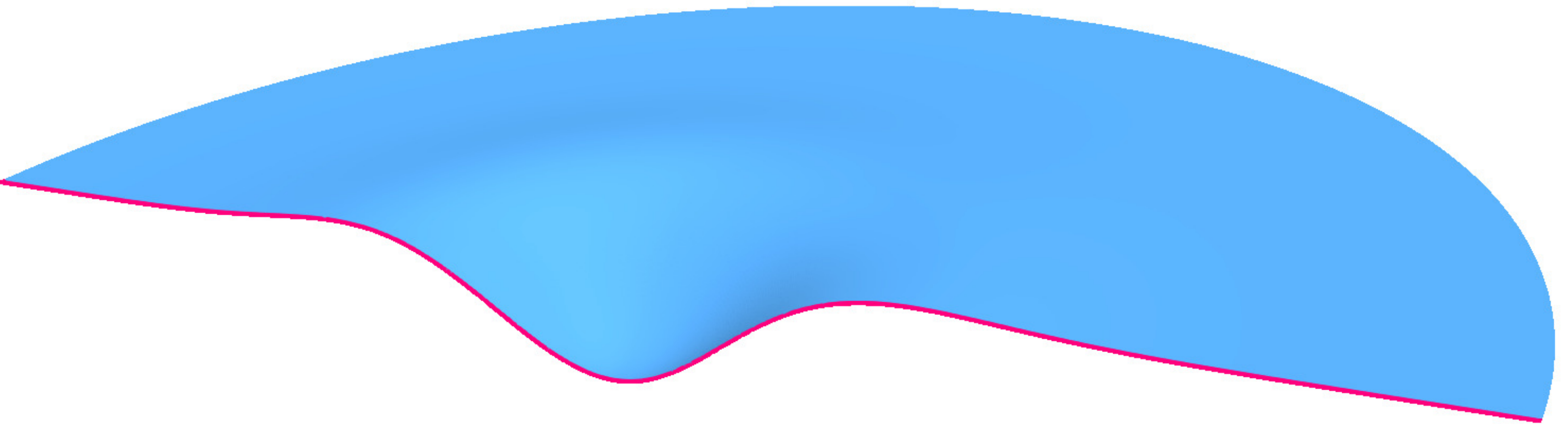}\label{fig4a}}\\
    \subfloat[]{\includegraphics[scale=0.13]{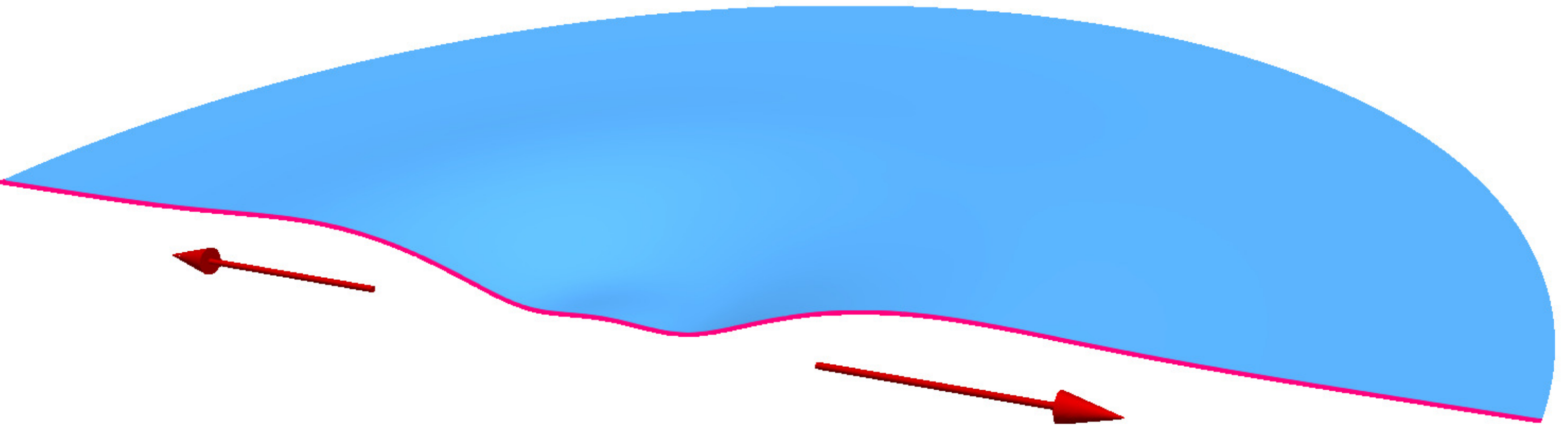}\label{fig4b}}
    \subfloat[]{\includegraphics[scale=0.13]{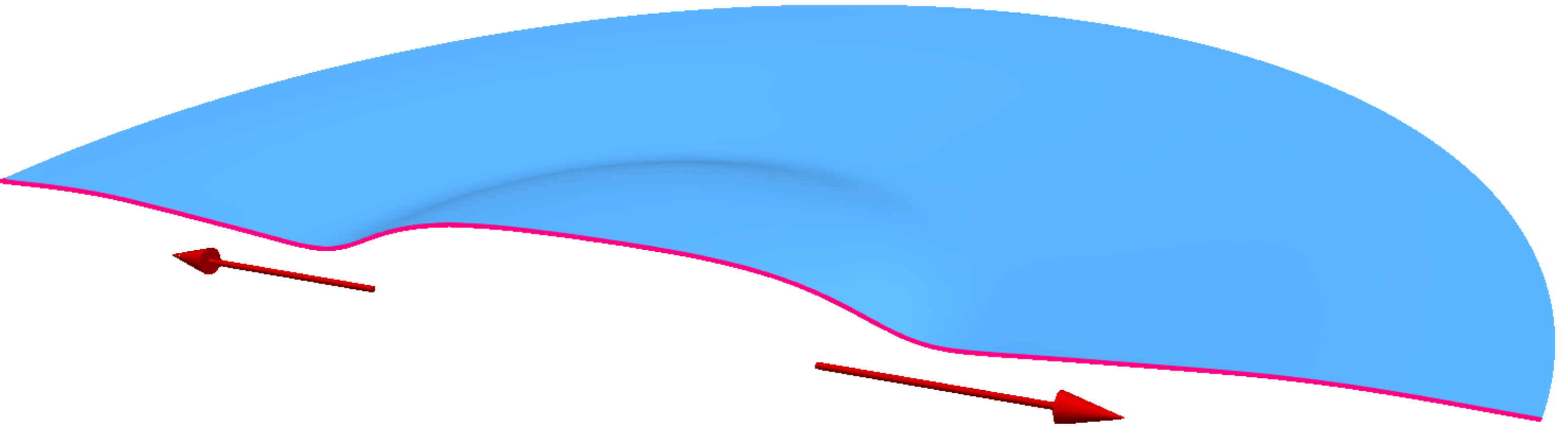}\label{fig4c}}\\
    \subfloat[]{\includegraphics[scale=0.13]{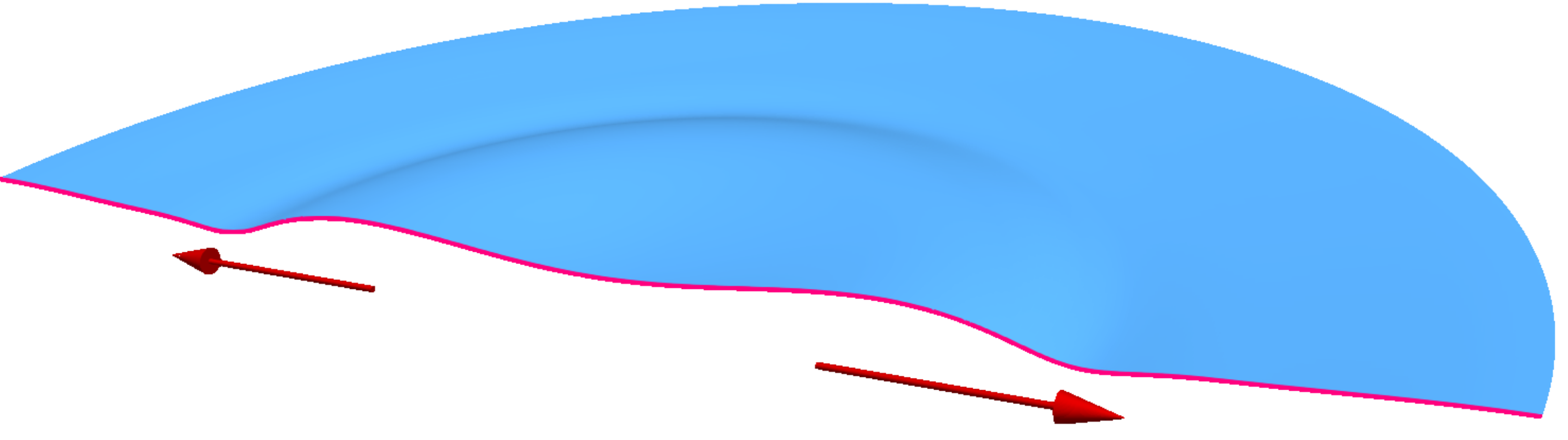}\label{fig4d}}
    \subfloat[]{\includegraphics[scale=0.13]{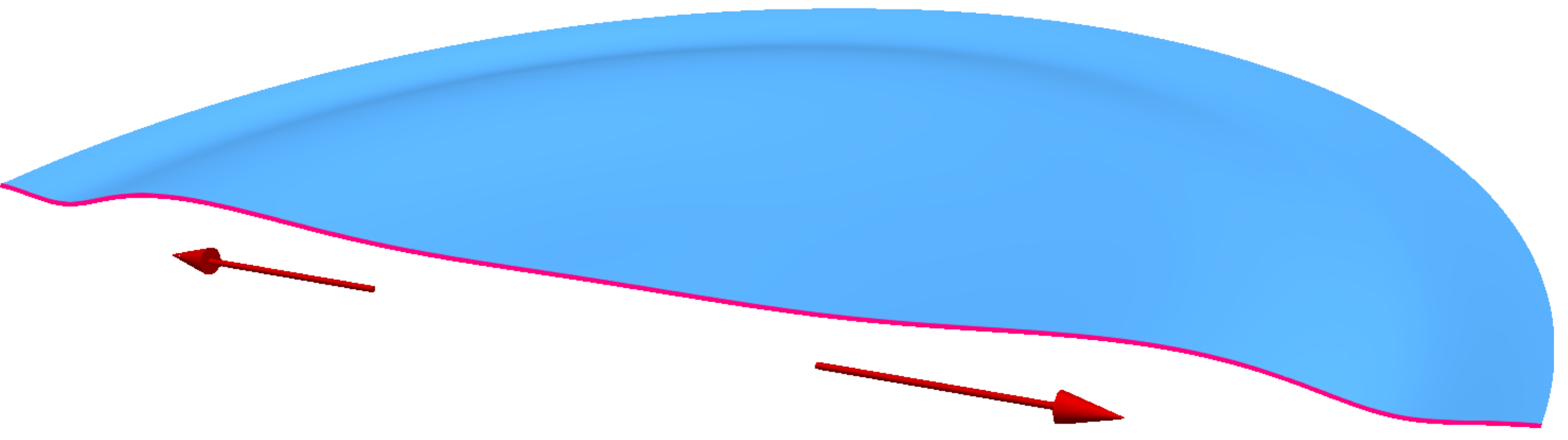}\label{fig4e}}\\
    \subfloat[]{\includegraphics[scale=0.13]{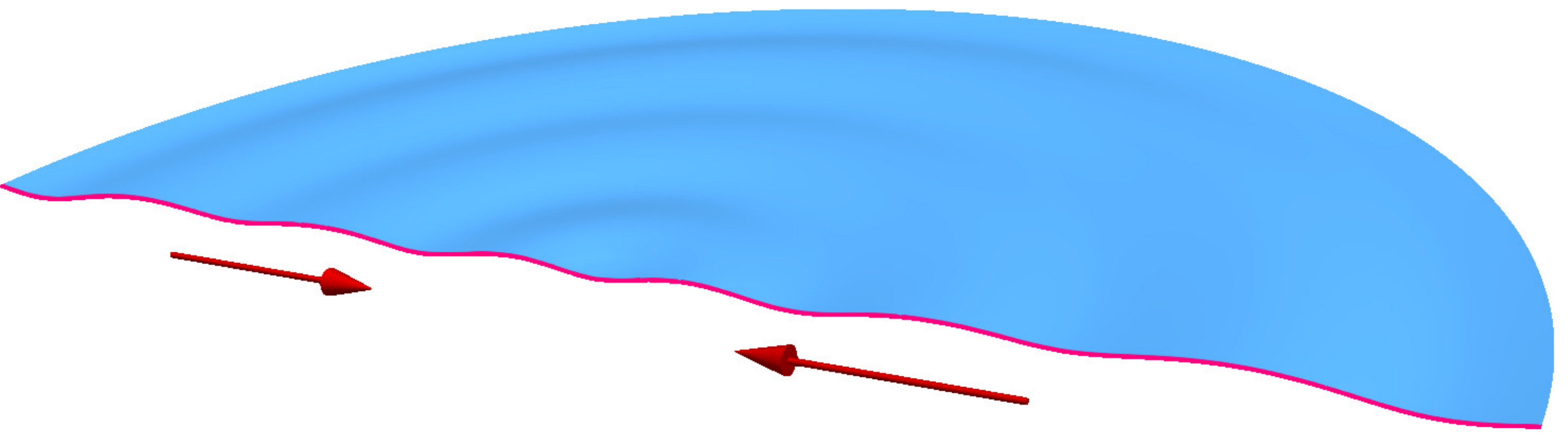}\label{fig4f}}
    \subfloat[]{\includegraphics[scale=0.13]{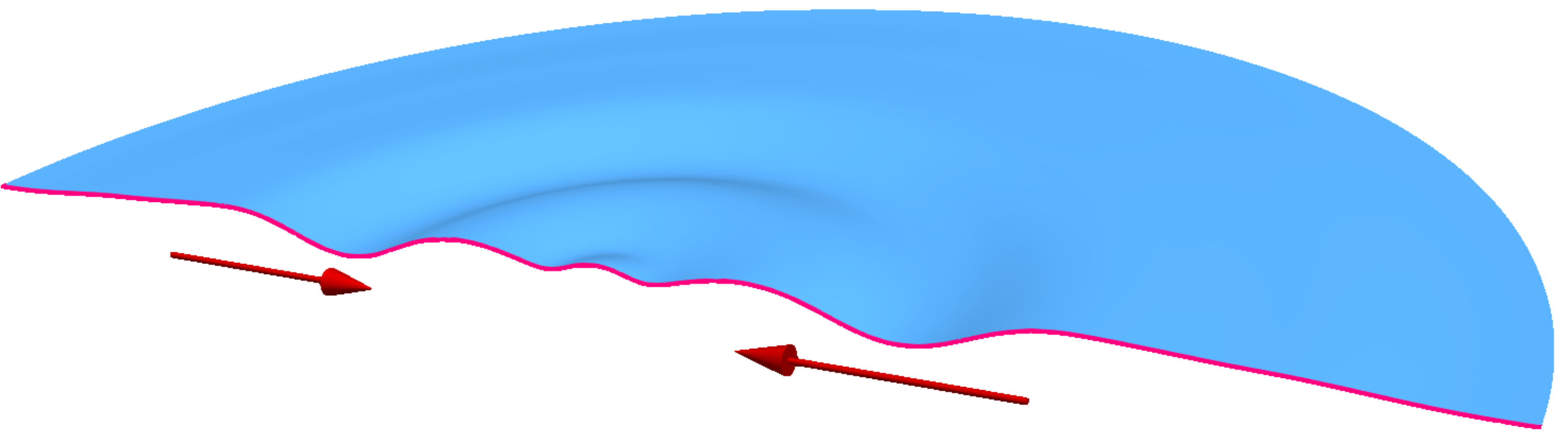}\label{fig4g}}\\
    \subfloat[]{\includegraphics[scale=0.13]{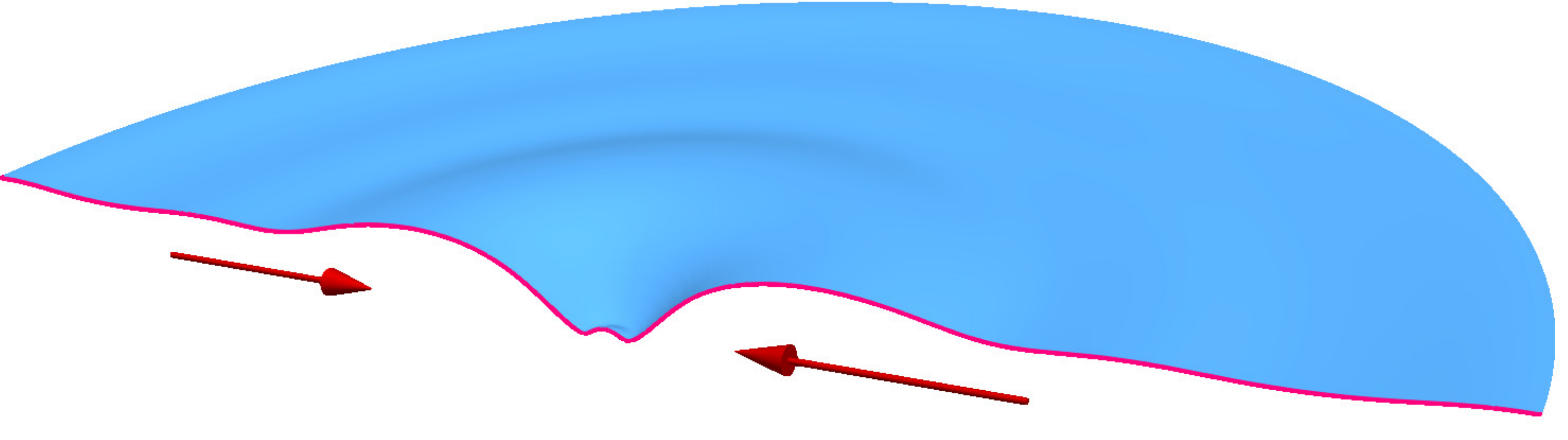}\label{fig4h}}
    \subfloat[]{\includegraphics[scale=0.13]{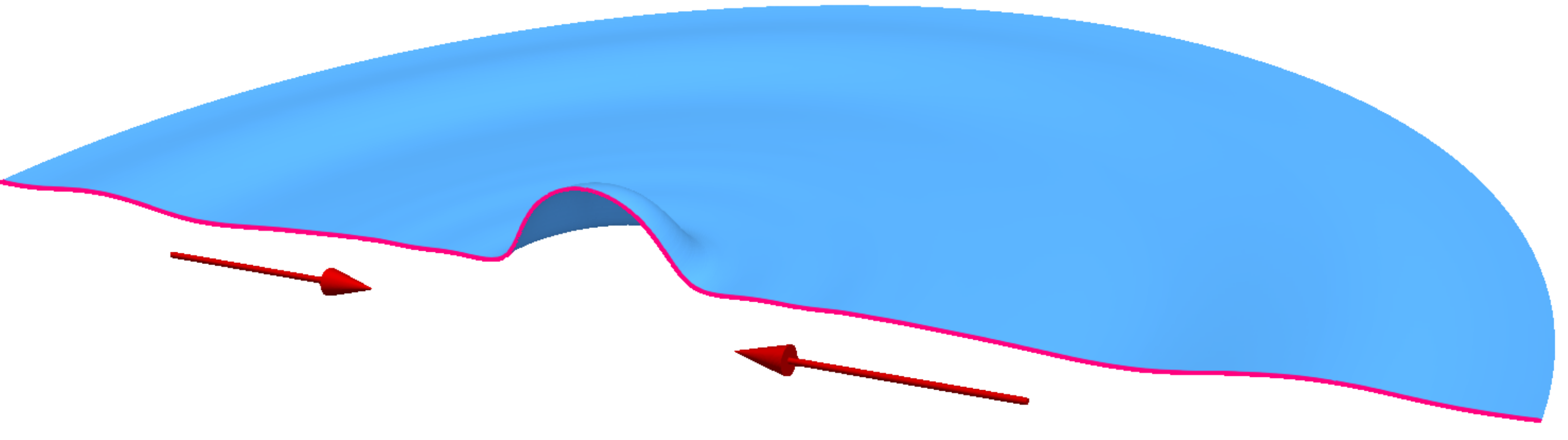}\label{fig4i}}
    \caption{Wave focussing observed in DNS from the cavity-shaped interface distortion at $t=0$ (panel (a)). The figure is to be read left to right and top to bottom for progression of time. After the waves reflect off the cylinder wall (between panels (e) and (f); the confining walls are not shown), they focus inwards towards $r=0$ producing strongly nonlinear oscillations of increasing amplitude. The arrows indicate the instantaneous direction of wave motion. The DNS parameters may be read from Case $1$ in table \ref{tab:sim_params}.}
    \label{fig4}
    \end{figure}

     We now turn to the theoretical analysis of the phenomenology illustrated in fig.~\ref{fig4}. In the base state, we consider a quiescent pool of liquid with density $\rho$ and surface tension $T$ contained in a cylinder of radius $\hat{R}$. For analytical simplicity, we assume this pool is infinitely deep compared to the wavelength of the excited interface waves.
     For further simplicity, we assume that the solid-liquid contact angle at the cylinder wall is always fixed at $\pi/2$ and the contact line is free to move ($\partial_nv_t = 0$). This is the simplest contact line condition which allows for reflection of waves at the boundary without complicating the analytical treatment of the problem \citep{snoeijer2013moving}. The variables $\hat{\eta}\left(\hat{r},\hat{t}\right)$ are used to represent the axisymmetric perturbed interface (see fig. \ref{fig1}) and $\hat{\phi}\left(\hat{r},\hat{z},\hat{t}\right)$ is the disturbance velocity potential; $\hat{r}$ and $\hat{z}$ being the radial and axial coordinates in cylindrical geometry respectively.
     Variables with the dimensions of length (e.g. $\hat{r},\hat{z},\hat\eta$) and time ($\hat{t}$) are scaled using length and time-scales $L \equiv \hat{R}$ and $T_0 \equiv \sqrt{\frac{\hat{R}}{g}}$, respectively. The velocity potential $\hat{\phi}$ is non-dimensionalised using the scale $L^2/T_0$. Under the potential flow approximation, the nondimensional governing equations and boundary conditions governing perturbed quantities are,
    \begin{subequations}\label{eq1}
    	\begin{align}
    	&\frac{\partial^2\phi}{\partial r^2}+\frac{1}{r}\frac{\partial\phi}{\partial r}+\frac{\partial^2\phi}{\partial z^2} = 0, \tag{\theequation a} \\
	   	&\frac{\partial\eta}{\partial t}+\left(\frac{\partial\eta}{\partial r}\right)\left(\frac{\partial\phi}{\partial r}\right)_{z=\eta}  - \left(\frac{\partial\phi}{\partial z}\right)_{z=\eta} = 0, \tag{\theequation b} \\
	   	&\left(\frac{\partial\phi}{\partial t}\right)_{z=\eta}+\eta+\frac{1}{2}\left\lbrace\left(\frac{\partial\phi}{\partial r}\right)^2+\left(\frac{\partial\phi}{\partial z}\right)^2\right\rbrace_{z=\eta} - \frac{1}{Bo}\left\lbrace\frac{\frac{\partial^2\eta}{\partial r^2}}{\left\lbrace 1+\left(\frac{\partial\eta}{\partial r}\right)^2\right\rbrace^{\frac{3}{2}}}+\frac{1}{r}\frac{\frac{\partial\eta}{\partial r}}{\left\lbrace1+\left(\frac{\partial\eta}{\partial r}\right)^2\right\rbrace^{\frac{1}{2}}}\right\rbrace =0, \tag{\theequation c} \\
    	&\int_{0}^{1}r\eta(r,t)dr=0,\quad \left(\frac{\partial\phi}{\partial r}\right)_{r=1}= 0,\tag{\theequation d,e} \\
	   	&\lim_{z\to-\infty}\phi \to \text{finite}\tag{\theequation f} \\
	   	& \eta(r,t=0) = -\varepsilon\left(1 - \frac{r^2}{b^2}\right)\exp\left(-\frac{r^2}{b^2}\right) = \sum_{m=1}^{N}\eta_m(0)\mj_{0}(k_m r),\;\; \frac{\partial\phi}{\partial n}(r,z = \eta(r,0),t=0) = 0,  \tag{\theequation g,h}
    	\end{align}
    \end{subequations}

    \noindent where $\varepsilon >0$ and $n$ in eqn.~\ref{eq1}h is a distance coordinate measured normal to the free-surface at $t=0$. The dimensionless parameters are defined as follows: $\frac{1}{Bo} \equiv \alpha \equiv \dfrac{T}{\rho g \hat{R}^2}$, representing the inverse Bond number (based on the cylinder radius);
    $b \equiv \dfrac{\hat{b}}{\hat{R}}$ is the dimensionless measure of cavity width; and $\varepsilon \equiv \dfrac{\hat{a}_0}{\hat{R}}$ is the dimensionless measure of cavity depth (see Fig. \ref{fig3} caption for the meaning of the symbols). Here onwards, we use $\alpha$ to represent the inverse Bond number.

    In cylindrical, axisymmetric coordinates. eqn.~\ref{eq1}a is the Laplace equation, \ref{eq1}b and \ref{eq1}c are the kinematic boundary condition and the Bernoulli equation applied at the free surface respectively. Eqn.~\ref{eq1}d restricts initial interfacial distortions to those which are volume conserving while \ref{eq1}e enforces no-penetration at the cylinder wall. Eqn.~\ref{eq1}f is the finiteness condition at infinite depth.

\begin{figure}
        \centering
        \subfloat[Cavity shape at $t=0$]{\includegraphics[scale=0.23]{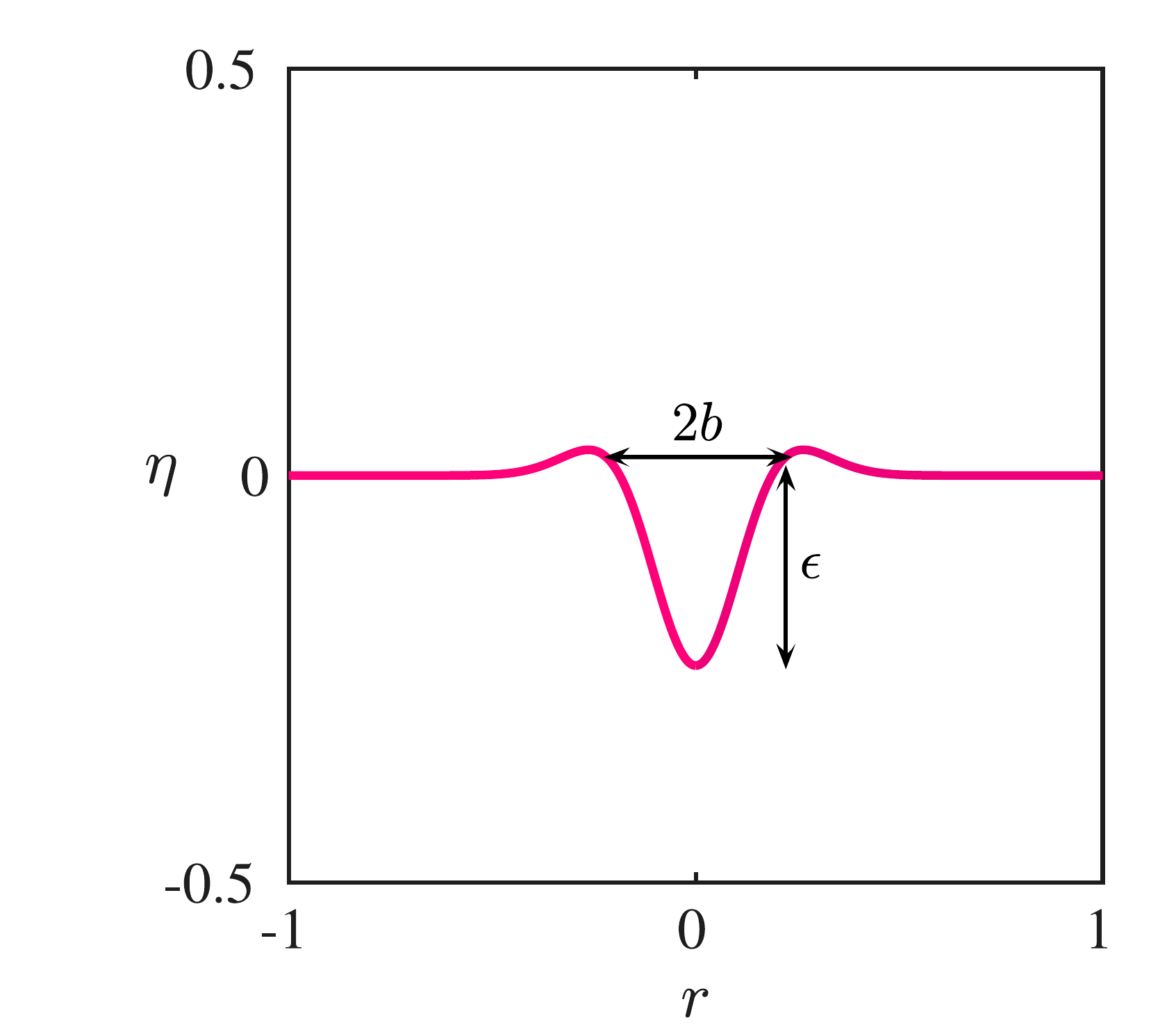}\label{fig5a}}\quad\quad
        \subfloat[Bessel function coefficients]{\includegraphics[scale=0.23]{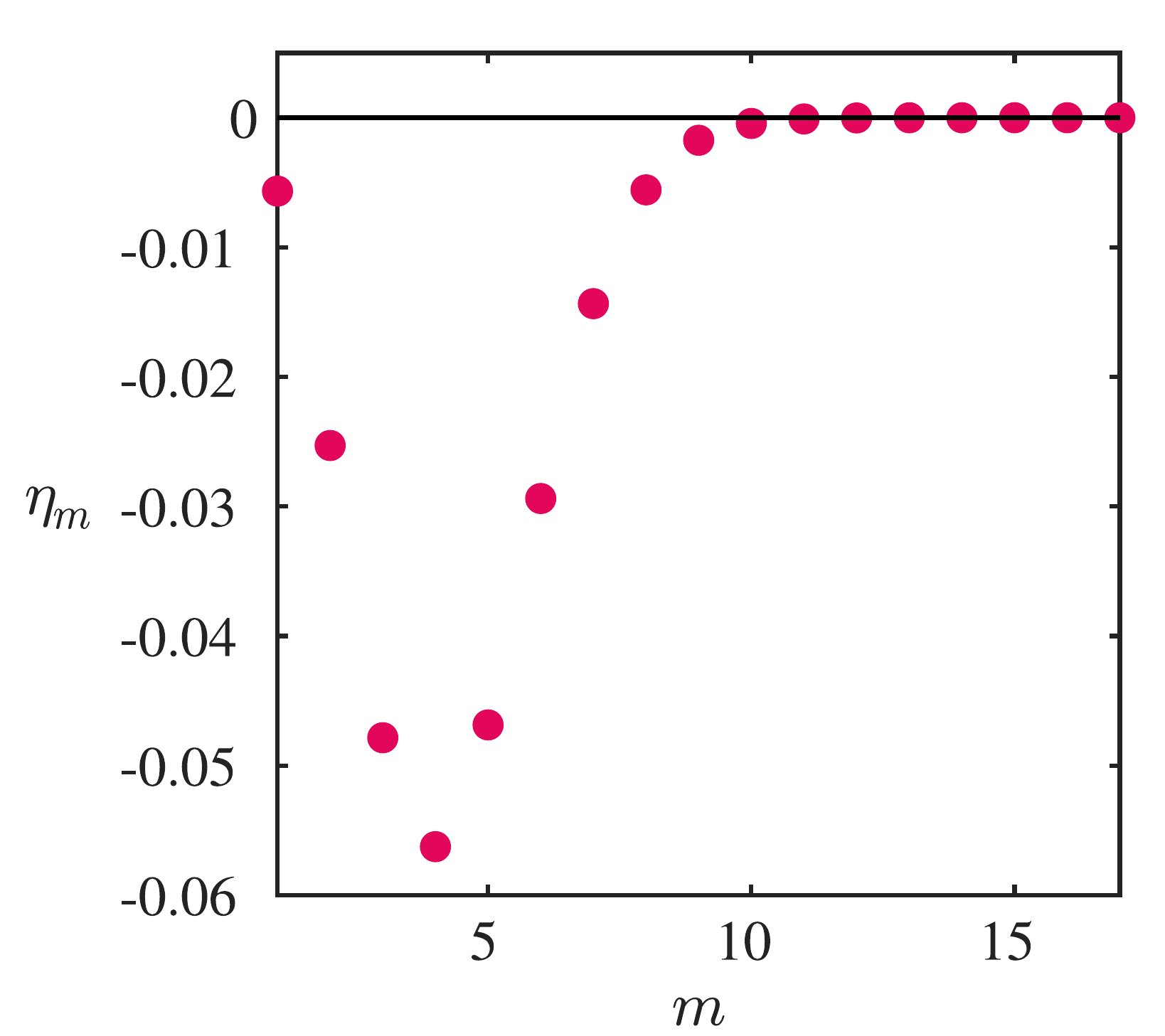}\label{fig5b}}
        \caption{Panel (a) The gas-liquid interface initially deformed as a cavity of half-width $b = \dfrac{\hat{b}}{\hat{R}}$ and depth $\varepsilon \equiv \dfrac{\hat{a_0}}{\hat{R}}$. Panel (b) The cofficients $\eta_m(0)$ are obtained by decomposing the initial distorted interface. For this initial distortion, $\varepsilon=0.091,b=0.187$. It is seen that only the first ten or so \textcolor{black}{Bessel functions  / wavenumbers} are excited initially. For accuracy, we consider the energy in the first seventeen  initially ($m=1,2,3\ldots17$).}
        \label{fig5}
   \end{figure}

    Eqns.~\ref{eq1} g \& h represent the initial conditions. We decompose the initial interface distortion i.e. $\eta(r,t=0)=-\varepsilon\left(1 - \frac{r^2}{b^2}\right)\exp\left(-\frac{r^2}{b^2}\right)$ \citep{miles1968cauchy}, into its \textcolor{black}{Fourier-Bessel} series as indicated by the second equality sign in eqn. \ref{eq1}g and $\mj_1(k_m)=0$ for $m \in \mathbb{Z}^{+}$ \textcolor{black}{(from eqn. \ref{eq1}e; note the identity $\mathrm{J}_0'(\cdot)=-\mathrm{J}_1(\cdot)$, prime indicating derivative)}. The numerical values of the coefficients $\eta_m$ at $t=0$ i.e. $\eta_m(0)  \,(m=1,2,3\ldots)$ in eqn. \ref{eq1}g are determined from the orthogonality relation between Bessel functions i.e. $\eta_m(0)=\frac{\int_0^{1}dr\;r\textrm{J}_0(k_m r)\eta(r,0)}{\int_0^{1}dr \;r\textrm{J}_0^2(k_m r)}$. A sample representation of the initial condition and its \textcolor{black}{Fourier-Bessel} coefficients is presented in fig. \ref{fig5}a and \ref{fig5}b respectively where it is seen that about $17$ \textcolor{black}{wavenumbers} are excited initially. Subject to these initial and boundary conditions presented in eqns. \ref{eq1} a-h, we need to determine the amplitudes $\eta_m(t),\, m=1,2,3\ldots$ as a function of time and this is carried out next.

    \subsection{Equations for $\eta_j(t)$}

    In this section we solve the initial, boundary-value problem posed in eqns.~\ref{eq2} a-h. We derive equations governing the time evolution of the coefficients $\eta_j(t)$ upto quadratic order (i.e. terms which are cubic or higher in the coefficients are neglected). The approach for doing this is classical and was laid out in \cite{hasselmann1962non} in Cartesian coordinates although their initial conditions were random functions in contrast to the deterministic initial distortion posed in eqn. \ref{eq1}g. The procedure below closely follows the approach of \cite{nayfeh1987surface}, who derived similar equations (his eqns. $14$ and $15$) in the context of the Faraday instability (i.e. with vertical oscillatory forcing) including gravity but not surface tension \citep{nayfeh1987surface} in his analysis. In contrast to forced waves being studied by \cite{nayfeh1987surface}, we consider free waves in our current study and include both surface-tension and gravity in the analysis. We first expand $\phi$ and $\eta$ in eqns. \ref{eq1} as

    \begin{subequations}\label{eq2}
        \begin{align}
        \phi(r,z,t)=\sum_{m=1}^\infty \phi_m(t)\textrm{J}_0(k_mr)\exp(k_mz), \quad\quad\eta(r,t)=\sum_{m=1}^\infty \eta_m(t)\textrm{J}_0(k_mr) \tag{\theequation a,b}
        \end{align}
    \end{subequations}

   \noindent By construction, each term in the expansion in \ref{eq2} satisfies the Laplace equation \ref{eq1}a, eqns. \ref{eq1} (d) (\textcolor{black}{the integral mass conservation condition evaluates to be numerically very small for the chosen paramters}) and (e) as well as the finiteness condition \ref{eq1}f. Taylor expanding eqns. \ref{eq1}b and c about $z=0$ we obtain

    \begin{subequations}\label{eq3}
    \begin{align*}
         &\frac{\partial\eta}{\partial t}-\left(\frac{\partial\phi}{\partial z}\right)_{z=0}-\left(\frac{\partial^2\phi}{\partial z^2}\right)_{z=0}\eta+\frac{\partial\eta}{\partial r}\left(\frac{\partial\phi}{\partial r}\right)_{z=0}+\text{H.O.T}=0 \tag{\theequation a} \\
         &\left(\frac{\partial\phi}{\partial t}\right)_{z=0}+\eta\left(\frac{\partial^2\phi}{\partial t \partial z}\right)_{z=0}+\eta + \frac{1}{2}\left\lbrace\left(\frac{\partial\phi}{\partial r}\right)^2+\left(\frac{\partial\phi}{\partial z}\right)^2\right\rbrace_{z=0} \\
         &-\alpha\left\lbrace\frac{\partial^2\eta}{\partial r^2}+\frac{1}{r}\frac{\partial\eta}{\partial r}\right\rbrace
         + \text{H.O.T}=0 \tag{\theequation b}
         \end{align*}
    \end{subequations}

    \noindent where $\text{H.O.T}$ represents higher order terms. Substituting expansions \ref{eq2}a \& b into \ref{eq3}a,b and using orthogonality relations between Bessel functions we obtain for $n,p,m \in \mathbb{Z}^{+}$

    \begin{subequations}\label{eq4}
        \begin{align}
        &\frac{d\eta_n}{d t}-k_n\phi_n(t)+\sum_{m,p}\left(D_{npm}-k_m^2C_{npm}\right)\phi_m(t)\eta_p(t)=0 \tag{\theequation a}\\
        &\frac{d\phi_n}{d t}+(1+\alpha k_n^2)\eta_n(t)+\sum_{m,p}k_m C_{npm}\left(\frac{d\phi_m}{d t}\right)\eta_p(t)+\frac{1}{2}\sum_{m,p}\left(D_{npm}+k_mk_pC_{npm}\right)\phi_m(t)\phi_p(t)=0 \tag{\theequation b} \\
        & n = 1,2,3 \ldots \nonumber
        \end{align}
    \end{subequations}

    \noindent The nonlinear interaction coefficients $C_{npm}$ and $D_{npm}$ in eqn. \ref{eq4} are related as \citep{nayfeh1987surface}:

    \begin{eqnarray}\label{eq5}
    D_{npm}=\frac{1}{2}\left(k_p^2+k_m^2-k_n^2\right)C_{npm}
    \end{eqnarray}

    \noindent and $C_{npm}=\dfrac{\int_{0}^1 r\textrm{J}_0(k_nr)\textrm{J}_0(k_pr)\textrm{J}_0(k_mr)dr}{\int_{0}^1r\textrm{J}^2_0(k_nr)dr}$.
    For the benefit of the reader, the detailed proof of \ref{eq5} is provided in Appendix A. Retaining self-consistently up to quadratic order terms, eqns. \ref{eq4} a and b may be combined into a second order equation for $\eta_n$ alone. This is:

    \begin{eqnarray}\label{eq6}
    &&\dfrac{d^2\eta_n}{d t^2}+\omega_n^2\eta_n+k_n\sum_{m,p}\left[1+\frac{k_p^2-k_m^2-k_n^2}{2k_mk_n}\right]C_{npm}\left(\dfrac{d^2\eta_m}{d t^2}\right)\eta_p \nonumber\\
    &+&\dfrac{1}{2}k_n\sum_{m,p}\left[1+\dfrac{k_p^2+k_m^2-k_n^2}{2k_mk_p}+\dfrac{k_p^2-k_m^2-k_n^2}{k_mk_n}\right]C_{npm}\left(\dfrac{d\eta_m}{d t}\right)\left(\frac{d\eta_p}{d t}\right)=0
    \end{eqnarray}

     \noindent Note that $\omega_n$ is the linear oscillation frequency of the $n^{\text{th}}$ mode, viz. $\omega_n\equiv \sqrt{k_n\left(1+\alpha k_n^2\right)}$ (\textcolor{black}{the effect of the nonlinear terms due to curvature in eqn. \ref{eq3}b and thus surface-tension, appears only through the linear order dispersion relation up to second order}). We solve the coupled ordinary differential eqns. \ref{eq6} numerically subject to the initial conditions discussed earlier for $n=1,2,3..\ldots 34$ (i.e. twice the initial number, see fig. \ref{fig5b}) using ‘DifferentialEquations.jl’, an open-source package by \cite{rackauckas2017differentialequations} and collaborators. The `DifferentialEquations.jl` automatically chooses an ODE solver based on stiffness detection algorithms as described by \cite{rackauckas2019confederated}. The Julia script file can be found in \cite{url}. We note that while numerically solving eqn. \ref{eq6}, we compute $\frac{d^2\eta_m}{d t^2}$ in the third term of the equation (the nonlinear term) via the linear estimate, viz, $\frac{d^2\eta_m}{d t^2}=-\omega_m^2 \eta_m$. Interestingly, the solution to eqn. \ref{eq6} shows instability \textcolor{black}{albeit only at large time (compared to the focussing time) when high wavenumbers ($k$) appear in our model. This instability could either be numerical or physical and possibly related to instability of finite-amplitude capillary waves. Further investigations are necessary to acertain the origin of this and is outside the scope of this study. As the instability occurs outside the time window of our study, it does not impact the results presented in this work. We thus restrict ourselves to numerical solutions to eqn. \ref{eq6} within the time period of our interest where this instability does not appear.}

\begin{figure}
		\centering
		\subfloat[t = 0.303]{\includegraphics[scale=0.13]{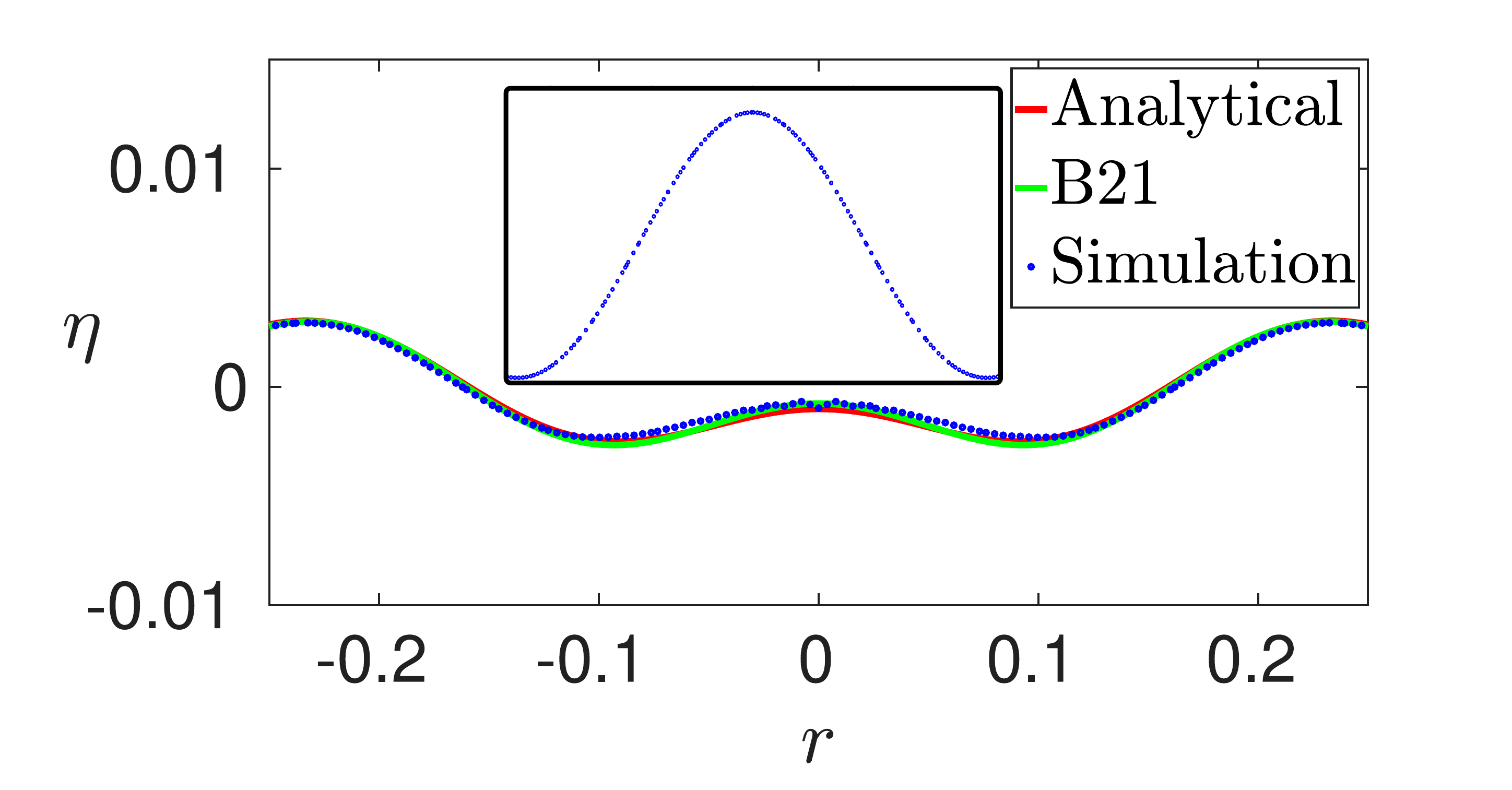}\label{fig6a}}
		\subfloat[t= 0.409]{\includegraphics[scale=0.13]{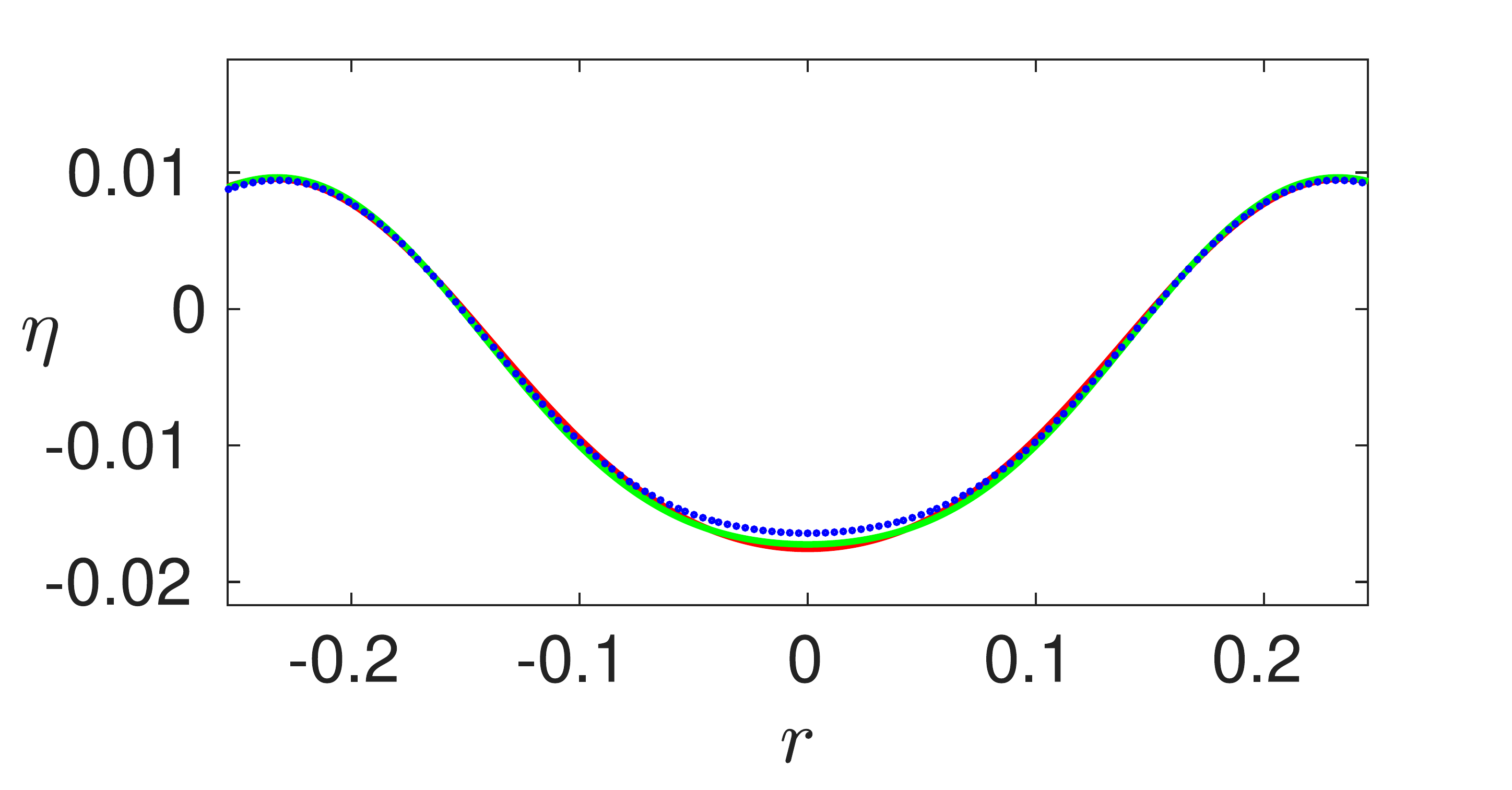}\label{fig6b}}\\
		\subfloat[t = 0.772]{\includegraphics[scale=0.13]{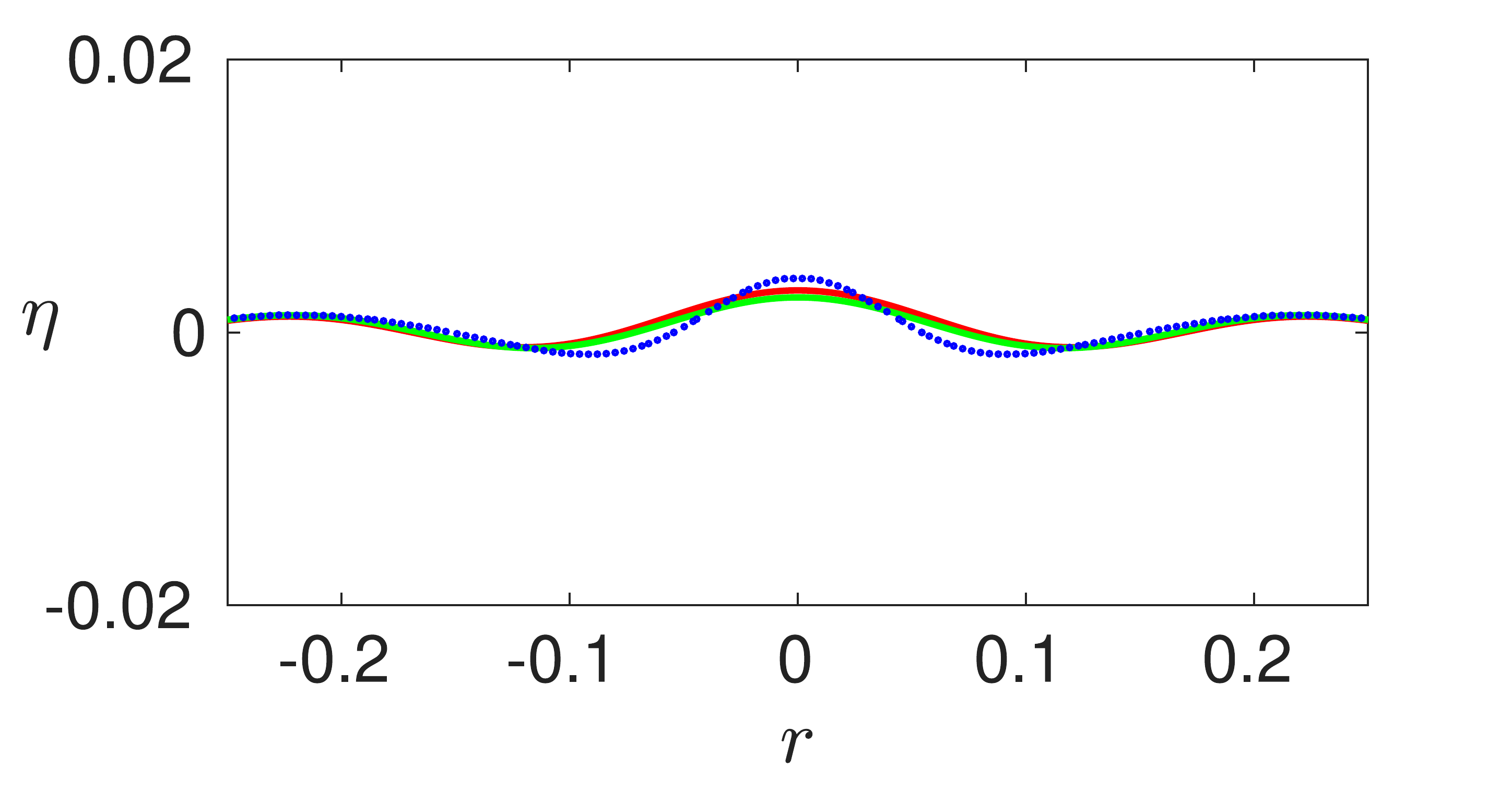}\label{fig6c}}
		\subfloat[t=1.029]{\includegraphics[scale=0.13]{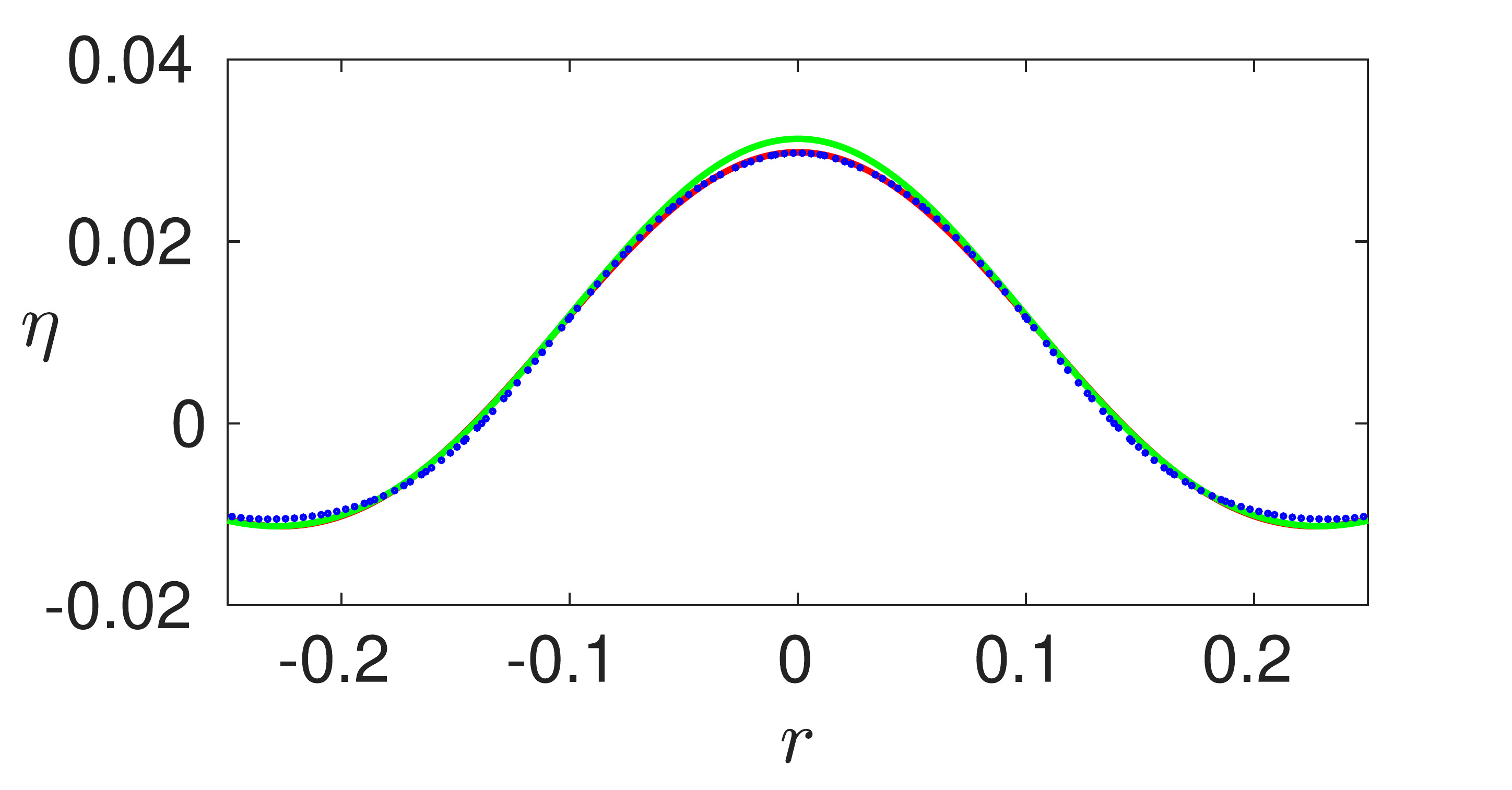}\label{fig6d}}
		\caption{Benchmarking of our solution procedure for solving the coupled O.D.E's in eqns. \ref{eq6} against inviscid DNS (indicated as `Simulation' in the legend of panel (a)) and analytical predictions by \cite{basak2021jetting}, indicated as `B21'. For DNS, the dimensionless parameters are $\varepsilon \equiv \dfrac{a_0}{\hat{R}} = \frac{0.5}{16.4706} = 0.03, \alpha = 0.004$ and $Oh=0$. Note that the initial condition here has a crest around $r=0$, see inset of panel (a).}
		\label{fig6}
	\end{figure}

     As benchmarking of our numerical solution procedure, we first solve eqns.~\ref{eq6} employing the single \textcolor{black}{Bessel function} initial surface distortion that was studied in \cite{basak2021jetting} i.e. in our current notation $\eta(r,t=0) = \varepsilon \mj_{0}(l_5\;r),\;\varepsilon > 0$ where $l_5=16.4706$ is the fifth non-trivial root of the Bessel function $\mj_1$. For this initial condition, the second-order accurate solution is expectedly of the form

     \begin{eqnarray}
     	\eta(r,t)= \varepsilon \eta_1(r,t) + \varepsilon^2 \eta_2(r,t)
     	\label{eq7}
     \end{eqnarray}

     \noindent where explicit expressions for $\eta_1$ and $\eta_2$ were provided in \cite{basak2021jetting} (we note the slight difference in non-dimensionalisation of length between the current study and the one by \cite{basak2021jetting} involving a factor of $l_q$). Fig. \ref{fig6}, demonstrates a comparison between the prediction of expression \ref{eq7} (indicated in the figure as `B21' for \cite{basak2021jetting}), the solution obtained from solving eqn. \ref{eq6} with the same initial condition (labelled in the figure as `Analytical') and the numerical simulation from Basilisk (depicted as `Simulation'). Fig. \ref{fig6} demonstrates good agreement between the three, thereby providing confidence on our numerical procedure for solving eqns. \ref{eq6}.

 \section{Comparison of direct numerical simulations (DNS) with theory}
 \label{sec:3}
 \subsection{Description}
\begin{figure}
 	\centering
 	\includegraphics[scale=1]{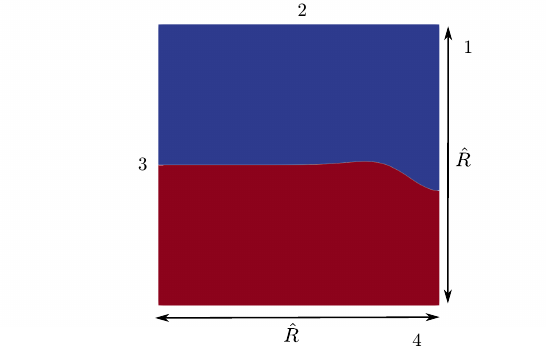}
 	\caption{\textcolor{black}{Simulation domain. Only half of the domain is depicted, due to the axis of symmetry (side $1$). For both viscous as well as inviscid simulations, the boundaries $2,3$ and $4$ are modelled as free-slip walls.}}
 	\label{fig7}
 \end{figure}

\textcolor{black}{ We have used the open source code Basilisk \citep{popinet-basilisk} to solve the Navier-Stokes equation with an interface viz.}
 \begin{eqnarray}
 	&\bm{\nabla.u}=0\label{nava}\\&
 	\frac{\partial\bm{u}}{\partial t}+\bm{\nabla}.(\mathbf{u\bigotimes u})=-\frac{\bm{\nabla}p}{\rho}+g+\frac{T}{\rho}\kappa \delta_s \bm{n}+\nu\nabla^2\bm{u}\label{navb}\\&
 	\frac{\partial f}{\partial t}+\bm{\nabla.}(f\bm{u})=0
 	\label{navc}
 \end{eqnarray}
 \textcolor{black}{Here, $\mathbf{u},\,p,\,\kappa$, $T$ and $f$ are the velocity field, pressure field, interface curvature, surface-tension and the color-function field, respectively. Basilisk is a one-fluid solver where the color function $f$ takes values $0$ and $1$ in the two phases with the interface being represented geometrically using the volume-of-fluid algorithm in cells where $0 < f < 1$. The density and viscosity are represented as weighted average of the respective values of the two phases, employing the color function as the weight.
 Fig. \ref{fig7} depicts the simulation domain, wall $\# 1$ is the symmetry-axis and the liquid and gas are indicated in different colors. We have solved eqns. \ref{nava}, \ref{navb} and \ref{navc} numerically in cylindrical axisymmetric coordinates, using an adaptive mesh based on temporal changes of the color function $f$, and velocity $\mathbf{u}$. Grid resolution for different cases are provided in table \ref{tab:sim_params}. In all the viscous simulations treated in the manuscript, we have used free-slip walls with a $90^{\circ}$ contact angle, in order to be compatible with a freely moving contact line and obviate the well-known contact line singularity \citep{snoeijer2013moving}. By using free-slip conditions, we maintain consistency with the analytical expressions used in our study and facilitate a more direct comparison between our numerical results and theoretical predictions.
This boundary condition naturally enforces a $90^{\circ}$ contact angle at the wall, setting a vanishing gradient for the color function close to the wall, which is consistent with the assumptions in the theoretical model far from the center of the cavity. For discussions, we refer to \citet{Wildeman2016} which shows that the free-slip condition with a $90^{\circ}$ contact angle effectively eliminates dissipation close to the contact line, allowing us to focus on the interfacial dynamics that are central to our study.}
 \subsection{Comparison}

\begin{figure}
        \centering
        \includegraphics[scale=0.65]{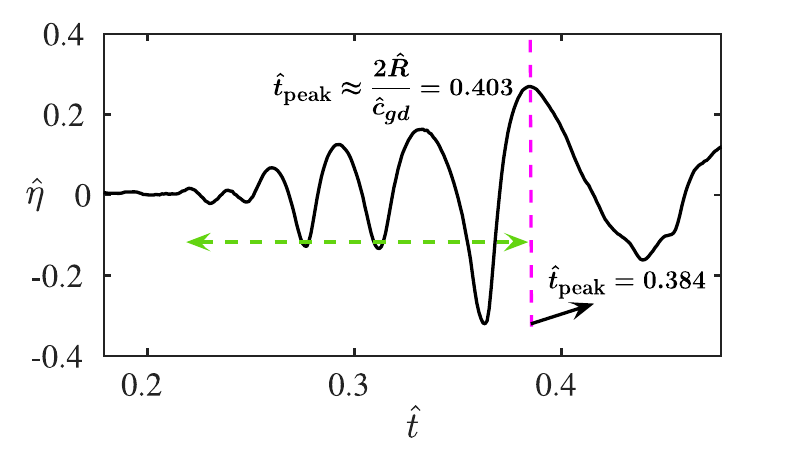}
        \caption{Time signal of the interface at $\hat{r}=0$. The green line indicates approximately the time window when focussing takes place at $\hat{r}=0$}
        \label{fig8}
    \end{figure}

 In this section, we compare results from our direct numerical simulations with the theory discussed in \S~\ref{sec:2}.
 Before this, it is instructive to rationalize the reflection process and estimate its duration. To do this, we observe that the Fourier-Bessel spectrum of the initial interface distortion prominently features a \textcolor{black}{peak} at $m=4$ (see Fig. \ref{fig5b}). A rough estimate of the time required for the energy associated with any \textcolor{black}{wavenumber} excited in the initial spectrum to complete a return trip (from $\hat{r}=0$ to the wall and back) can be derived from linear theory. When this return time is estimated for the dominant wavenumber in the initial spectrum, we expect the numerical value to roughly coincide with the generation time of the largest amplitude oscillation at $\hat{r}=0$ during the focussing process. This is illustrated in Fig. \ref{fig8}, where the time signal from tracking the interface at $\hat{r}=0$ is presented (case 2 in Table \ref{tab:sim_params}). Note that this figure uses dimensional variables, denoted with hats. After the outward travelling waves move away, the interface at $\hat{r}=0$ remains relatively quiescent, as indicated by the nearly flat time signal around $\hat{t}=0.2$ s.
 As a result of reflection, the energy associated with every wavenumber $k$ present initially focusses back to $\hat{r}=0$, this return trip is carried out with its group-velocity $\hat{c}_{g} = \frac{g + 3\left(T/\rho\right) k^2}{2\sqrt{gk + Tk^3/\rho}}$. In fig. \ref{fig5b}, the dominant wavenumber is $k_d = \frac{l_4}{\hat{R}}$ and the largest oscillation at $\hat{r}=0$ during the focussing process is seen to be generated at $\hat{t}_{\text{peak}} = 0.384$ s from fig. \ref{fig8}. Using the linear estimate $\hat{t}_{\text{peak}} \approx \frac{2\hat{R}}{\hat{c}_{gd}}$ where $\hat{c}_{gd}$ is the group-velocity of the dominant wavenumber, we obtain the value $0.403$ s which is reasonably close to the observed  $\hat{t}_{\text{peak}}=0.384$ s.

    \begin{table}
    \begin{center}
        \def~{\hphantom{0}}
        \begin{tabular}{cccccc}
        $\mathrm{Case}$ &  $\varepsilon \equiv \frac{\hat{a}_0}{\hat{R}}$      & $Oh \equiv \frac{\mu}{\sqrt{\rho T \hat{b}}}$ & $\hat{a}_0$ &   $\mu$ & Grid (Maximum)\\\\[3pt]
        1 & 0.061  & 0 & 0.26  & 0 & 10 \\
        2 & 0.091 & 0 & 0.39  & 0 &9,10,11 \\
        3 & 0.091  & $1.17\times 10^{-5}$ & 0.39 & $8.9\times 10^{-5}$ &10\\

        4 & 0.091  & $1.17\times 10^{-4}$ & 0.39 & $8.9\times 10^{-4}$&9,10,11\\
        5 & 0.091 & $1.17\times 10^{-3}$ & 0.39 & $8.9\times 10^{-3}$&10\\
        6 & 0.091 & $1.17\times 10^{-2}$ & 0.39 & $8.9\times 10^{-2}$&10	\\
        7 & 0.006 & 0 & 0.026 & 0\\
        8 & 0.006 & $1.17\times 10^{-5}$ & 0.026 & $8.9\times 10^{-5}$&10\\
        9 & 0.006 & $1.17\times 10^{-4}$ & 0.026 & $8.9\times 10^{-4}$&10\\
        10 & 0.006 & $1.17\times 10^{-3}$ & 0.026 & $8.9\times 10^{-3}$&10\\
        11 & 0.006 & $3.7\times 10^{-3}$ & 0.026 & $2.81\times 10^{-2}$&10\\
        12 & 0.006 & $1.17\times 10^{-2}$ & 0.026 & $8.9\times 10^{-2}$&10\\
        13 & 0.091  & $2.34\times 10^{-4}$ & 0.39 & $1.78\times 10^{-3}$&10\\
        14 & 0.091  & $4.68\times 10^{-4}$ & 0.39 & $3.56\times 10^{-3}$&10 \\
        15 & 0.091 &$2.92\times 10^{-6}$ & 0.39 & $2.22\times 10^{-5}$&10\\
       16 &0.091& $5.85\times 10^{-6}$ & 0.39 & $4.45\times 10^{-5}$&10
        \end{tabular}
        \caption{All dimensional lengths are indicated with a hat. Values are quoted in CGS units. In all of the cases we have used $\hat{R}=4.282$ cm, $\hat{b}=0.8$ cm, $T=72$ dyne$/$cm, $g=-981$ cm$/\text{s}^2$, $\rho=1$ gm$/\textrm{cm}^3$. These imply dimensionless values $b\equiv \frac{\hat{b}}{\hat{R}}=0.187$, $\alpha\equiv\frac{T}{\rho g \hat{R}^2}=0.004$. \textcolor{black}{$Oh$ has been defined using $\hat{b}$, in order to be comparable to its value for a bursting bubble where radius of the bubble is considered for defining $Oh$. One may obtain a new Ohnesorge number $Oh^{'}$ based on $\hat{R}$ by using the formulae $Oh' \equiv \frac{\mu}{\sqrt{\rho T \hat{R}}}=Oh\times b^{1/2}$ with $b\equiv \frac{\hat{b}}{\hat{R}}$. } \textcolor{black}{The maximum grid resolution reported here are in powers of two. The conditions for adaptivity may be found in further detail in the script files \citep{url}}.}
        \label{tab:sim_params}
    \end{center}
    \end{table}

\begin{figure}
    \centering
        \subfloat[$t=0.166$]{\includegraphics[scale=0.12]{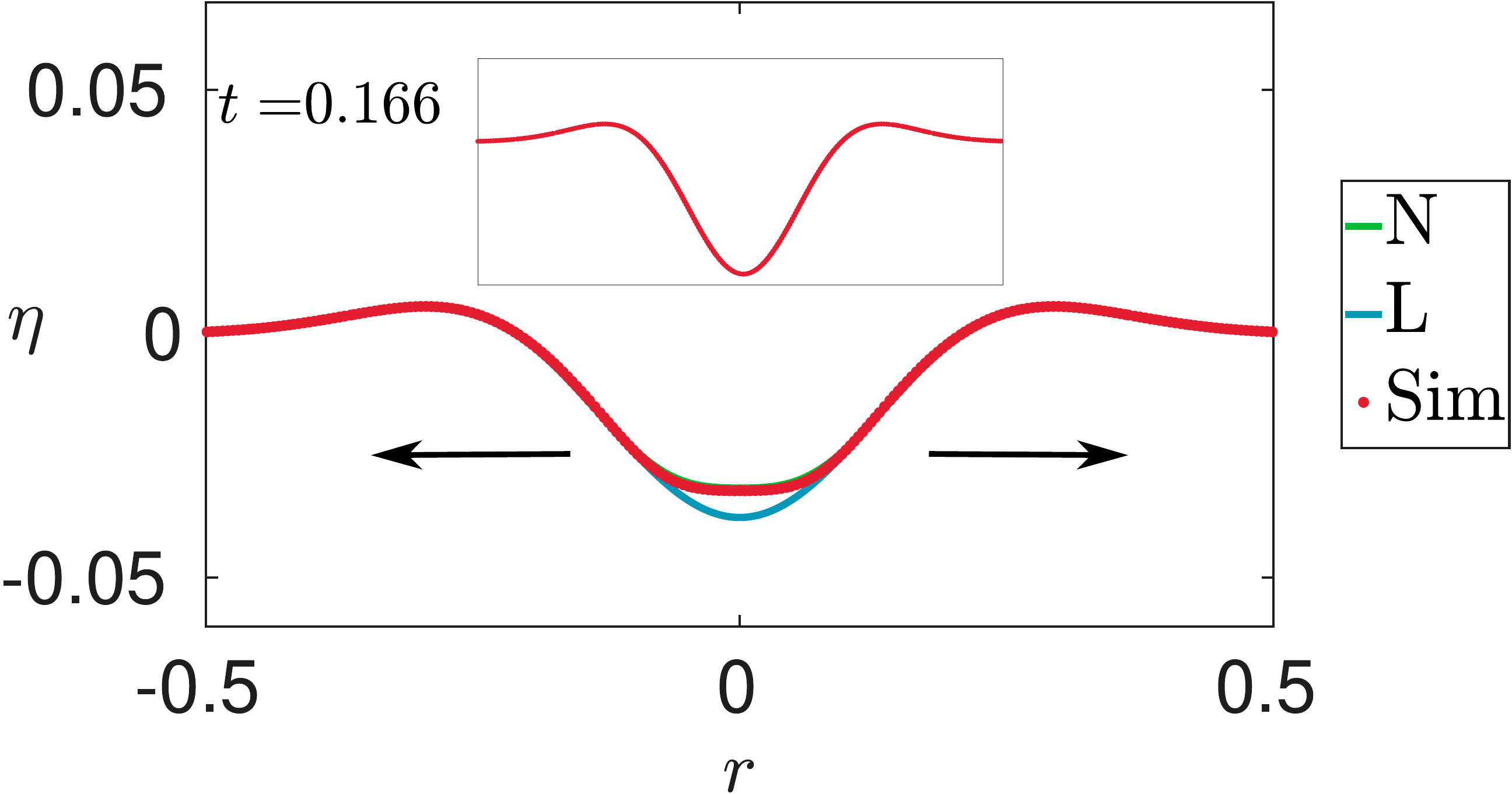}\label{fig9a}}\quad
        \subfloat[$t=0.439$]{\includegraphics[scale=0.12]{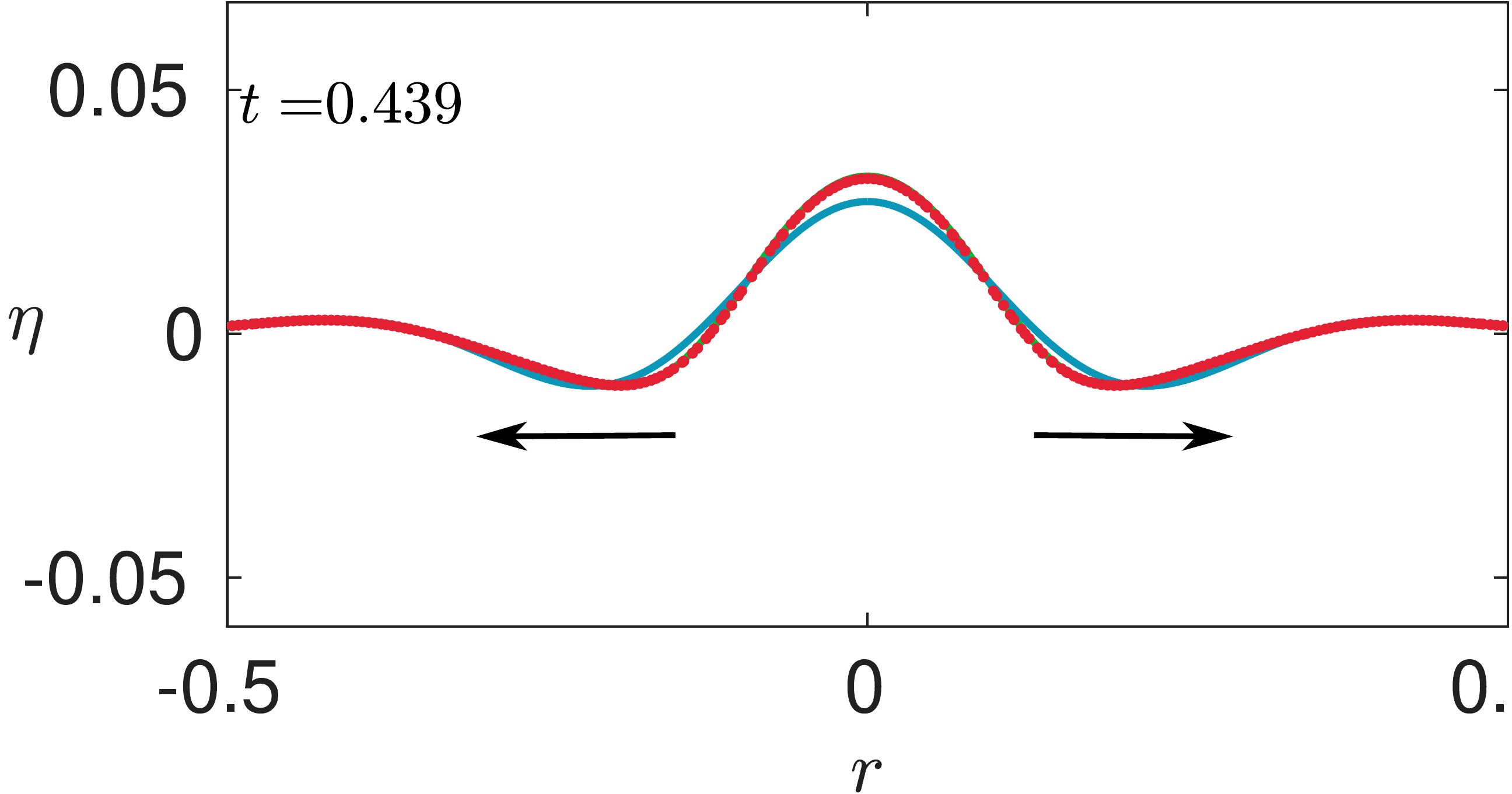}\label{fig9b}}\\
        \subfloat[$t=1.075$]{\includegraphics[scale=0.12]{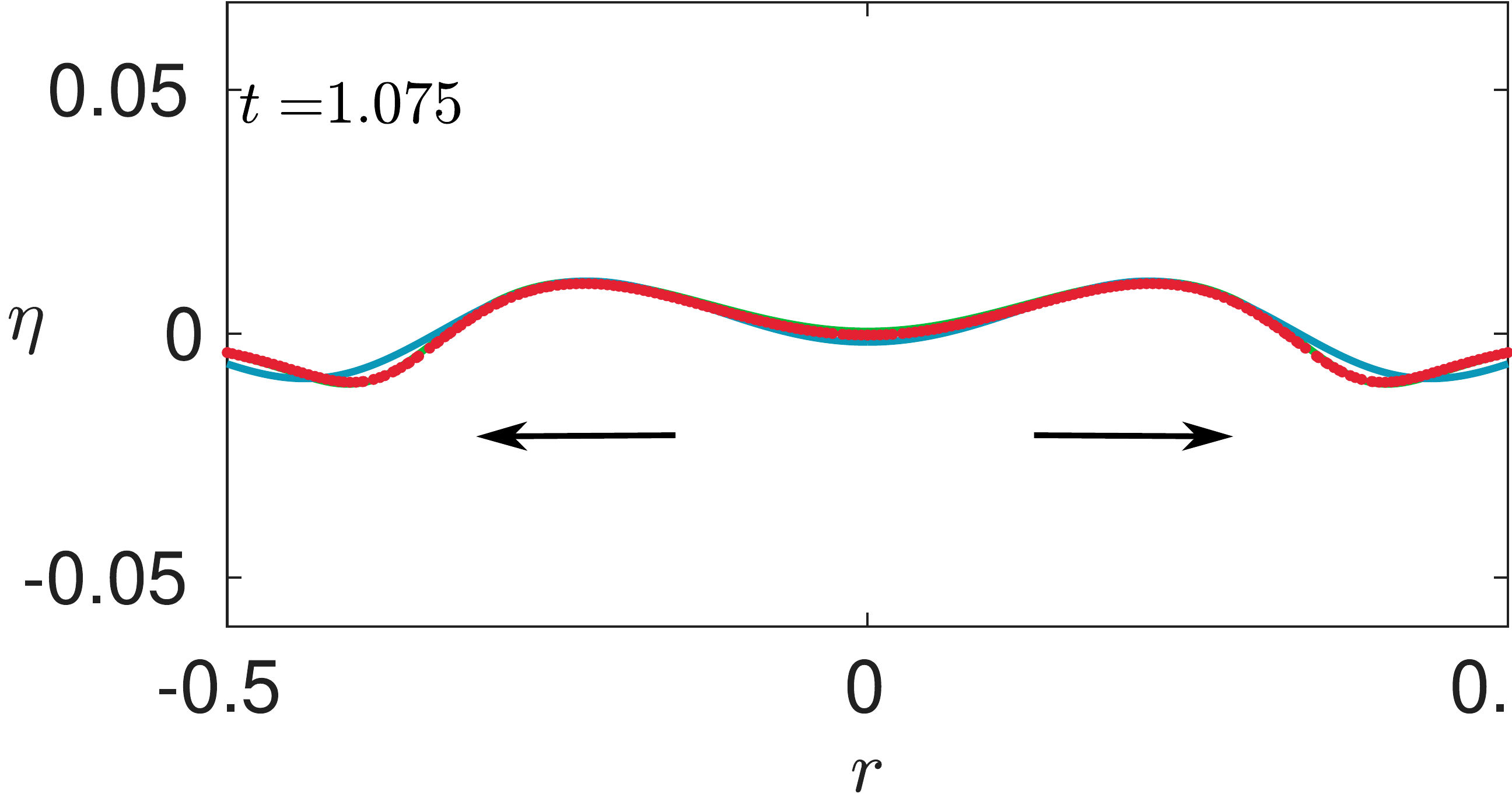}\label{fig9c}}\quad
        \subfloat[$t=4.056$]{\includegraphics[scale=0.12]{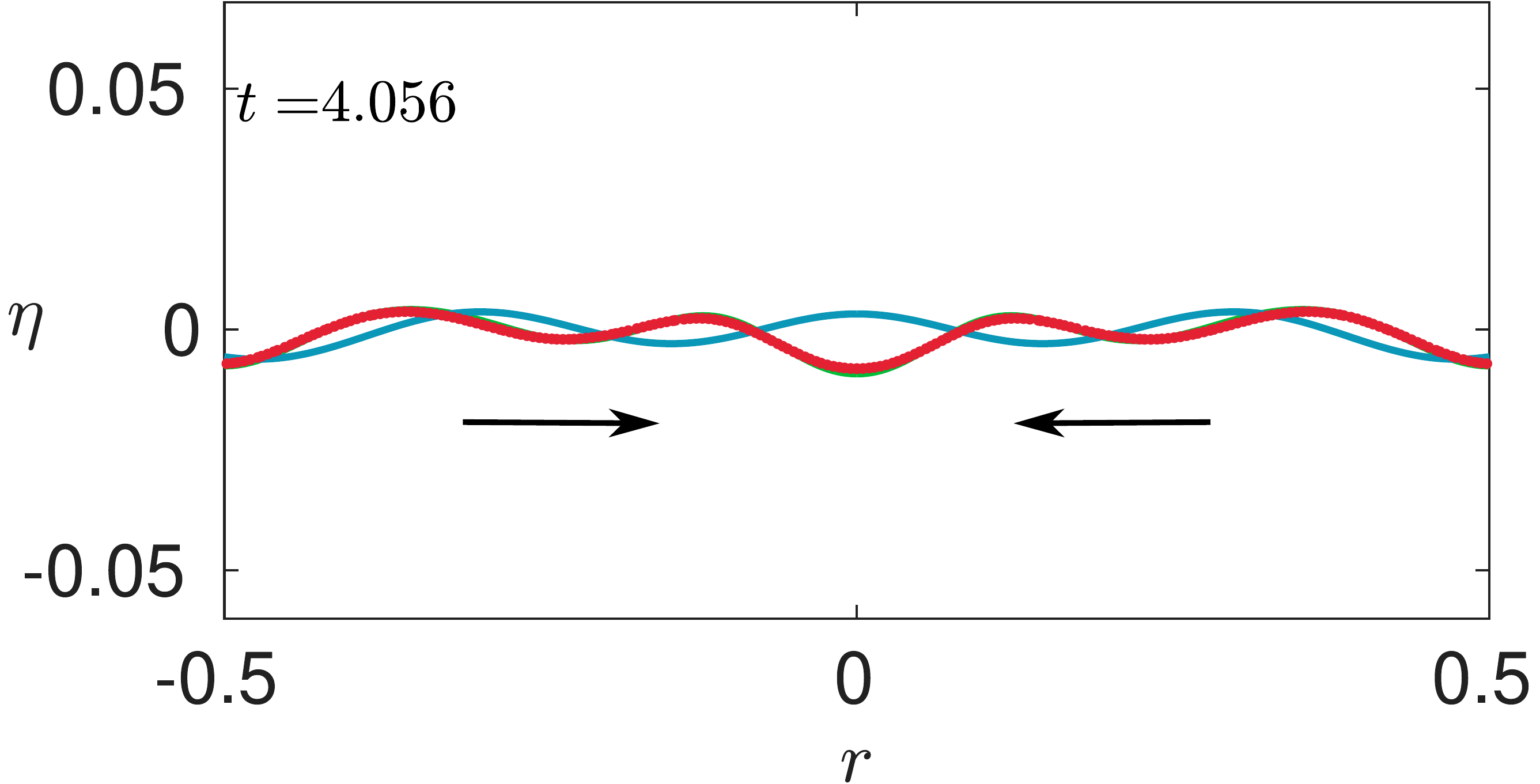}\label{fig9d}}\\
        \subfloat[$t=4.117$]{\includegraphics[scale=0.12]{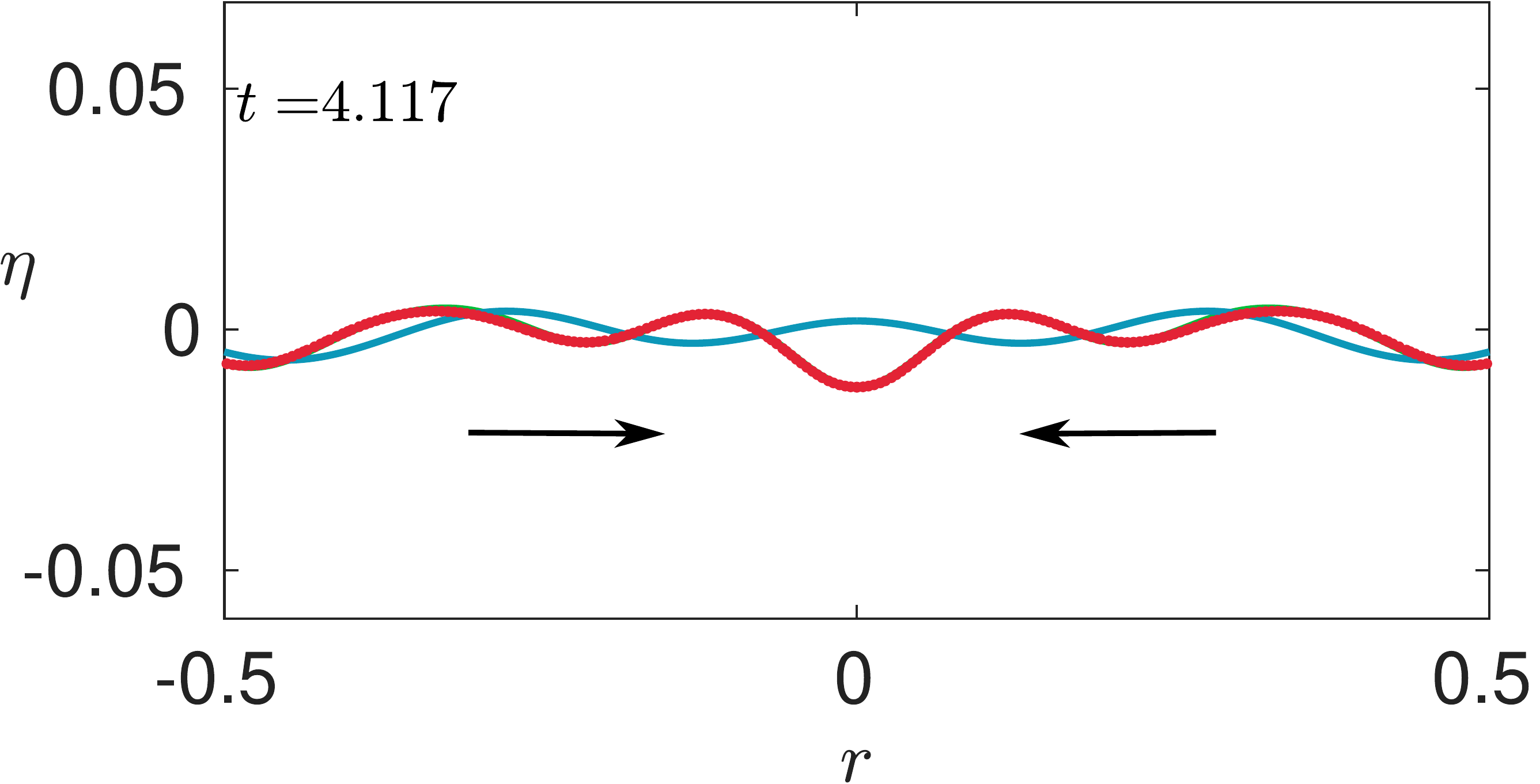}\label{fig9e}}\quad
        \subfloat[$t=4.435$]{\includegraphics[scale=0.12]{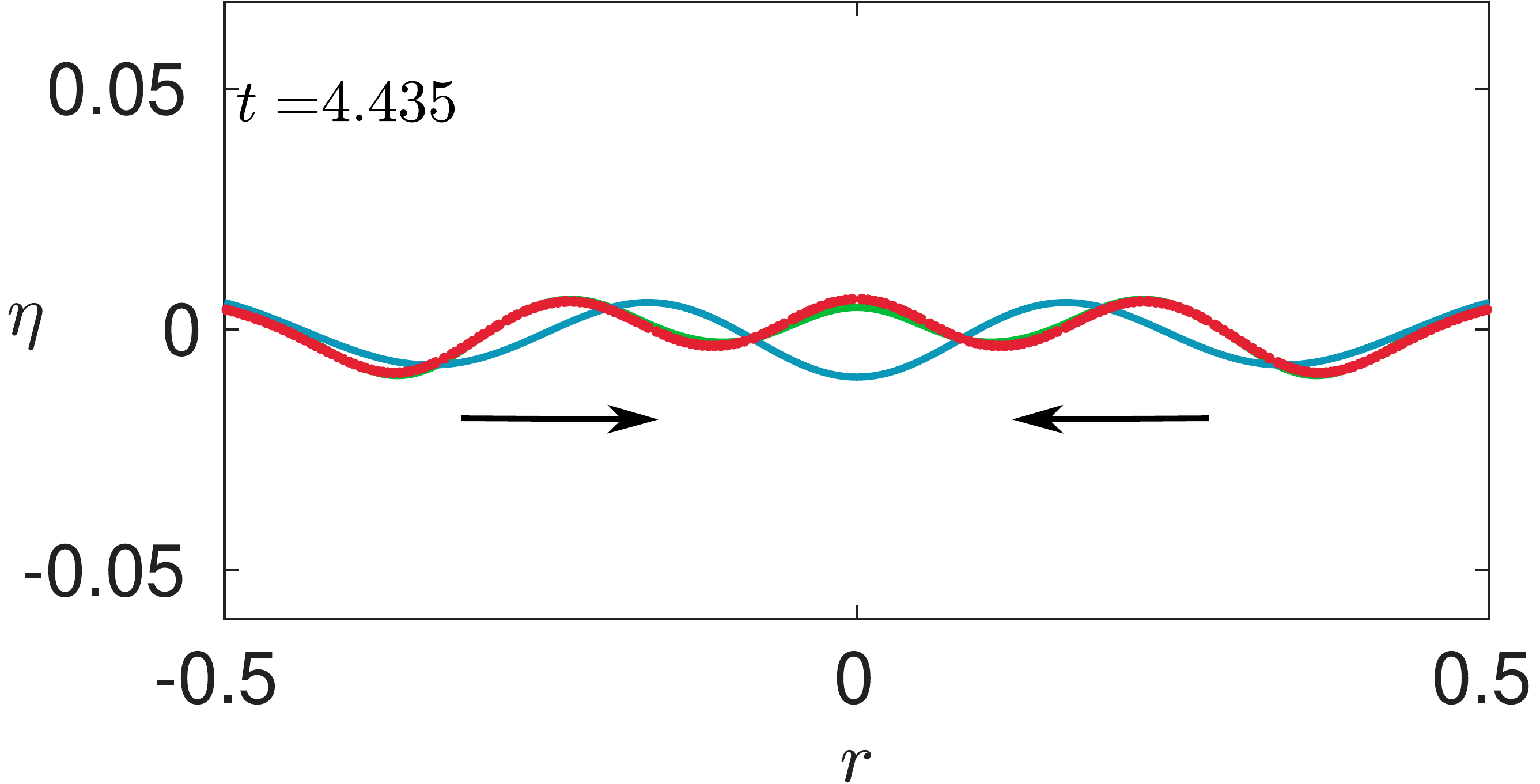}\label{fig9f}}\\
        \subfloat[$t=4.753$]{\includegraphics[scale=0.12]{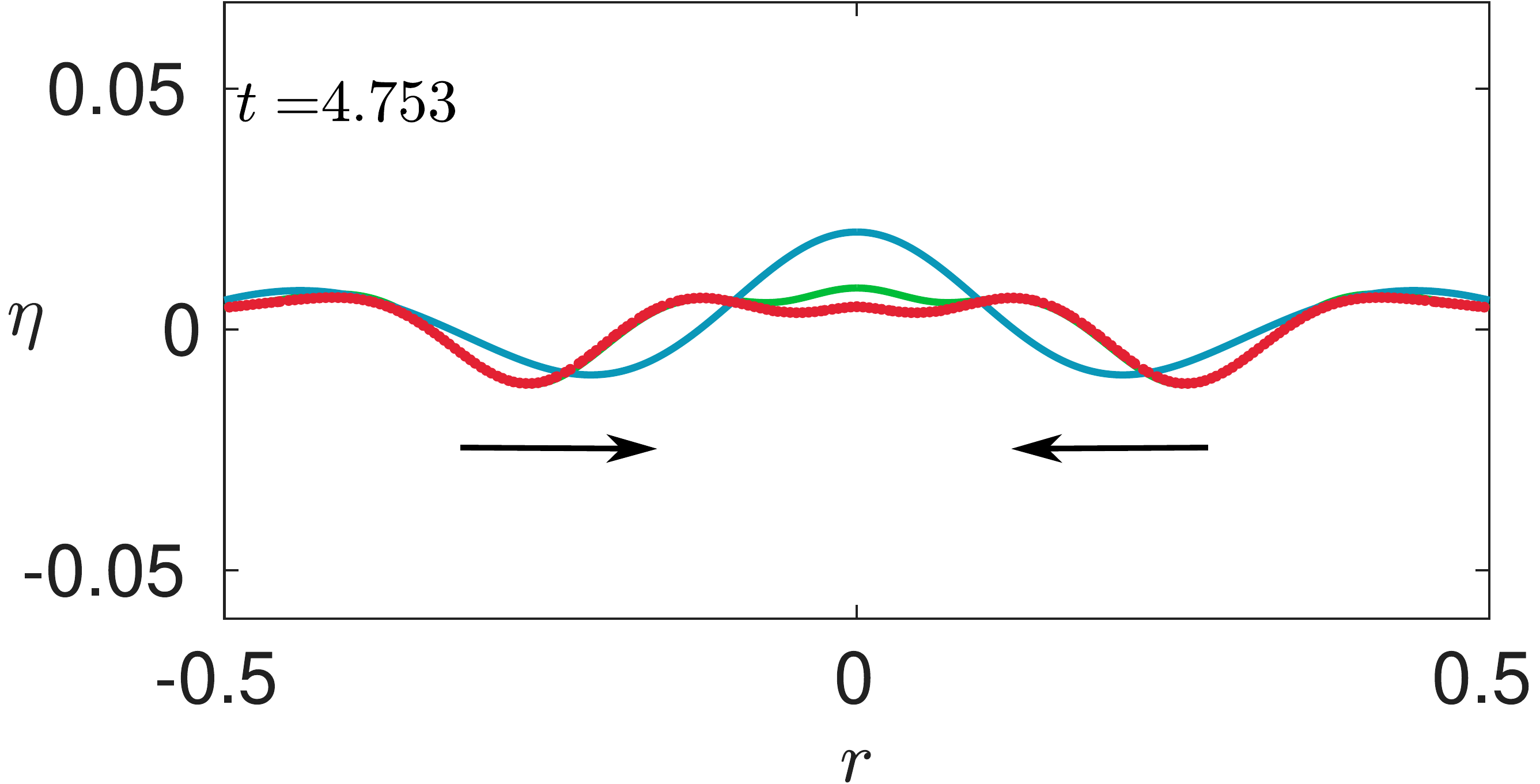}\label{fig9g}}\quad
        \subfloat[$t=5.358$]{\includegraphics[scale=0.12]{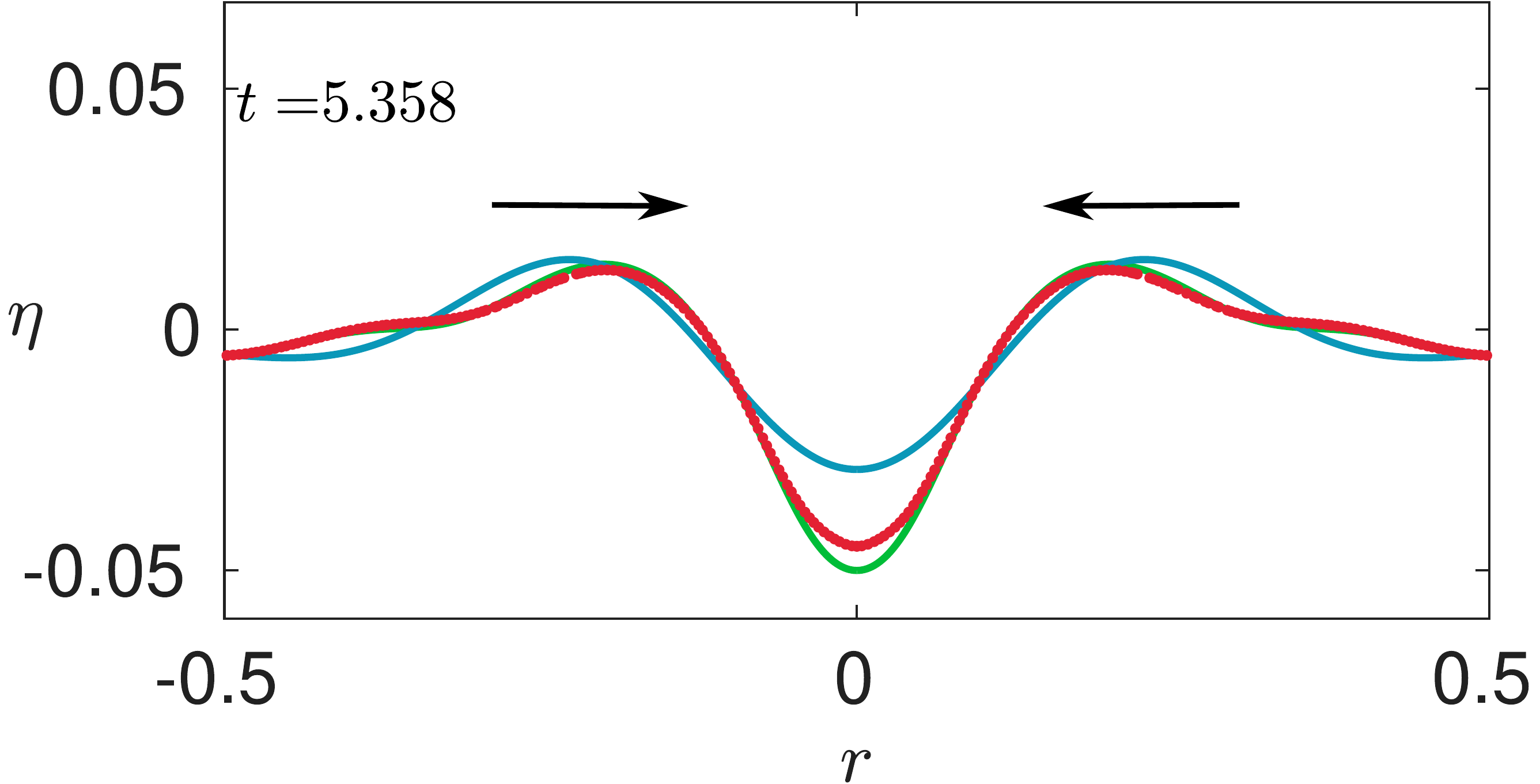}\label{fig9h}}\\
        \subfloat[$t=5.615$]{\includegraphics[scale=0.12]{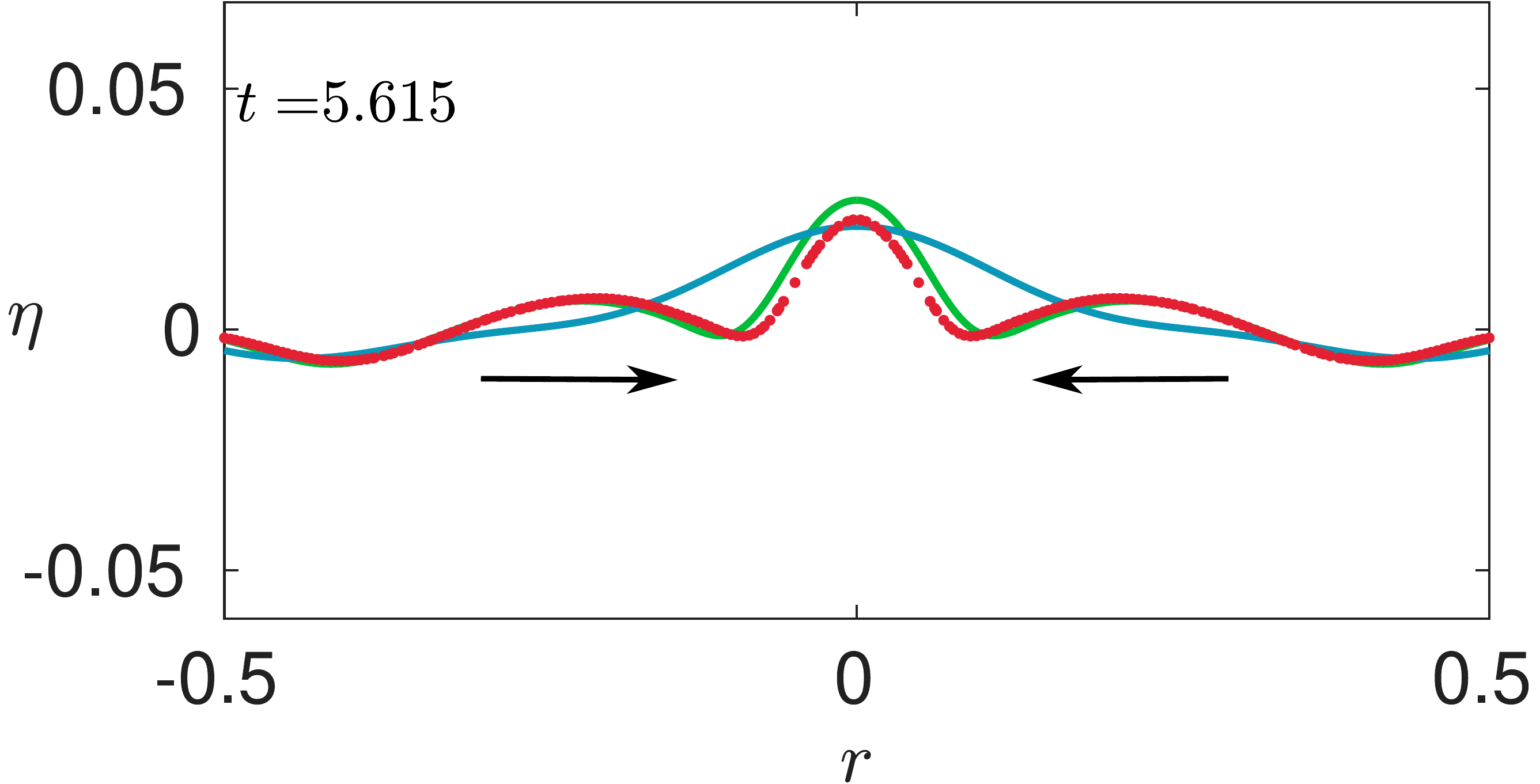}\label{fig9i}}\quad
        \subfloat[$t=5.842$]{\includegraphics[scale=0.12]{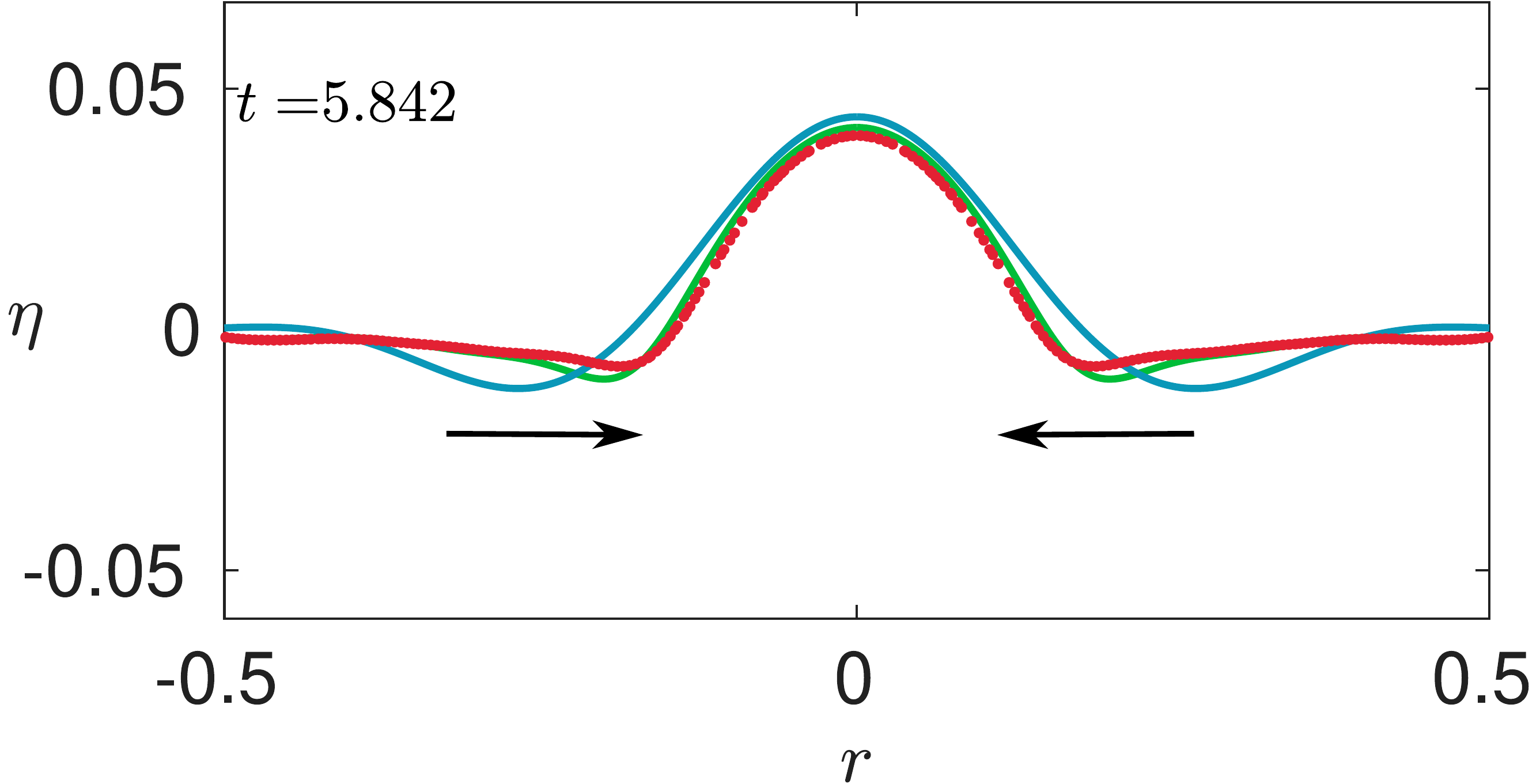}\label{fig9j}}
        \caption{Waves generated from the cavity shaped interface distortion at $t=0$ (inset of panel (a)). We compare the interface shape as a function of time as predicted by linear theory (L, solid blue line), second-order nonlinear theory (N, solid green line) and (inviscid) DNS (Sim, red symbols). The waves reflect-off the cylinder wall at $r=1$ (not shown) and focus back towards $r=0$ generating oscillations of increasing amplitude. This corresponds to case $1$ of table \ref{tab:sim_params} with $\varepsilon = 0.061$. To highlight the difference between linear and nonlinear predictions, the figures have been plotted upto $r=0.5$ instead of the entire radial domain up to $r=1$. The arrows depict the instantaneous direction of motion of the waves.}
        \label{fig9}
    \end{figure}

\begin{figure}
		 	\centering
		 	\subfloat[$t=0.166$]{\includegraphics[scale=0.12]{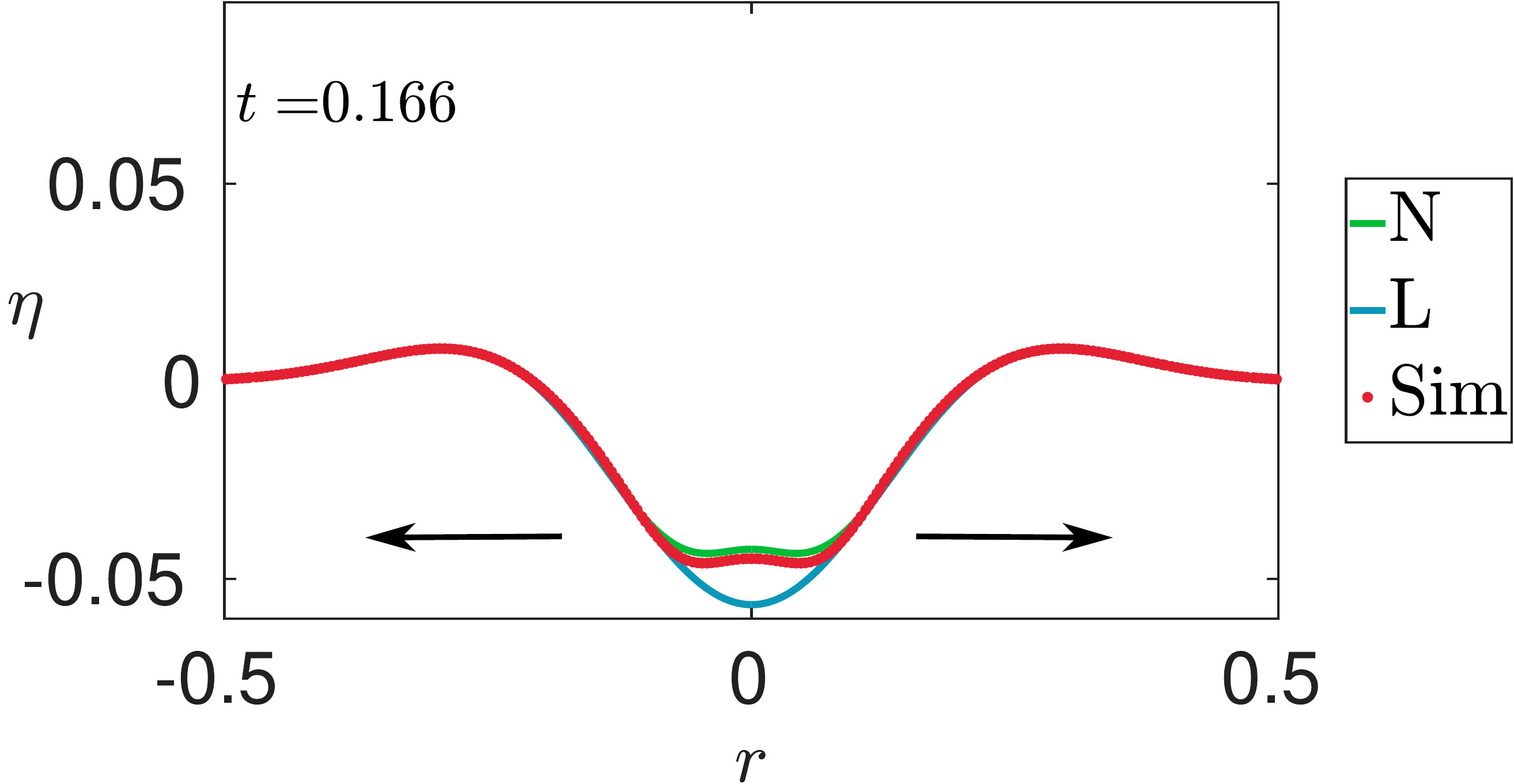}\label{fig10a}}\quad
		 	\subfloat[$t=0.439$]{\includegraphics[scale=0.12]{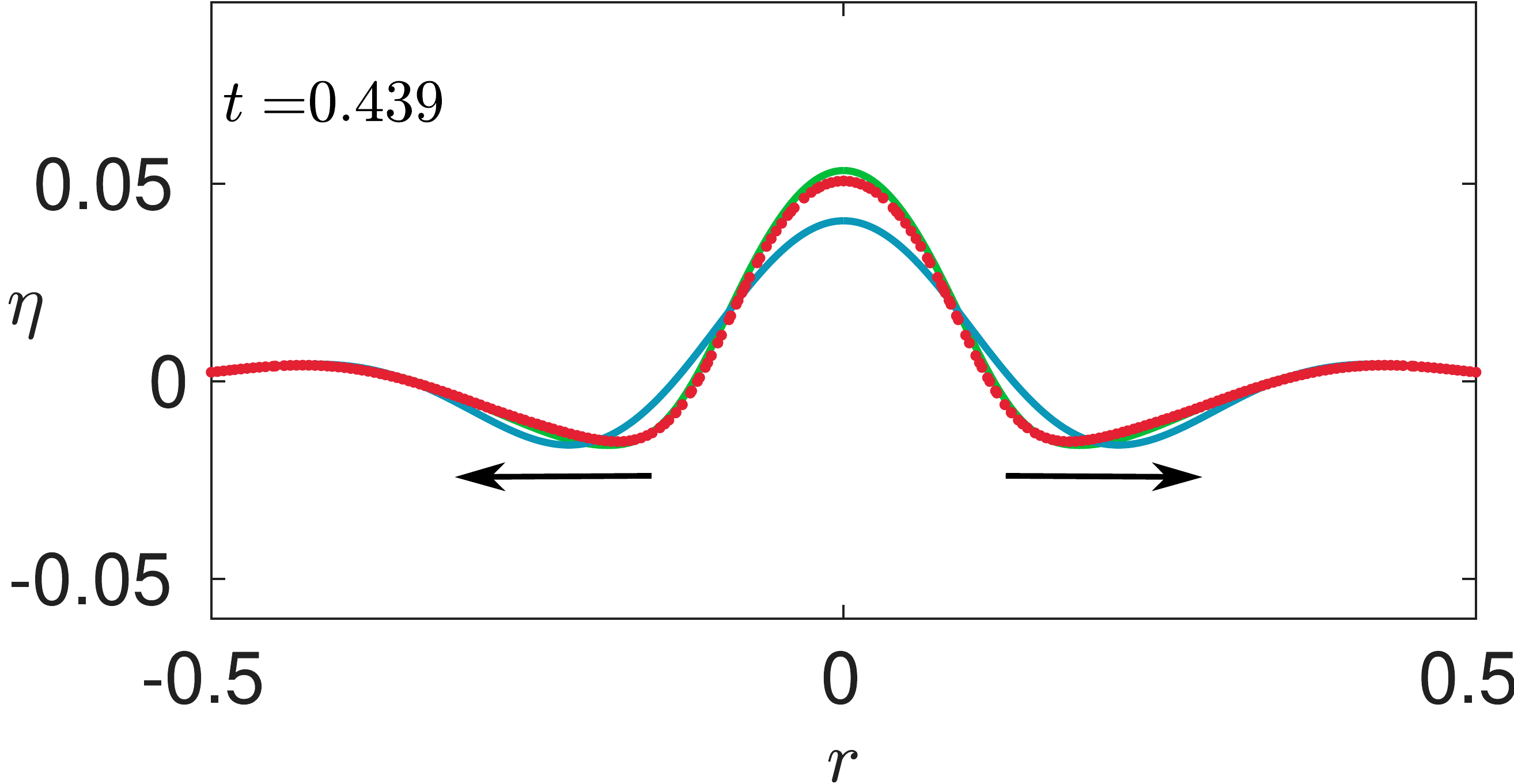}\label{fig10b}}\\
		 	\subfloat[$t=1.075$]{\includegraphics[scale=0.12]{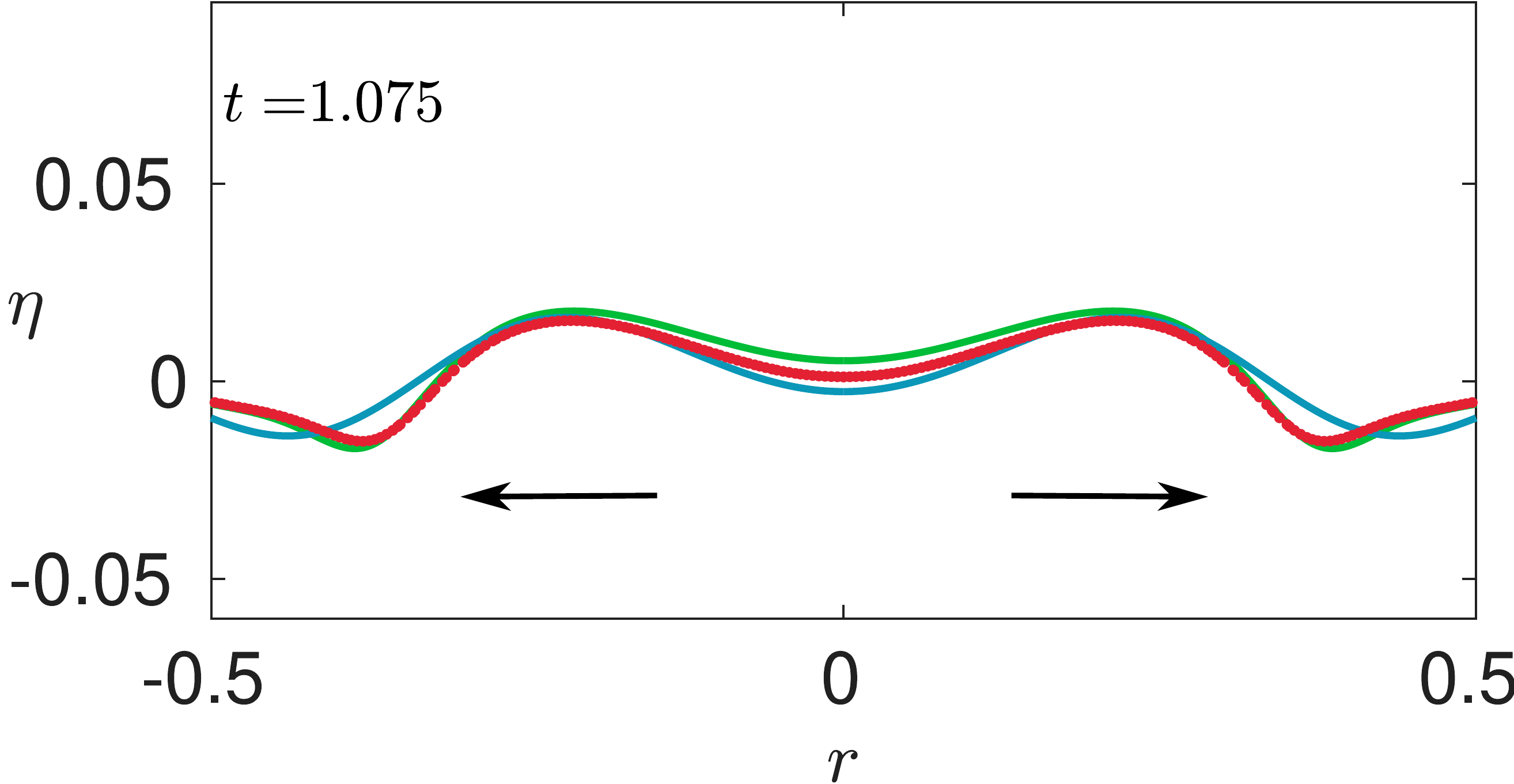}\label{fig10c}}\quad
		 	\subfloat[$t=4.056$]{\includegraphics[scale=0.12]{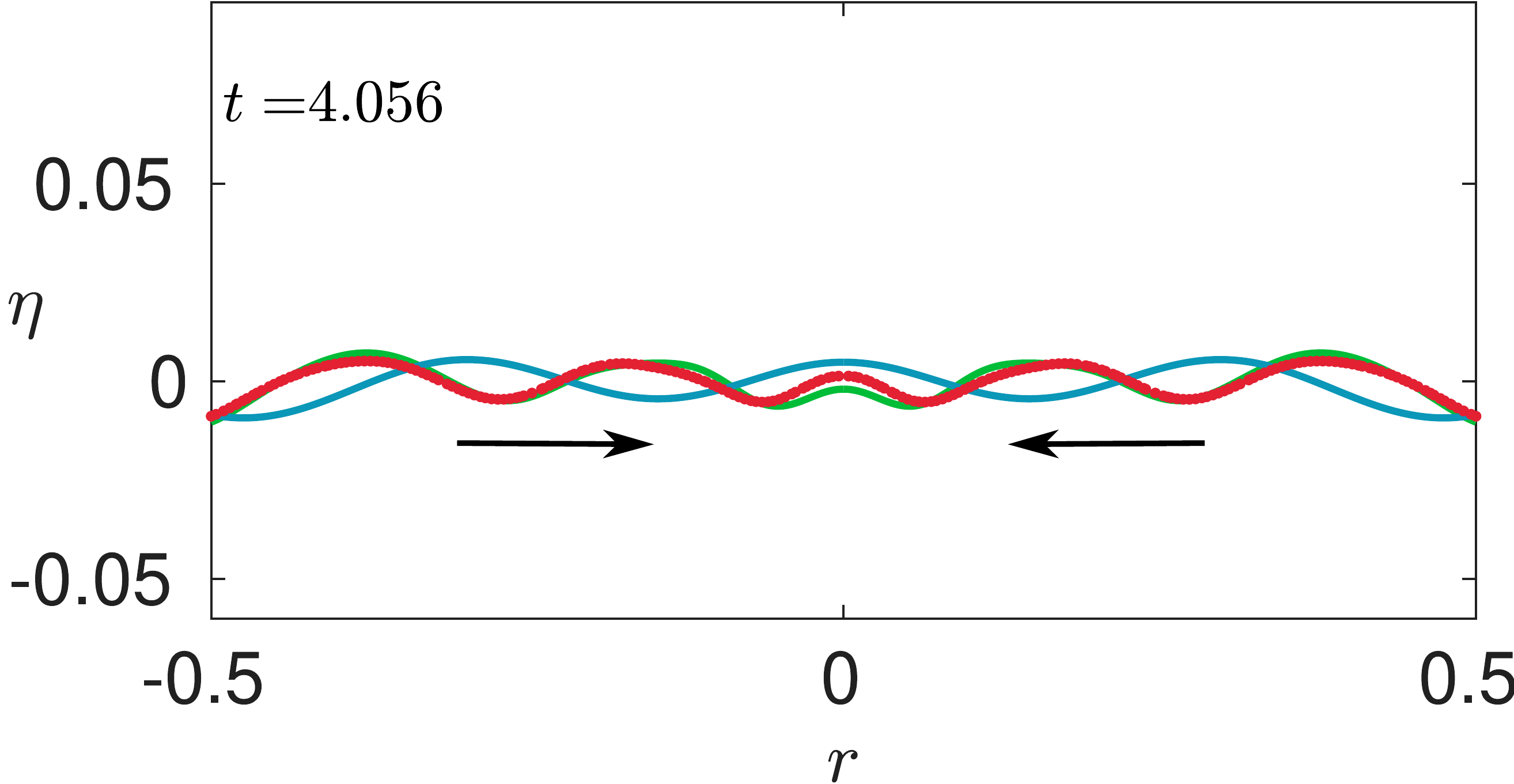}\label{fig10d}}\\
		 	\subfloat[$t=4.117$]{\includegraphics[scale=0.12]{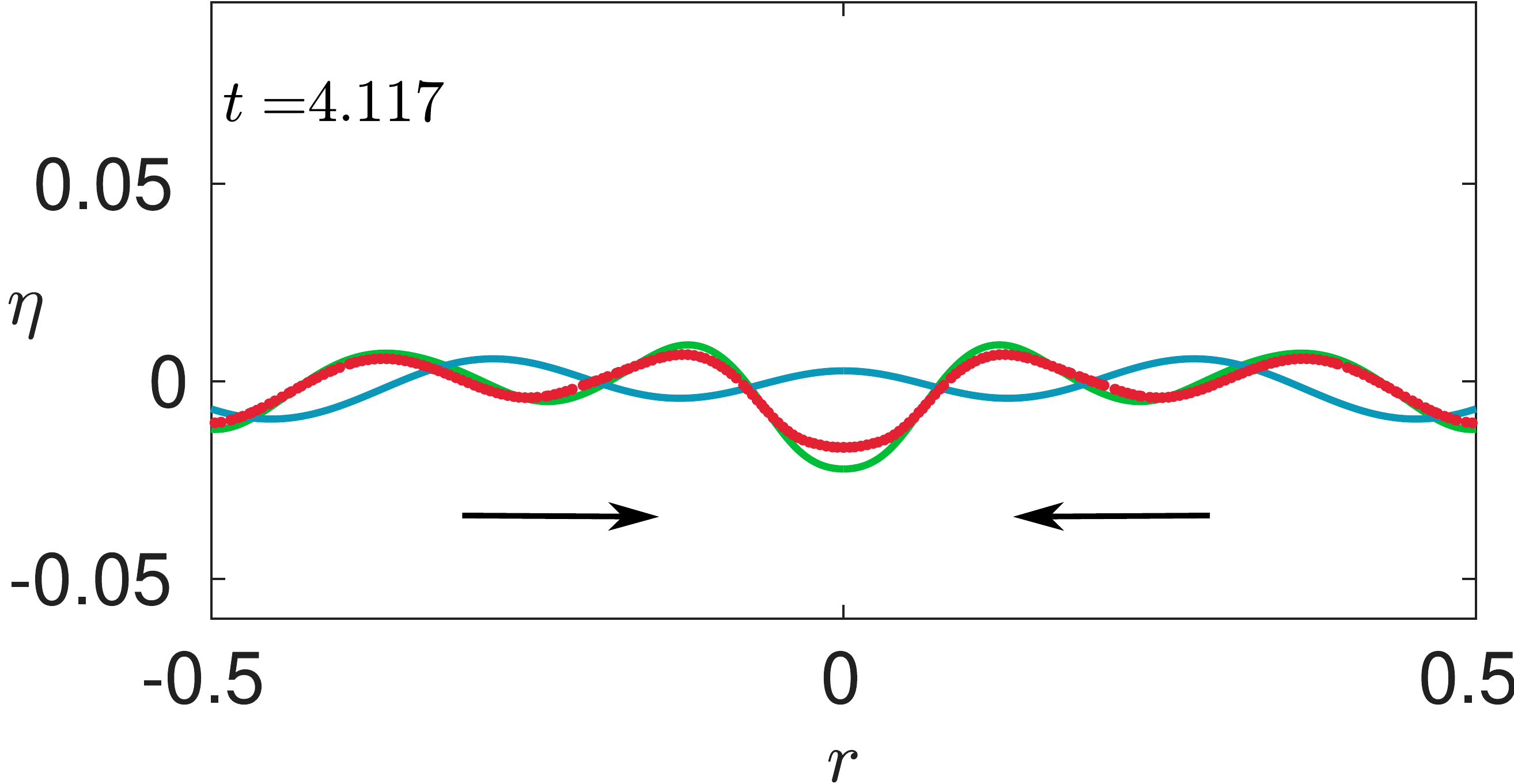}\label{fig10e}}\quad
		 	\subfloat[$t=4.435$]{\includegraphics[scale=0.12]{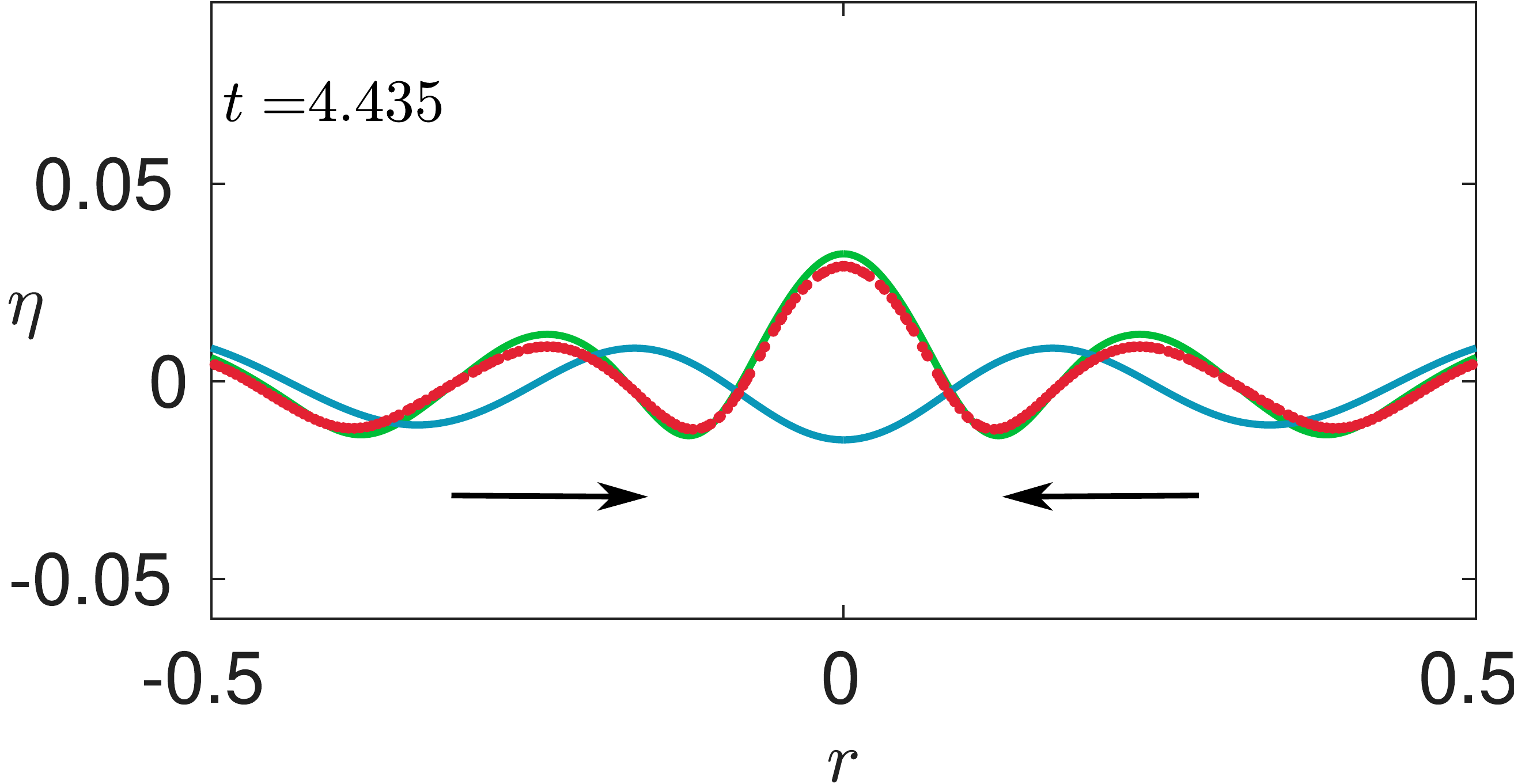}\label{fig10f}}\\
		 	\subfloat[$t=4.753$]{\includegraphics[scale=0.12]{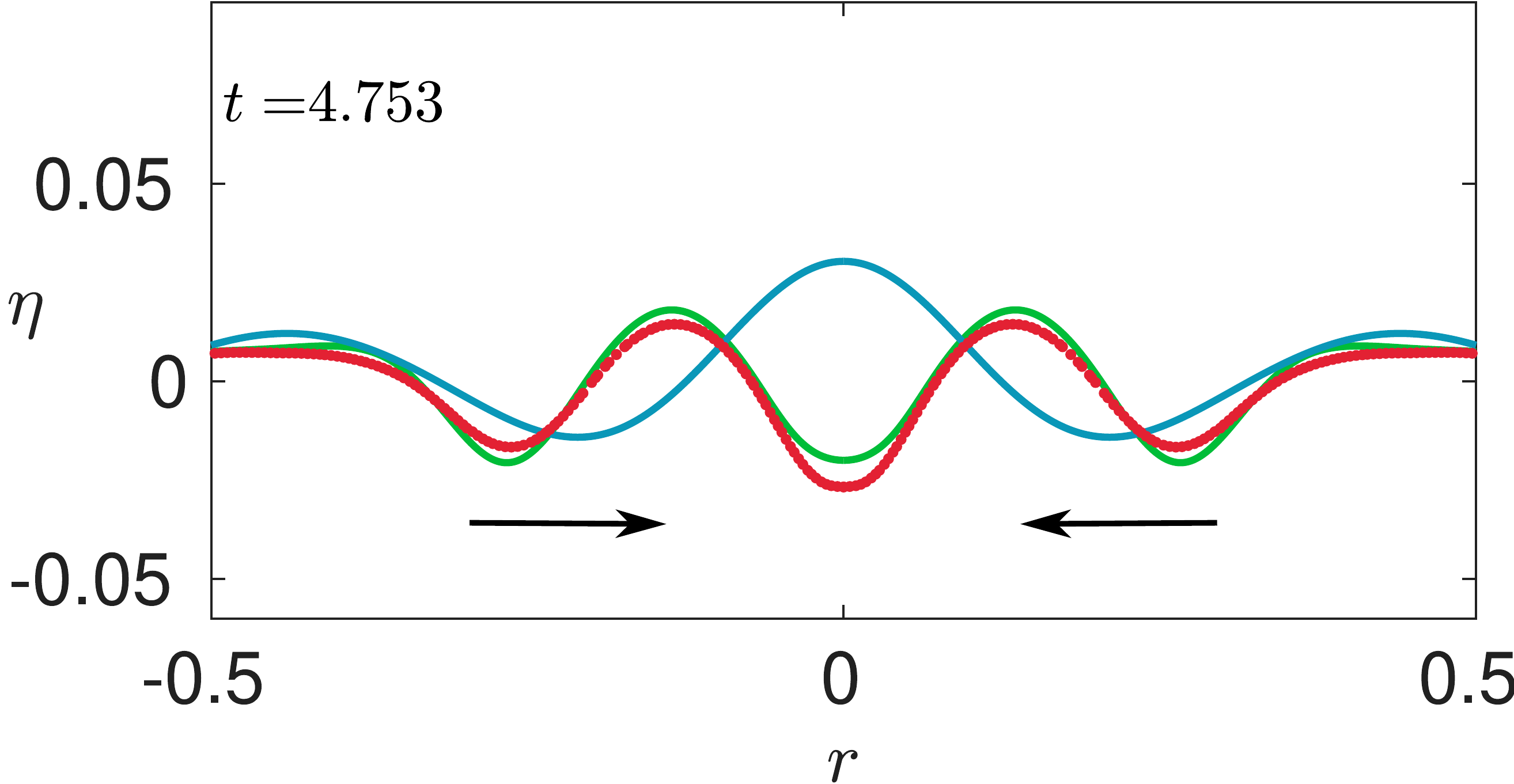}\label{fig10g}}\quad
		 	\subfloat[$t=5.358$]{\includegraphics[scale=0.12]{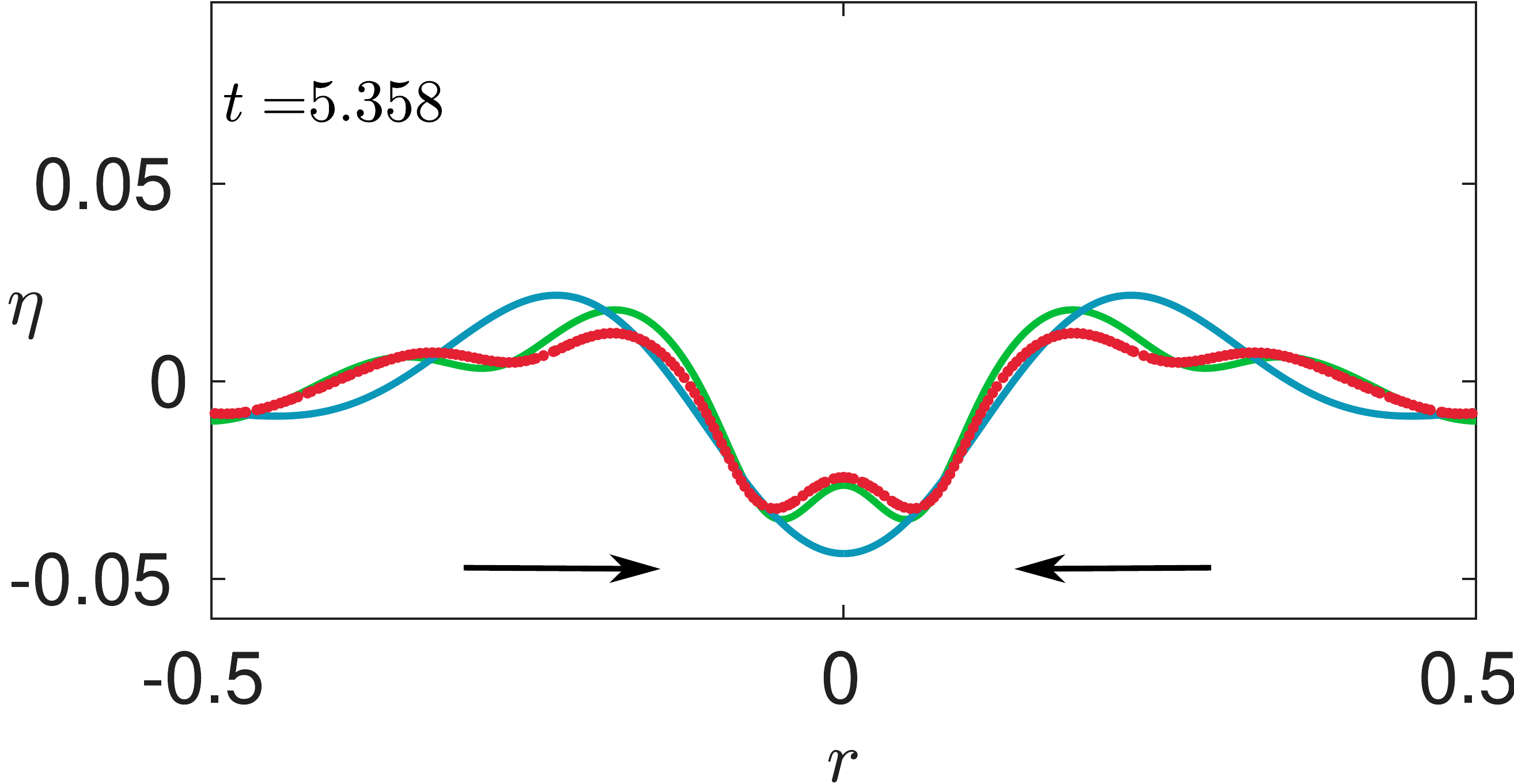}\label{fig10h}}\\
		 	\subfloat[$t=5.165$]{\includegraphics[scale=0.12]{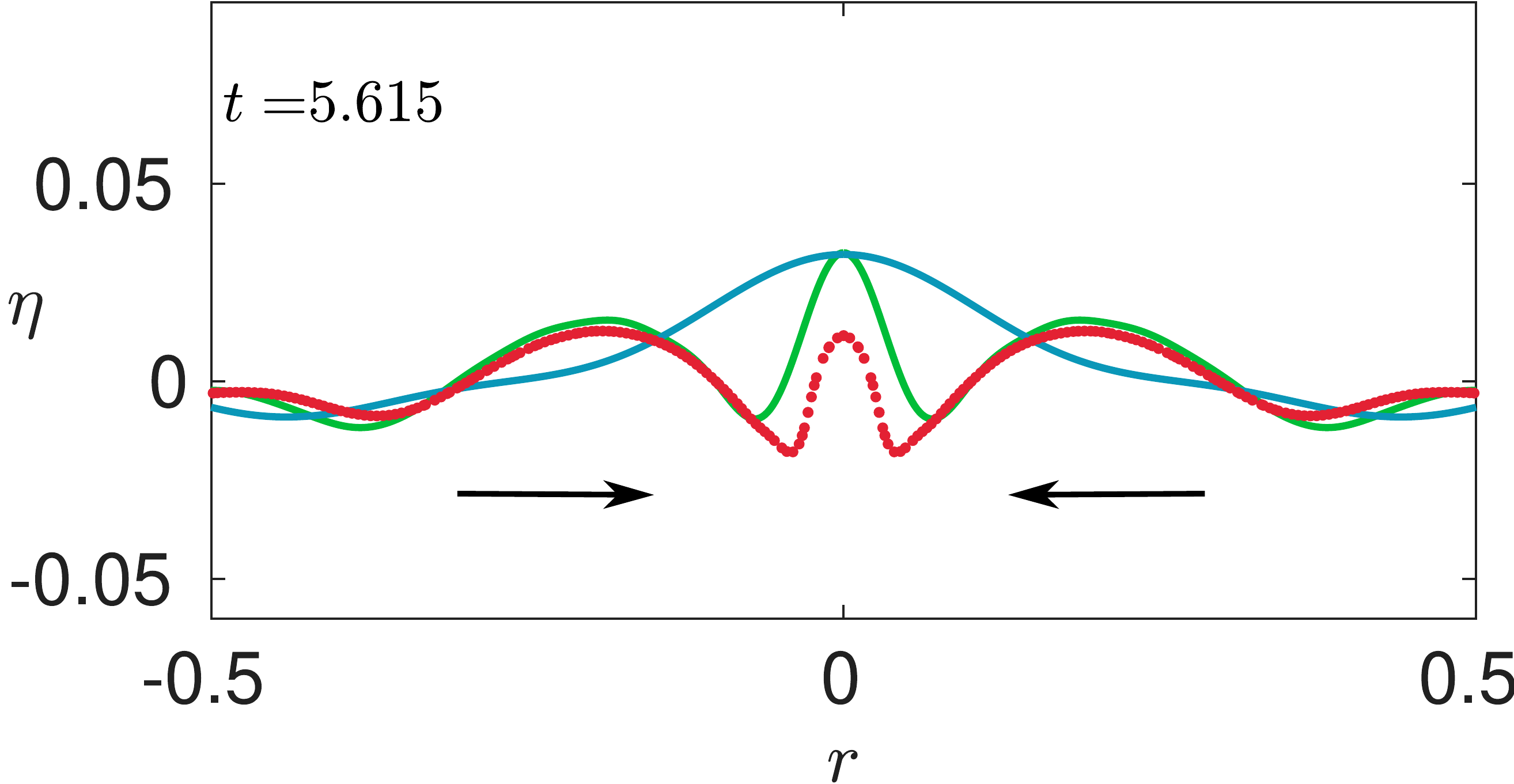}\label{fig10i}}\quad
		 	\subfloat[$t=5.842$]{\includegraphics[scale=0.12]{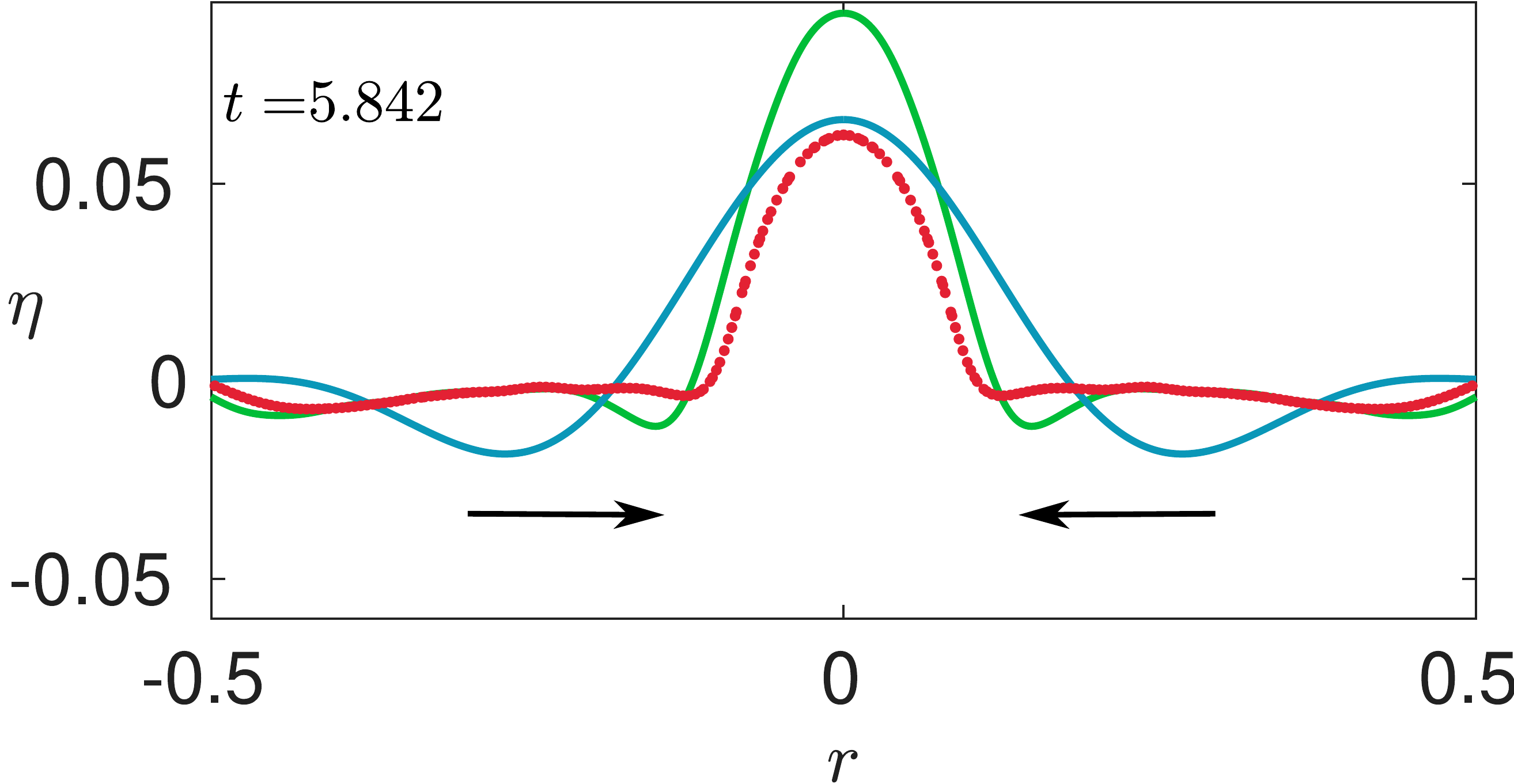}\label{fig10j}}
		 	\caption{The same as fig. \ref{fig9}, but for $\varepsilon=0.091$ corresponding to case $2$ in table \ref{tab:sim_params}. Note the good qualitative agreement between nonlinear theory and (inviscid) DNS but not linear theory, in capturing the dimple in panel (h). Also note the large amplitude oscillations at $r=0$ with a tendendency to generate narrow jet-like structures (panels (i) and (j)) \textcolor{black}{although no jets are seen}.}
			\label{fig10}
 \end{figure}

   In the collage of images in figs. \ref{fig9} and \ref{fig10}, we present the shape of the interface as a function of time for case $1$ and $2$ in table \ref{tab:sim_params} respectively, comparing this to linear and nonlinear theoretical predictions. The only difference between these two figures is in the value of $\varepsilon$, all other dimensionless numbers remaining the same. Here linear theory implies solution to eqn. \ref{eq6} without the nonlinear terms. Note that this is equivalent to superposition of the form $\eta(r,t) = \sum_{m=1}^{17}\eta_m(0)\mj_{0}(k_mr)\cos(\omega_m t)$ where $\omega_m(k_m)$ satisfies the gravito-capillary dispersion relation for deeep water. In fig. \ref{fig9}, the transition from outward propagating waves to inward propagating ones occur between panels (c) and (d). For panels (a), (b) and (c) it is evident that linear theory represents the outgoing waves accurately. However as focussing commences from panel (d) onwards, we notice significant differences between linear theory and (inviscid) DNS. Interestingly, second order theory seems to predict the shape of the interface around $r=0$ quite well.
   Fig. \ref{fig10} shows a more intense scenario than Fig. \ref{fig9}, featuring a larger $\varepsilon=0.091$. The transition from predominantly linear to nonlinear behavior occurs between panels (c) and (d), representing outgoing and incoming waves, respectively. Notably, sharp dimple-like structures emerge around $r=0$, as seen in panel (h), which are well described by nonlinear theory. Additionally, the tendency to form jets (\textcolor{black}{although no clear jet is visible}), as seen in the final panel, is noteworthy, although the nonlinear theory is only qualitatively accurate in this context. We refer the reader to the accompanying Movie $\#1$ ($\varepsilon = 0.061$) and $\#2$ ($\varepsilon = 0.091$), see additional supplementary material which visualises these.


\subsection{Role of nonlinearity at $r=0$}
\textcolor{black}{Figs. \ref{fig9} and \ref{fig10} show that although the linear solution is a reasonable model for the interface evolution before reflection, it shows deviation from the fully nonlinear simulation at the axis of symmetry during radial convergence of the wave-train. Towards understanding this better, we provide two sets of analysis in the following subsections. In subsection~\ref{sec:3.1.1}, we analyse the time-periodic solution by \cite{mack1962periodic}, investigating the role of nonlinearity generated bound components around $r=0$. In subsection~\ref{sec:3.1.2}, we analyse the initial deformation as a Bessel function, \textcolor{black}{akin to \cite{basak2021jetting}}. It will be seen from both analysis that bound components play an important role in the interface deformation around $r=0$.}
\subsubsection{Comparison with time-periodic solution}\label{sec:3.1.1}
\textcolor{black}{Unlike the initial interface distortion studied so far which leads to aperiodic behaviour, there also exist finite-amplitude deformations which generate time-periodic oscillations. Such finite-amplitude, time-periodic solutions are the standing-wave counterparts of the well-known Stokes travelling wave. In rectangular coordinates, such a standing-wave solution was first developed by Rayleigh \citep{strutt1915deep} and in further detail by \cite{penney1952part}. This was extended to radially bounded, cylindrical geometry for finite liquid depth in \cite{mack1962periodic}. In the deep-water limit, Mack's solution contains three parameters, all appearing in the `free-wave' (see below) part of the solution represented by $\tilde{a}_0 J_0(k_q \frac{\hat{r}}{\hat{R}}),\; q=1,2,3\ldots$. These in turn lead to two non-dimensional parameters viz. $\tilde{\epsilon} \equiv \frac{\tilde{a}_0}{\hat{R}}$ and a positive integer $q=1,2,3\ldots$ specifying the number of zero crossings of $J_0$ within the radial domain, a measure of crest to crest distance of the perturbation ($J_0$ is not periodic but becomes so asymptotically). In non-dimensional form the time-periodic solution of \cite{mack1962periodic} may be written as}
\begin{eqnarray}
	\eta(r,\tilde{t};\tilde{\epsilon},q) &=& T_0(r;\tilde{\epsilon},q) + T_1(r;\tilde{\epsilon},q)\cos(2\pi\tilde{t}) + T_2(r;\tilde{\epsilon},q)\cos(4\pi\tilde{t}) + T_3(r;\tilde{\epsilon},q)\cos(6\pi\tilde{t}) \nonumber \\
	&& \label{mack-1a}
\end{eqnarray}
\textcolor{black}{where $\tilde{t}\equiv \frac{\omega \hat{t}}{2\pi}$. \cite{mack1962periodic} obtained expressions for $T_0(r), T_1(r)$ and $T_2(r)$ for $q=1$ employing $\tilde{\epsilon}$ as perturbative parameter (up to $\mathcal{O}(\tilde{\epsilon}^3)$) and expressions for these alongwith the nonlinear frequency $\omega(\tilde{\epsilon},q=1)$ are provided in the Appendix A, adapted to our notation.
}
\begin{figure}
	\centering
	\includegraphics[scale=0.55]{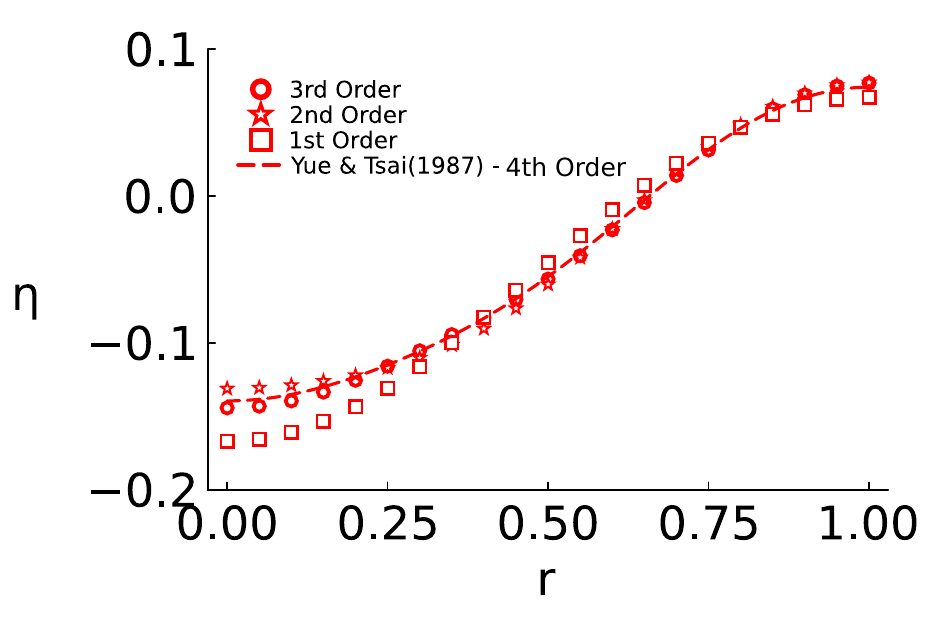}
	\caption{\textcolor{black}{The shape of the interface calculated from eqn. \ref{mack-1a} by retaining terms up to various orders in $\tilde{\epsilon}$ in the expressions for $T_0(r), T_1(r),T_2(r), T_3(r)$. We choose $\tilde{\epsilon}=0.16703$ and $q=1$ and plot the interface at $\tilde{t}=0.5$ when the velocity field everywhere is zero and the shape around $r=0$ has a depression. A fourth-order interface shape for the same $\tilde{\epsilon}$ is also presented here, obtained following the numerical
			procedure given in \cite{tsai1987numerical}.}}
	\label{fig11}
\end{figure}

\begin{figure}
	\centering
	\includegraphics[scale=0.5]{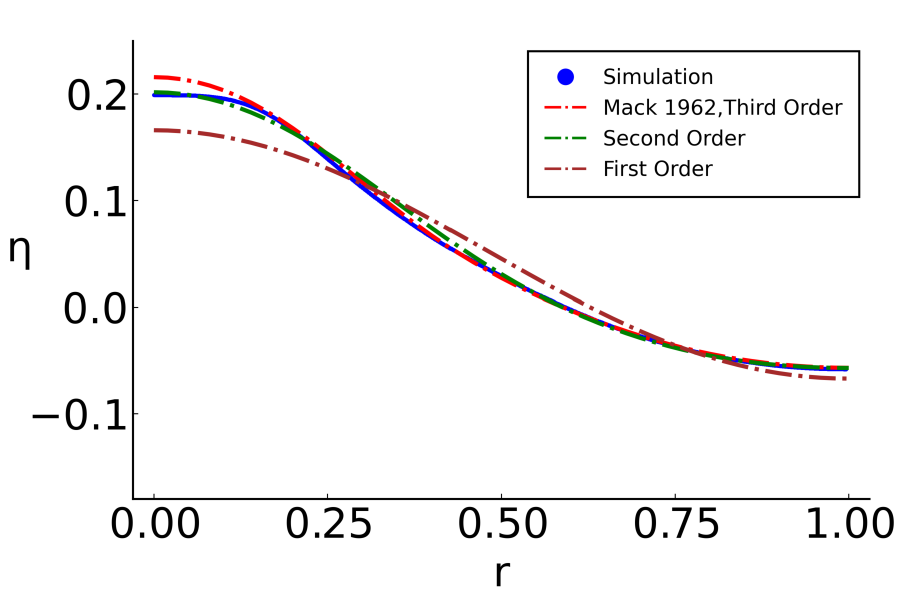}\label{fig11b}
	\caption{\textcolor{black}{Interface of various orders for $q=1$ and $\tilde{\epsilon}=0.16703$. The first, second and third order solutions are plotted at $\tilde{t}=1$ using eq. \ref{mack-1a} of \cite{mack1962periodic}. The numerical solution (indicated in blue as `Simulation') is initialised using the third-order solution of \cite{mack1962periodic} evaluated at $\tilde{t}=0.5$. Note that $\tilde{t}=0.5$ in eqn. \ref{mack-1a} is used to initialise the DNS and hence corresponds to $t=0$ for the latter.}}
	\label{fig12}
\end{figure}
\begin{figure}
	\centering
	\subfloat[Initial interface shape]{\includegraphics[scale=0.42]{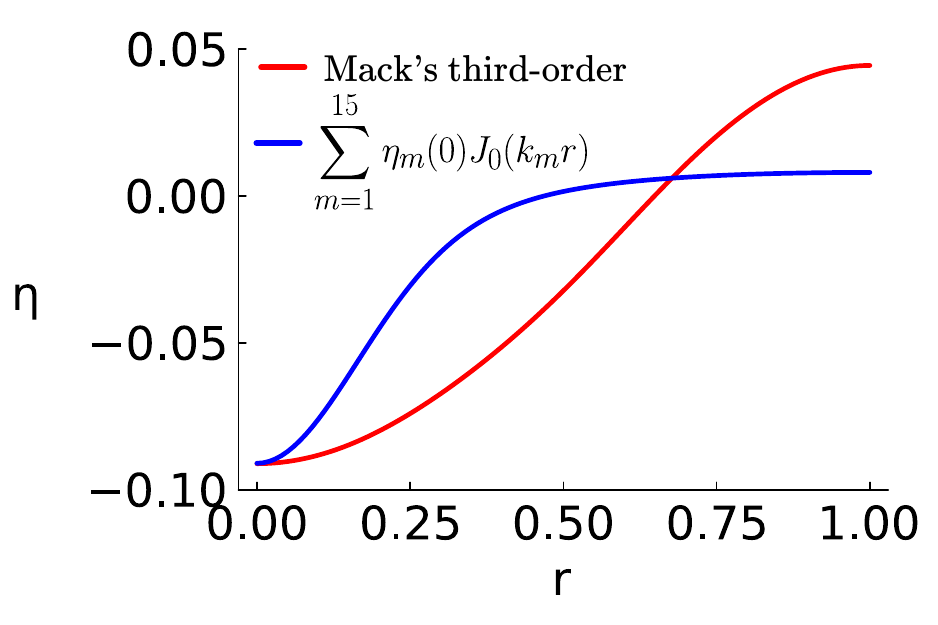}\label{fig13a}}
	\subfloat[$\eta_{m}$ for the cavity ($t=0$) and for eqn. \ref{mack-1a}]{\includegraphics[scale=0.42]{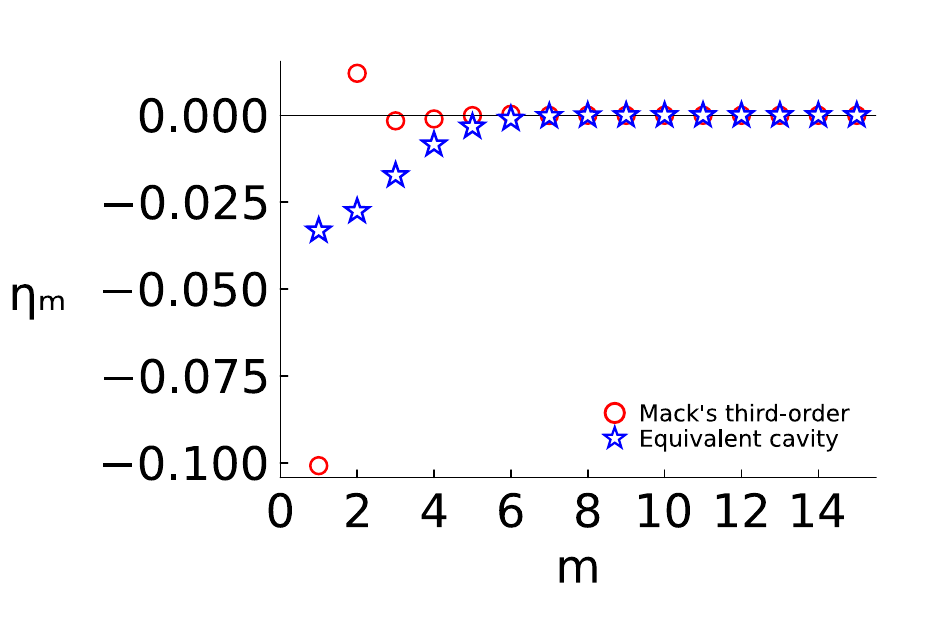}\label{fig13b}}
	\caption{\textcolor{black}{(a) A \textit{localised} cavity shaped deformation (blue) plotted against the \textit{de-localised} third-order, time-periodic solution (red) by \cite{mack1962periodic} plotted at a time when it is shaped as a depression around $r=0$. The Fourier-Bessel series for both shapes are $\eta_{m}(0)J_0(k_m r)$ where $\eta_m$ are provided on the right panel. For the time-periodic solution, $\tilde{\epsilon}=0.1014$ (third order). The two profiles have been depth matched at $r=0$. The cavity shape profile has the same dominant Bessel function ($k_1$) as the free wave in the third-order time periodic solution from eqn. \ref{mack-1a} \citep{mack1962periodic}. Unlike the cavity, the time-periodic solution is spatially de-localised as it has significant interface displacement at $r=1$, see left panel. (b) The deformations on the left panel are expressed as Fourier-Bessel series with coefficients $\eta_{m}$ presented on the right panel. The color scheme is the same in both panels.}}
	\label{fig13}
\end{figure}
\begin{figure}
	\centering
	\includegraphics[scale=1.0]{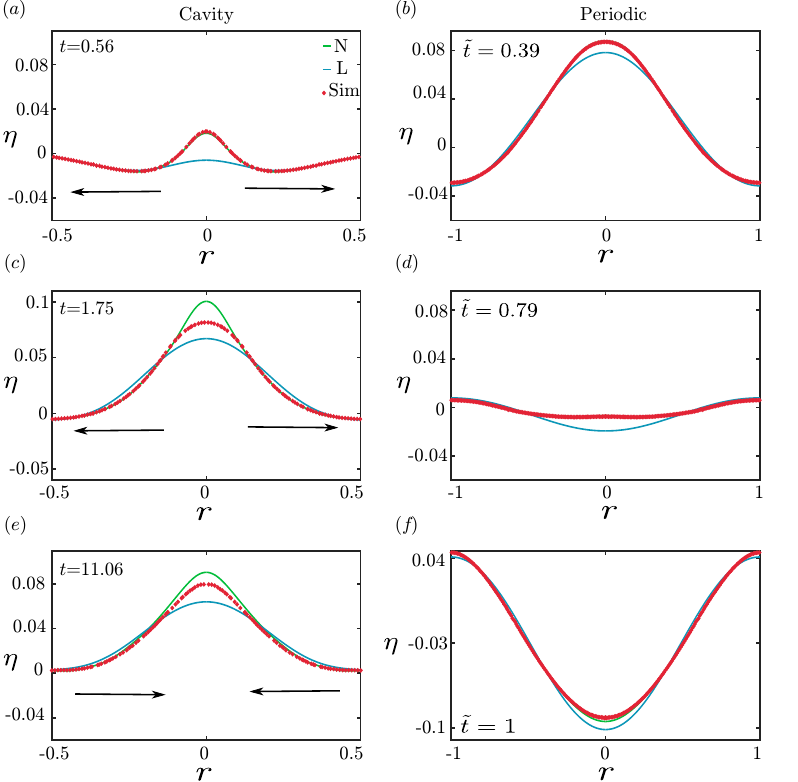}
	\caption{\textcolor{black}{Time evolution starting from the two deformations (and zero velocity in the liquid) shown in left panel of fig. \ref{fig13a}. The left column shows snapshots of evolution of the cavity at $t=0.56,1.75, 11.06$ from numerical simulations (Sim), nonlinear theory (N) obtained from the numerical solution to eq. \ref{eq6} and linear theory. In all cases, the nonlinear theory does significantly better than linear theory. The inward and outward propagating arrows show the instantaneous direction of wave propagation. On the right column, is the time evolution of the third-order interface shape depicted in the left panel of fig. \ref{fig13} (time-periodic solution) at $\tilde{t}=0.39, 0.79$ and $1.0$. One notes the excellent agreement between nonlinear theory and simulations while the difference at $r=0$ between the linear and nonlinear predictions are maintained. The color scheme is identical for both columns. Note that air-water surface tension has been used for the simulations. For staying consistent with \cite{mack1962periodic} where there is no surface-tension, we have considered a much larger cylindrical domain here compared to the earlier case. For the simulations, we have used (CGS units) $T=80$, $g=981$, $\hat{R}_0=100$ , $\nu=0$ (both fluids) with air-water density ratio.}}
	\label{fig14}
\end{figure}

\textcolor{black}{Note that the solution by \cite{mack1962periodic} excludes capillary effects. Referring to Appendix A, we note that $T_0(r)$ is of $\mathcal{O}(\tilde{\epsilon}^2)$ while $T_1(r), T_2(r)$ and $T_3(r)$ are of $\mathcal{O}(\tilde{\epsilon})$, $\mathcal{O}(\tilde{\epsilon}^2)$ and $\mathcal{O}(\tilde{\epsilon}^3)$ respectively. As a first step, we evaluate the accuracy of expression \ref{mack-1a} at a relatively high steepeness of $\tilde{\epsilon}\approx 0.16703$. This value is to be compared to its maximum possible value viz. $\tilde{\epsilon}_{\text{max}}=0.208$ (for $q=1$) computed by \cite{mack1962periodic} ($\tilde{\epsilon} \equiv k_1A_{11}$ in the notation by \cite{mack1962periodic}). In fig. \ref{fig11}, we plot the shape of the interface at various orders in $\tilde{\epsilon}$. The first-order approximation (leading order term in $T_1(r;\tilde{\epsilon})$) is $\eta(r,\tilde{t};\tilde{\epsilon})=\tilde{\epsilon} J_{0}(k_1r)\cos(2\pi \tilde{t})$ and represents the so-called `free-wave', as the wave-number $k_1$ and frequency $\omega$ satisfy the dispersion relation. However, all other corrections to $\eta(r,\tilde{t};\tilde{\epsilon})$ in eqn. \ref{mack-1a}, including those in $T_1(r)$, represent `bound components' as these do not satisfy the dispersion relation. In fig. \ref{fig11} comparing the third-order approximation by \cite{mack1962periodic} with the numerically computed fourth-order solution, indicates that the former is accurate at this chosen value of $\tilde{\epsilon}$. It is also apparent from fig  \ref{fig11} that the effect of systematically adding bound-components (nonlinear contribution) in determining the interface shape, has the largest effect at $r=0$. This is further established in fig. \ref{fig12}. In this fig., the third order interface depicted in fig. \ref{fig11} ($\tilde{t}=0.5$) is provided as an initial condition to the simulation. Half a time-period later ($\tilde{t} = 1$), we see that the analytical approximations (i.e. formulae \cite{mack1962periodic}) and the numerical simultion produce a higher elevation at $r=0$, compared to the first order approximation (free wave). We particularly highlight the asymmetry at $r=0$ between the elevation and depressions for the higher order approximations. For example, the third-order interface and the numerical simulation commence from a depression at $r=0$ in fig. \ref{fig11}, which is visibly less than that for the first order solution. At $\tilde{t} = 1$ in fig. \ref{fig12}, the elevation at $r=0$ is now significantly more for the solutions which include bound components compared to the free-wave (the first order solution). This behaviour, typical of nonlinear oscillators, should be contrasted against that of the free wave (a linear oscillator) which generates an elevation at $\tilde{t} = 1$ of the same magnitude as the depression at $\tilde{t} = 0.5$.}\\\\
\textcolor{black}{In order to facilitate comparison of the localised initial deformation of current interest, against the time-periodic solution by \cite{mack1962periodic}, it is useful to express eqn. \ref{mack-1a} as a linear superposition over Bessel functions. For this, we need to express the $T_i(r),\;i=0,1,2,3$ as Fourier-Bessel series.  Note that for time-periodic solution, $\eta_{m}(t;\tilde{\epsilon})$ in eqn. ~\ref{mack-1b} are also time-periodic and hence may be expressed as Fourier series} i.e.
\begin{eqnarray}
\eta(r,\tilde{t};\tilde{\epsilon}) &=& \sum_{m=1}^{N}\eta_m(\tilde{t};\tilde{\epsilon})J_0(k_mr) = \sum_{m=1}^{N}\left(\sum_{j=0}^{3}C_m^{(j)}(\tilde{\epsilon})\cos(2\pi j\tilde{t})\right)J_0(k_mr)   \nonumber \\
&=& \sum_{j=0}^{3}\left(\sum_{m=1}^{N}C_m^{(j)}(\tilde{\epsilon}) J_0(k_mr)\right)\cos(2\pi j\tilde{t}) \equiv \sum_{j=0}^{3}T_j(r;\tilde{\epsilon})\cos(2\pi j\tilde{t}) \label{mack-1b}
\end{eqnarray}
\textcolor{black}{where $\eta_{m}(\tilde{t};\tilde{\epsilon}) \equiv C_m^{(j)}(\tilde{\epsilon})\cos(2\pi j\tilde{t})$ and the $C_m^{(j)}$ are determined from orthogonality conditions by expressing the $T_j(r)$ in Fourier-Bessel series. Fig. \ref{fig13a} and \ref{fig13b} present a comparison of the coefficients $\eta_{m}$ for \cite{mack1962periodic} versus $\eta_{m}$ for a localised cavity. It is seen that the initial cavity shape whose time evolution has been studied here, have $\eta_{m}$ which are significantly different, especially for the lowest wavenumbers. Notably, for the time-periodic solution the $\eta_{m}$ change sign, whereas they are all negative for the cavity. In fig. \ref{fig14}, we compare the time evolution of the profiles in fig. \ref{fig13a}, provided as initial conditions. We refer the reader to the caption of this figure for analogous conclusions about the importance of nonlinearity at $r=0$.}

\subsubsection{Comparison with \cite{basak2021jetting}}\label{sec:3.1.2}

In this subsection, we explain the apparent significance of nonlinearity around the symmetry axis. To do this, we revisit results for the single \textcolor{black}{Bessel function} interface distortion described by $\eta(r,0)=\varepsilon\textrm{J}_0(k_5 r)$ (where $\varepsilon > 0$ corresponds to an initial crest at $r=0$ and $q=5$ is the \textcolor{black}{wavenumber excited at $t=0$}), as studied in \cite{basak2021jetting}. For this initial condition, the expression for $\eta(r,t)$ was analytically derived up to $\mathcal{O}(\varepsilon^2)$ in \cite{basak2021jetting} as:

\begin{equation}
    \eta(r,t)=\underbrace{\varepsilon \textrm{J}_0(k_5\;r)\cos(\omega_5t)}_{\text{Primary wave}}\;+\;\varepsilon^2\sum_{j=1}^\infty \left[\underbrace{\zeta_1^{(j)}\cos (\omega_{j}t)}_{\text{Free waves}}+\overbrace{\zeta_2^{(j)}\cos (2\omega_5 t)+\zeta_3^{(j)}}^{\text{Bound waves}}\right]\textrm{J}_0(k_j r),
    \label{basak}
\end{equation}

\noindent where $\zeta_1^{(j)} + \zeta_2^{(j)} + \zeta_3^{(j)} = 0, \forall j \in \mathbb{Z}^{+}$ to ensure that the initial condition is satisfied. Note that that expression \ref{basak} has been suitably modified from \cite{basak2021jetting} to make this compatible with the length and time scales in the present analysis.
Here $\varepsilon=\frac{\hat{a}_0}{\hat{R}}$, frequency $\omega_j=\sqrt{k_j(1+\alpha k_j^2)}$ and expressions for $\zeta_1^{(j)},\,\zeta_2^{(j)}$ and $\zeta_3^{(j)}$ are provided in the Appendix of \cite{basak2021jetting}.

\textcolor{black}{\footnote{We gratefully acknowledge an anonymous referee for several technical clarifications in this section.} As highlighted in eqn.~\ref{basak}, the expression for $\eta(r,t)$ comprises of three qualitatively different parts. The first term on the right hand side of eqn. \ref{basak} represents the primary wavenumber which is excited at $t=0$. This has wavenumber $k_5$ and oscillates harmonically with frequency $\omega_5$. For $\epsilon$ sufficiently large, the initial condition $\eta=\epsilon J_0(k_5r)$ represents an interface distortion which is significantly different in shape from that of the corresponding time-periodic solution by \cite{mack1962periodic} having its free wave as $k_5$. Due to this mismatch in initial shape, other ``free waves'' are generated at $t>0$ in the formula \ref{basak} and their frequency satisfy the dispersion relation i.e. Bessel functions with wavenumber $k_j$ have frequency $\omega_{j}$. Another kind of waves viz. the ``bound waves'' also appear at $\mathcal{O}(\varepsilon^2)$ and these do not satisfy the dispersion relation. These are necessary to cancel out the contribution from the free waves at $t=0$. Note that the amplitudes of the free waves viz. $\epsilon^2\zeta_1^{(j)}$ in eqn. \ref{basak}, do not evolve in time unlike that in the recent study on triadic resonant interactions among surface waves in \cite{durey2023resonant}; see the multiple scale analysis around their eqn. $4.1$. An important difference between this initial condition \citep{basak2021jetting} and the third-order solution by \cite{mack1962periodic} is that for the latter, there is only one free component and the rest are all bound-components at all $t$ whereas in the former, infinite free and bound components are generated at $t>0$.}

\begin{figure}
	\centering
	\includegraphics[scale=0.2]{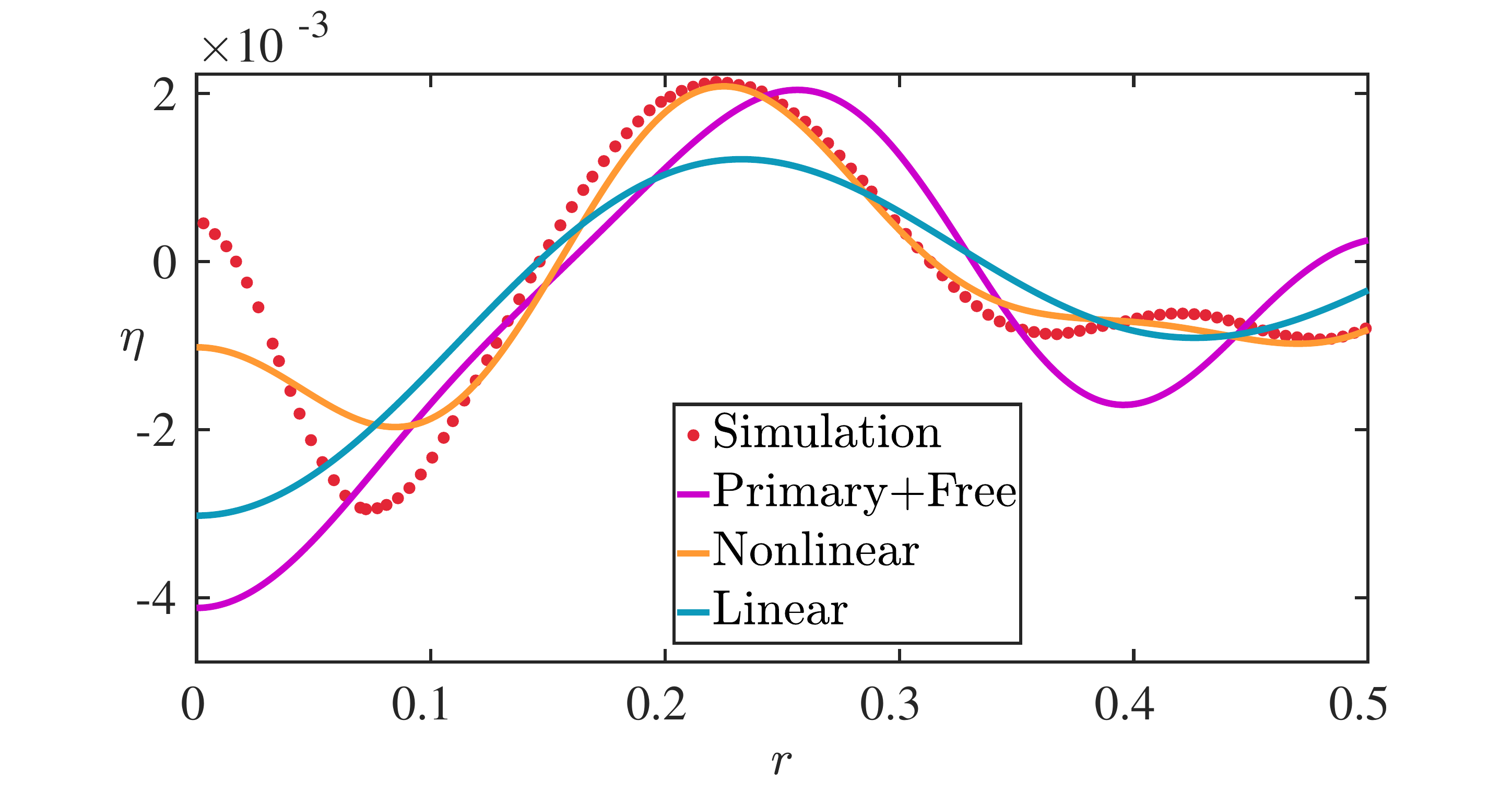}
	\caption{Various approximations for describing the dimple produced from a single \textcolor{black}{Bessel function} initial perturbation with moderately large amplitude.}
	\label{fig15}
\end{figure}

In Fig. \ref{fig15}, the interface from inviscid DNS with the initial condition $\eta(r,0) = \varepsilon \textrm{J}_0(k_5 r), \;\varepsilon = 0.03 > 0$ is shown at an instant when it forms a dimple-like protrusion at $r=0$. This is represented by the curve with red dots, labelled as `Simulation'. In the same figure, we also plot the formula from \cite{basak2021jetting}, excluding the bound components (labelled as `Primary + Free'), i.e., setting $\zeta_2^{(j)}=\zeta_3^{(j)}=0$ in equation \eqref{basak}. It is evident that this approximation does not capture the dimple, which is otherwise predicted by the full nonlinear expression (indicated as `Nonlinear' in the figure caption and referring to eqn. \ref{basak}).

\begin{figure}
	\centering
	\includegraphics[scale=0.2]{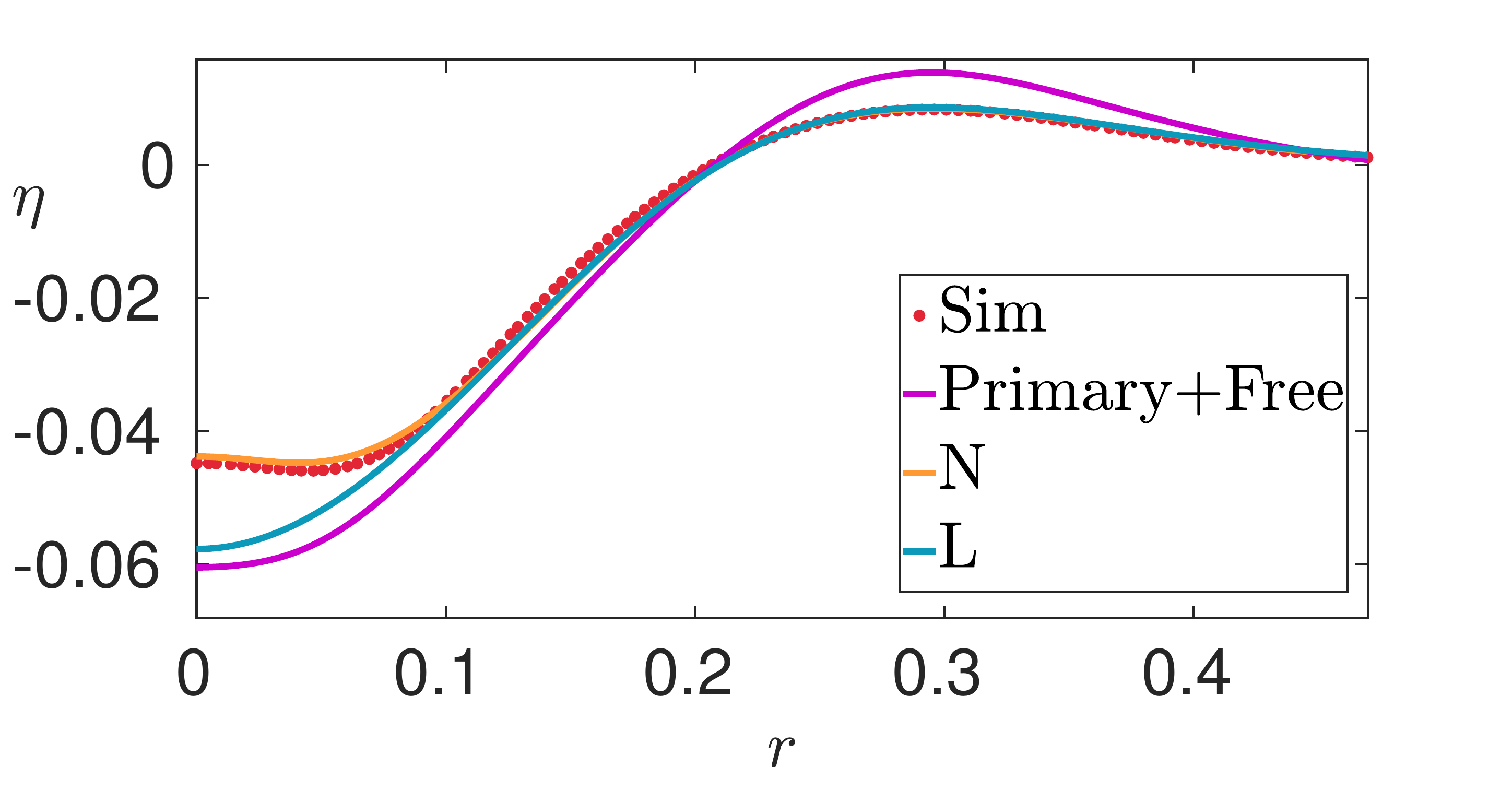}
	\caption{Shape of a dimple for a cavity with $\varepsilon = 0.091$} \label{fig16}
\end{figure}

The above exercise can also be carried out when the initial interface deformation takes the shape of a cavity. For this initial condition, $\eta(r,0) = \sum_{m=1}^{\infty} \eta_m(0) \textrm{J}_0(k_m r)$, as previously shown in fig. \ref{fig5a}. From the numerical solution to eqns. \ref{eq6}, the temporal frequency spectrum at $r=0$ is obtained. We track the time series generated by $\eta_m(t)$ and eliminate the frequencies $2\omega_m$ and $0$ from its Fourier spectrum. Fig. \ref{fig16} demonstrates that after the removal of these bound modes, the interface (labeled `Primary + Free') fails to capture the dimple shape. In contrast, the full numerical solution to eqns.~\ref{eq6} faithfully reproduces the dimple.

\subsection{Viscous effects: comparison with linear theory}
\label{subsec:linVisc}
In this section, we analyze viscous effects for the chosen initial condition. Using cylindrical coordinates, \cite{miles1968cauchy} solved the problem of free-surface waves on a viscous liquid in the linear regime within a radially unbounded domain for a continuous spectrum of wavenumbers in the radial direction. \cite{farsoiya2017axisymmetric} extended this theory to internal waves, considering viscosity and density due to both upper and lower fluids, for a single wavenumber in the initial spectrum.
Due to the availability of superposition in the linear regime, the results of \cite{farsoiya2017axisymmetric} are easily extended to initial excitations with multiple wavenumbers. In Cartesian geometry, the single wavenumber initial excitation case was first explicitly studied by \cite{prosperetti1976viscous} treating free-surface waves and by \cite{prosperetti1981motion} treating internal waves. In the Laplace domain and in cylindrical axisymmetric coordinates, the solution to the evolution of a single initial wavenumber $k_m$ was shown in \cite{farsoiya2017axisymmetric} to be given by,

\begin{equation}
	\tilde{\eta}_{m}(s)=\hat{\eta}_{m}(0)\dfrac{s+\left(4\tilde{k}_m^2\nu-\frac{4\tilde{k}_m^3\nu}{\tilde{k}_m+\sqrt{\tilde{k}_m^2+s/\nu}}\right)}{s^2+\left(4\tilde{k}_m^2\nu-\frac{4\tilde{k}_m^3\nu}{\tilde{k}_m+\sqrt{\tilde{k}_m^2+s/\nu}}\right)s+\hat{\omega}_m^2}, \quad \hat{\omega}_m^2 \equiv g\tilde{k}_m + T\tilde{k}_m^3/\rho,\quad \tilde{k}_m\equiv\frac{k_m}{\hat{R}} \label{eqn_visc1}
\end{equation}

\noindent Employing linear superposition, the corresponding (dimensional) expression for the interface evolution in the time domain for the current case becomes

\begin{equation}
	\hat{\eta}(\hat{r},\hat{t})= \sum_{m=1}^{17}\hat{\eta}_m(\hat{t})\mj_{0}\left(k_m\frac{\hat{r}}{\hat{R}}\right),\quad \hat{\eta}_m(\hat{t})\equiv \mathcal{L}^{-1}\left[\tilde{\eta}_m(s)\right]
 \label{eqn_visc2}
\end{equation}

\begin{figure}
    \centering
        \subfloat[]{\includegraphics[scale=0.28]{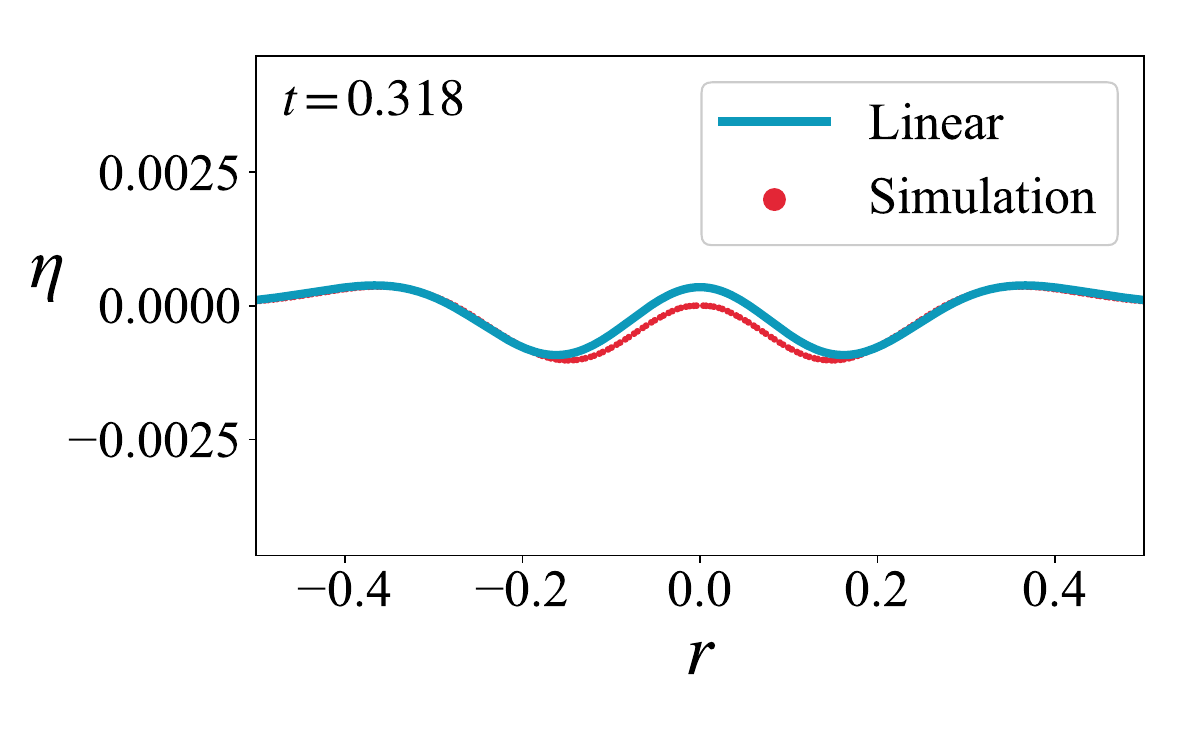}\label{fig17a}}\quad
        \subfloat[]{\includegraphics[scale=0.28]{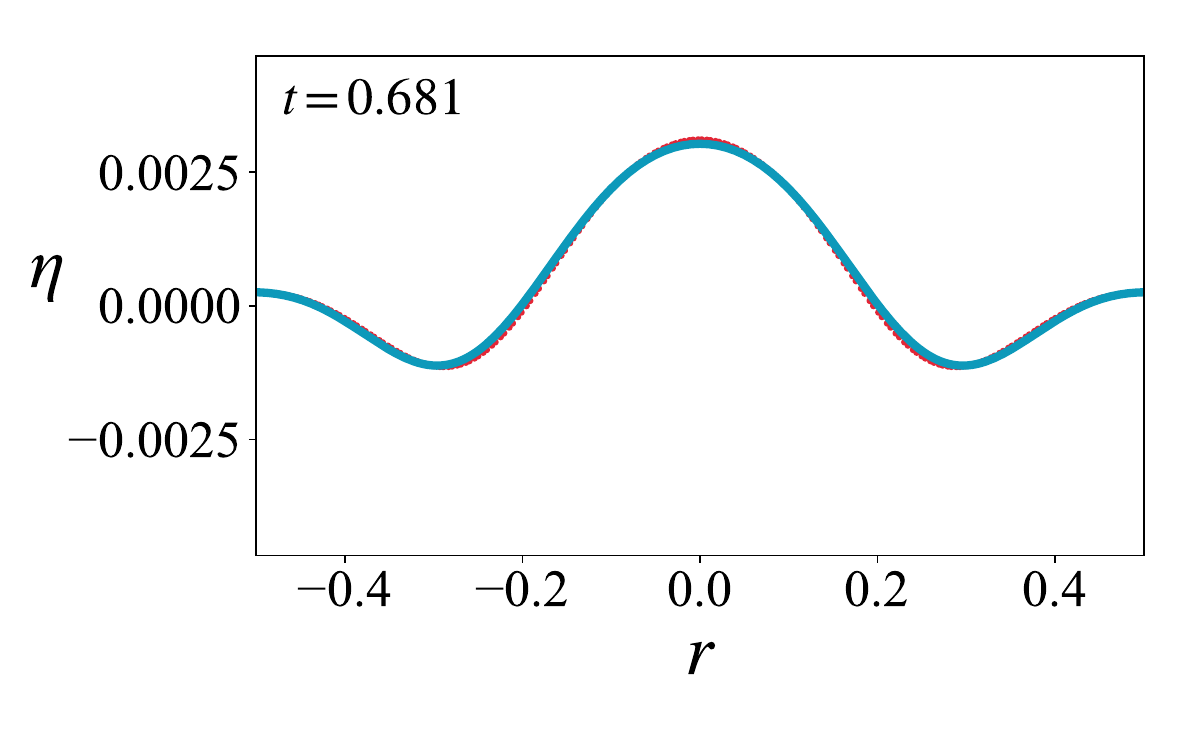}\label{fig17b}}\\
        \subfloat[]{\includegraphics[scale=0.28]{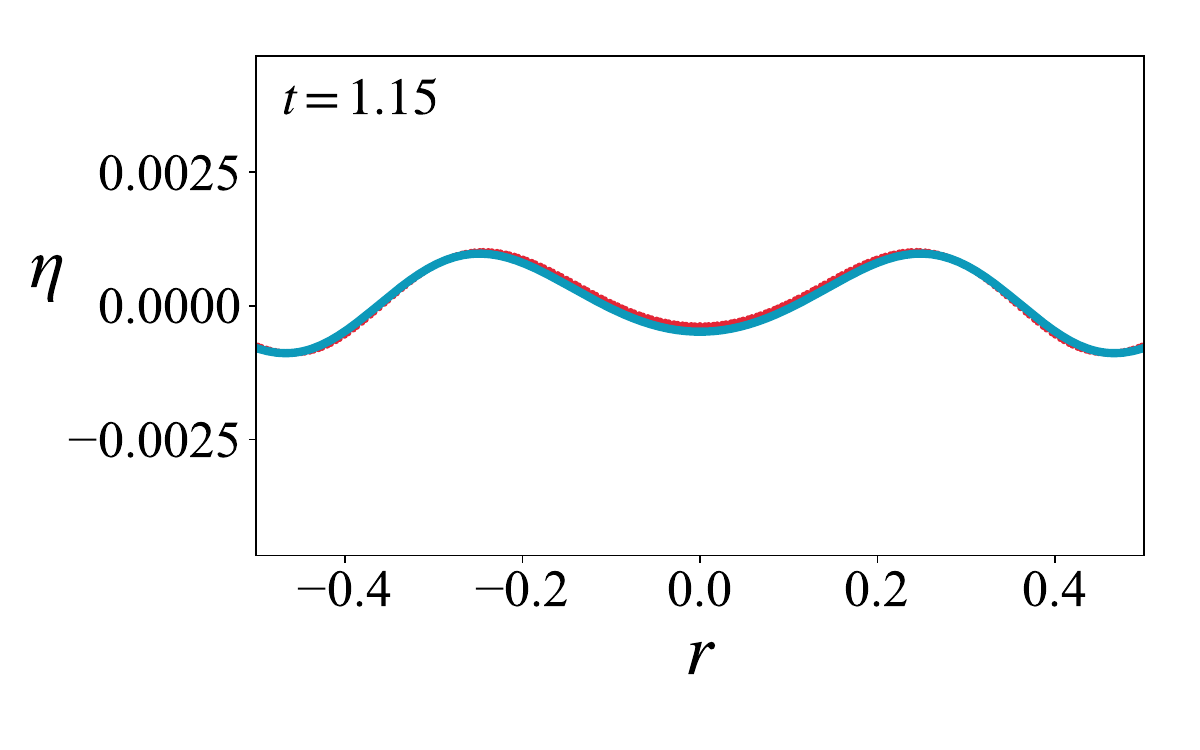}\label{fig17c}}\quad
        \subfloat[]{\includegraphics[scale=0.28]{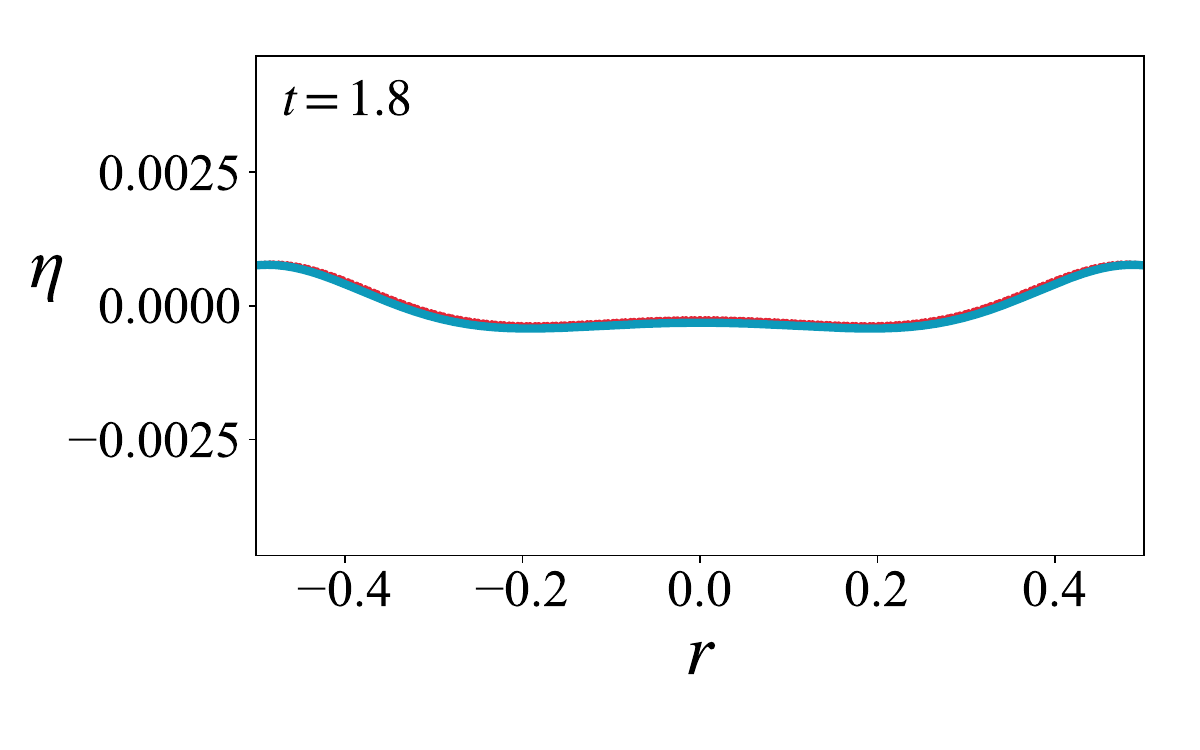}\label{fig17d}}\\
        \subfloat[]{\includegraphics[scale=0.28]{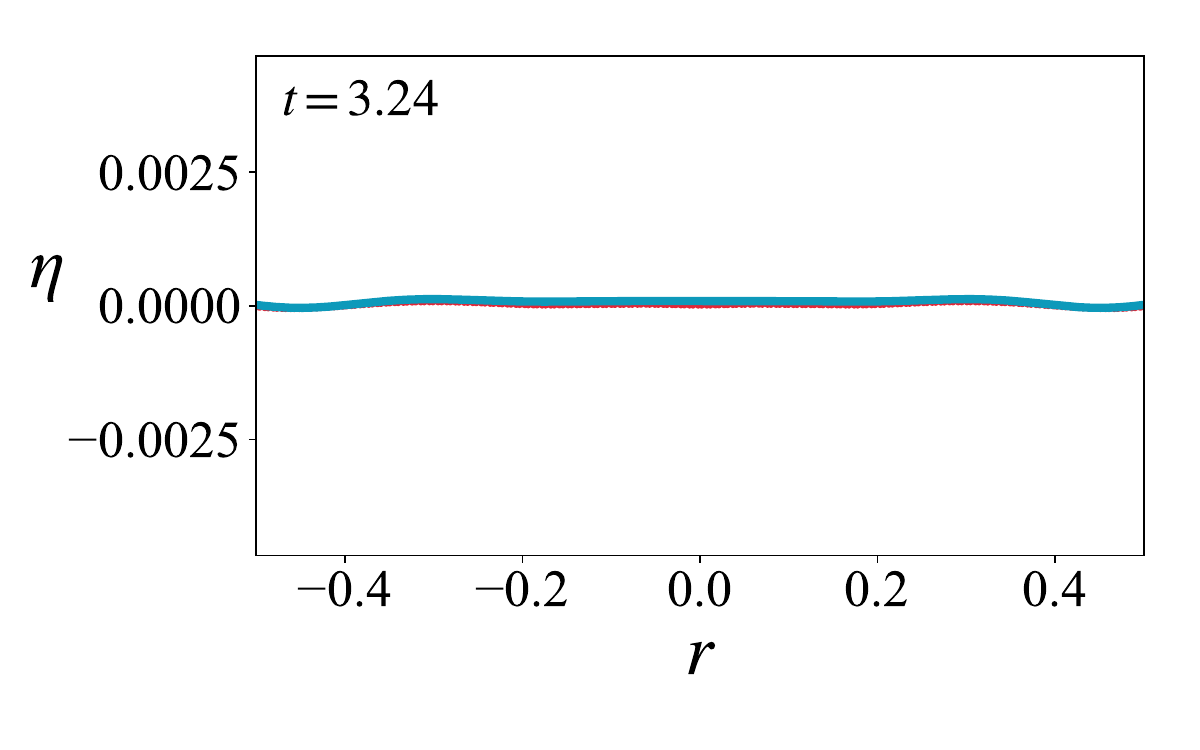}\label{fig17e}}\quad
        \subfloat[]{\includegraphics[scale=0.28]{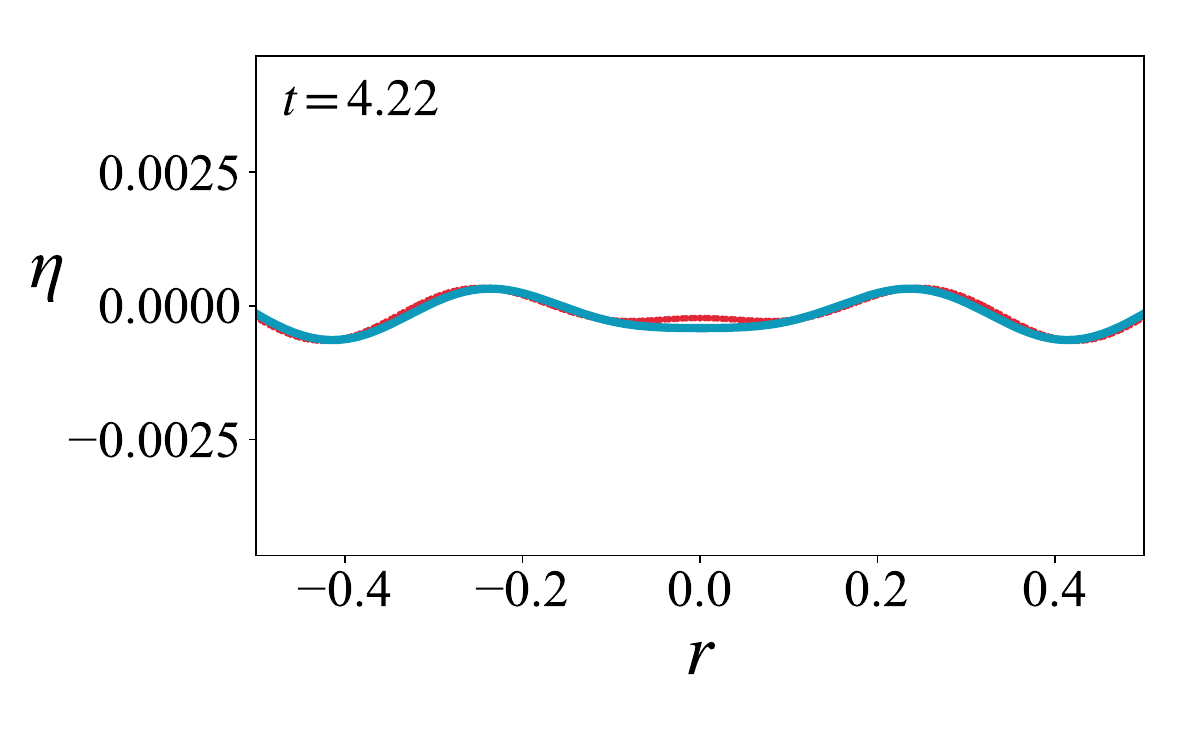}\label{fig17f}}\\
        \subfloat[]{\includegraphics[scale=0.28]{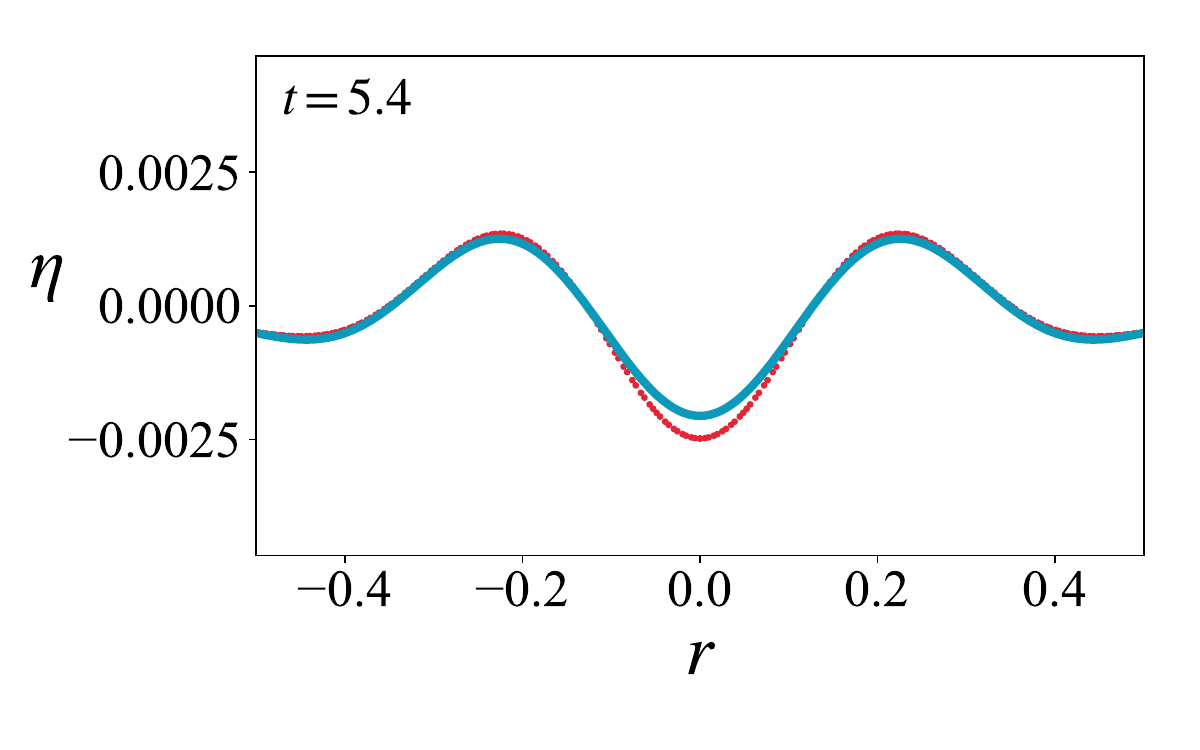}\label{fig17g}}\quad
        \subfloat[]{\includegraphics[scale=0.28]{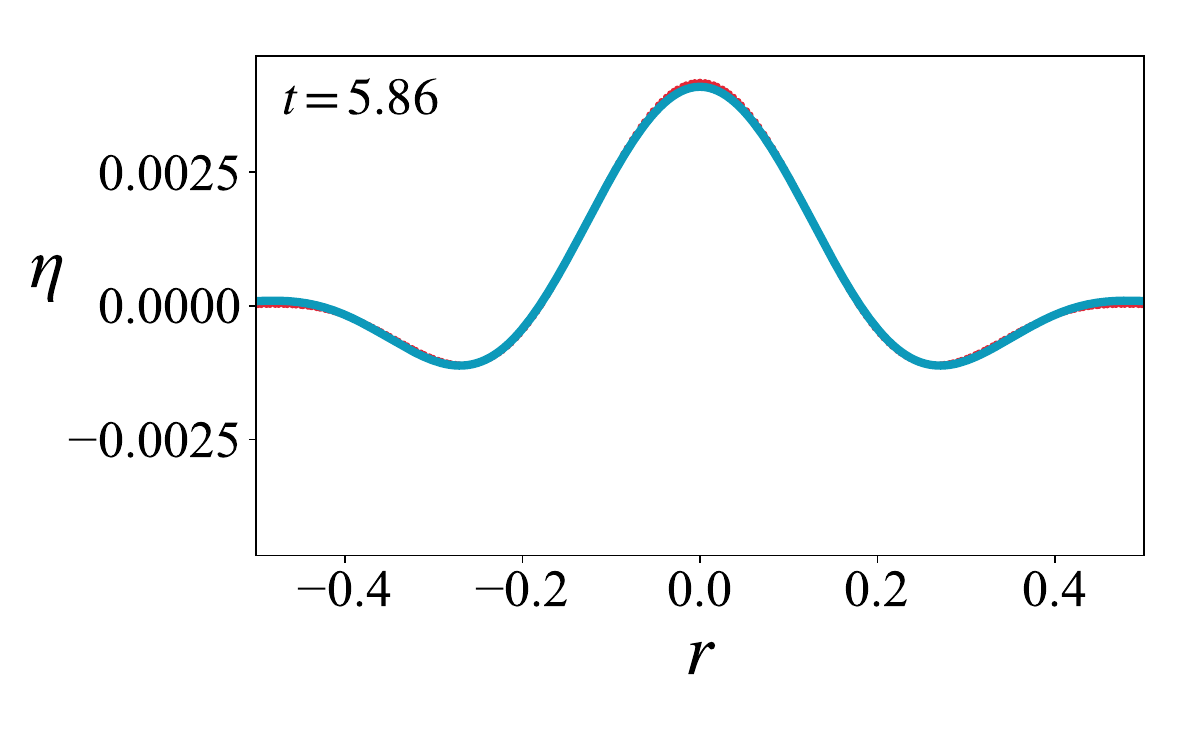}\label{fig17h}}\\
        \caption{Viscous DNS (indicated as `Simulation' with red dots in the legend to panel (a)}) with $\varepsilon=0.006$ and $Oh = 1.17\times10^{-3}$ corresponding to case $10$ in table \ref{tab:sim_params}. One notes the excellent agreement with linear, viscous theory (blue line, 'Linear', eqn. \ref{eqn_visc2} in text) with hardly any nonlinear contribution.
        \label{fig17}
    \end{figure}

\begin{figure}
        \centering
        \subfloat[]{\includegraphics[scale=0.28]{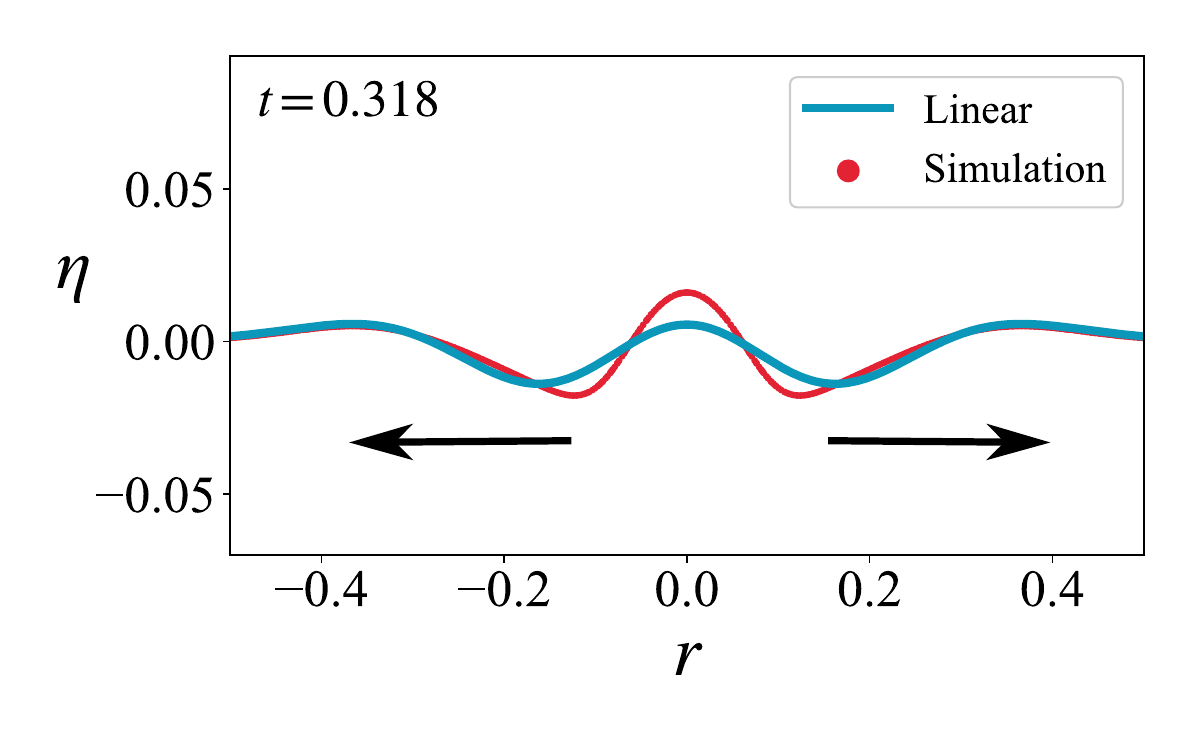}\label{fig18a}}\quad
        \subfloat[]{\includegraphics[scale=0.28]{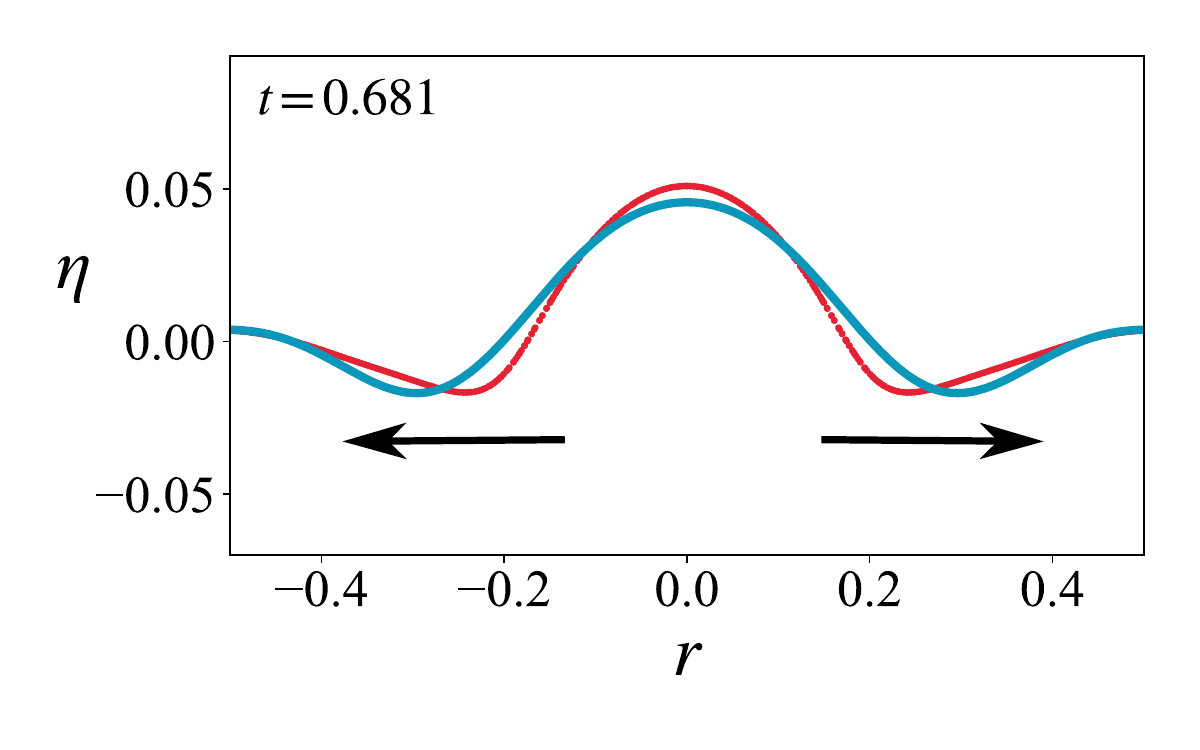}\label{fig18b}}\\
        \subfloat[]{\includegraphics[scale=0.28]{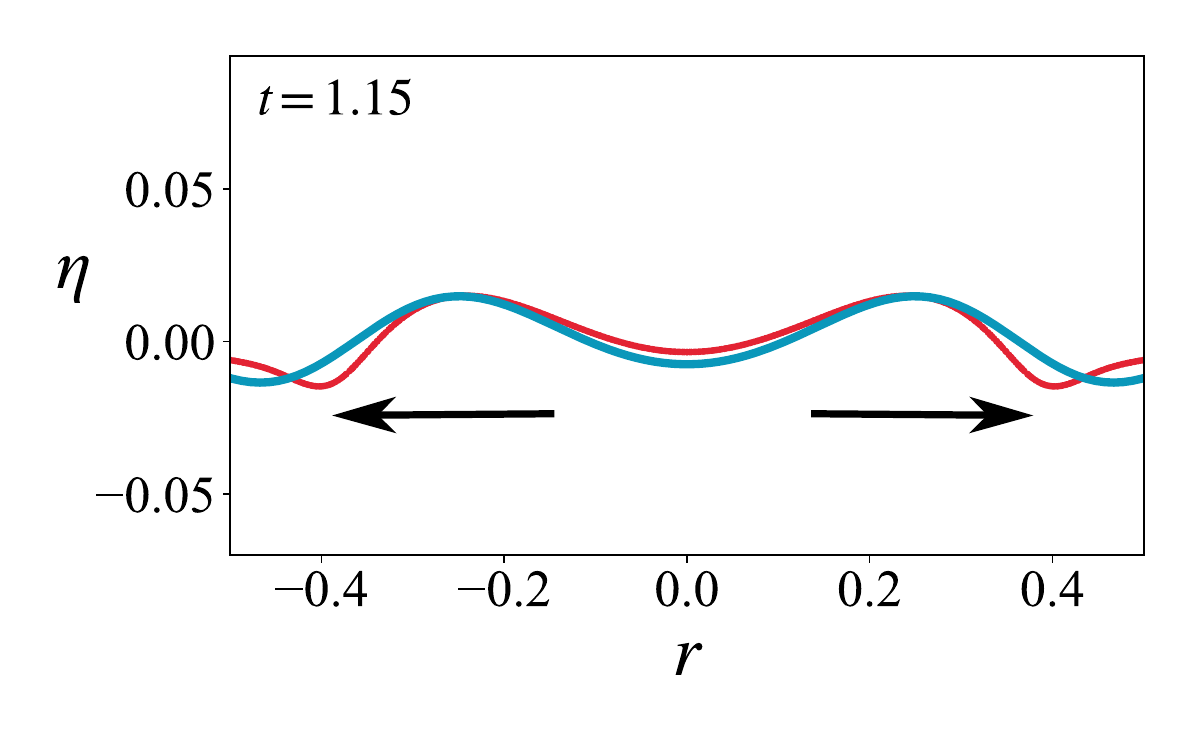}\label{fig18c}}\quad
        \subfloat[]{\includegraphics[scale=0.28]{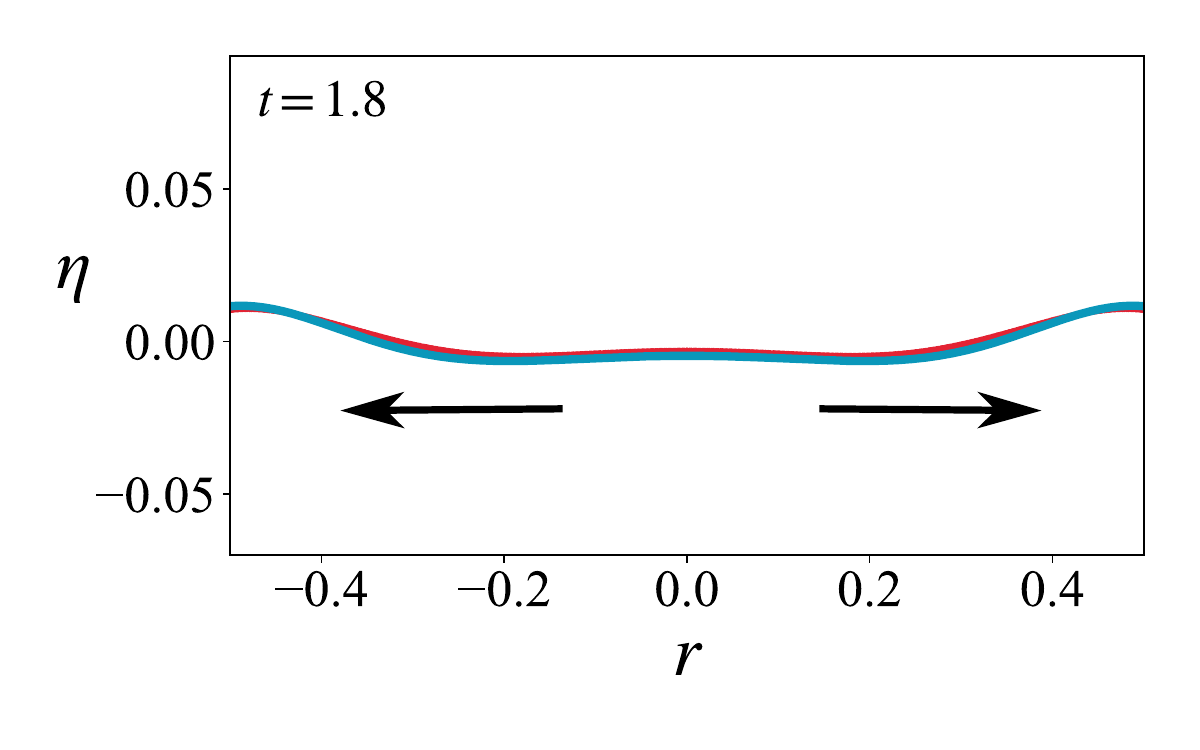}\label{fig18d}}\\
        \subfloat[]{\includegraphics[scale=0.28]{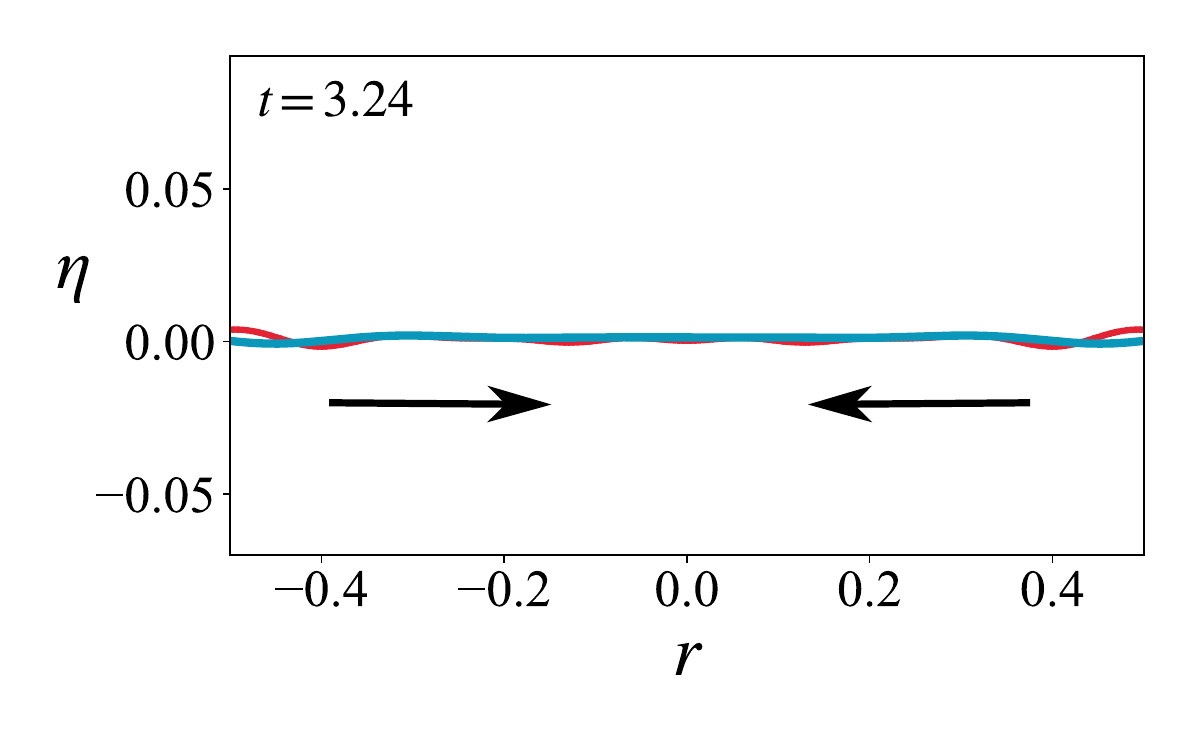}\label{fig18e}}\quad
        \subfloat[]{\includegraphics[scale=0.28]{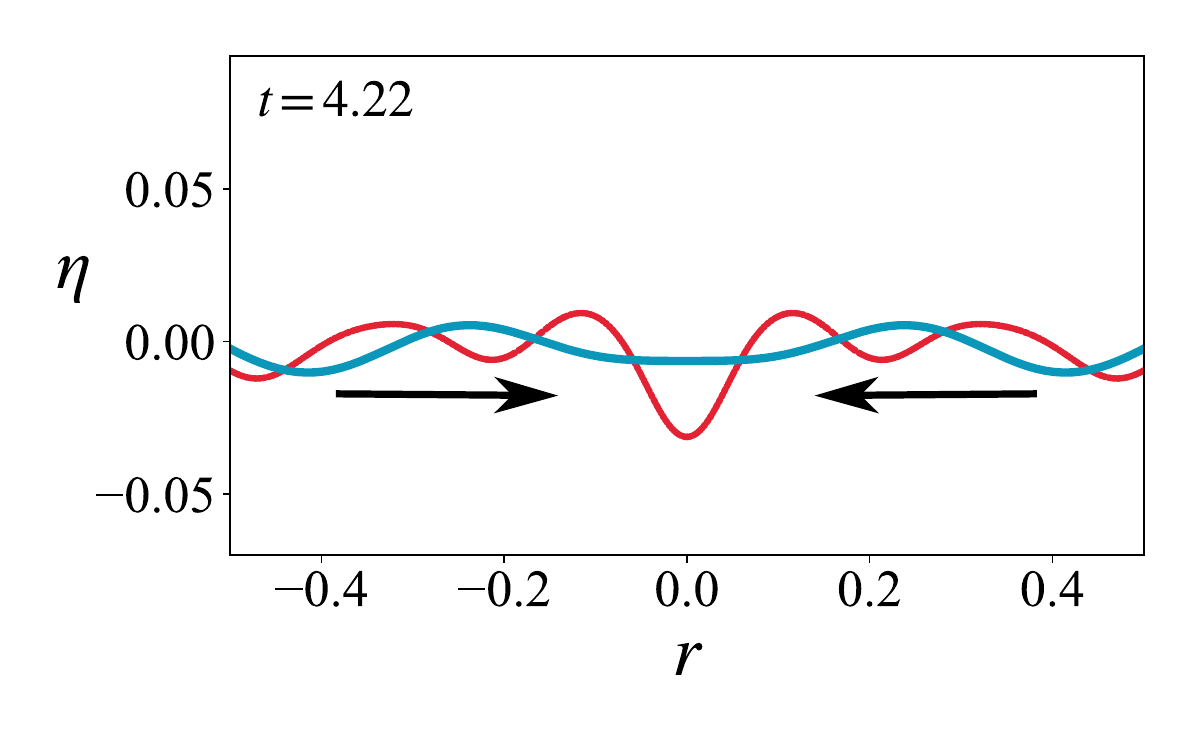}\label{fig18f}}\\
        \subfloat[]{\includegraphics[scale=0.28]{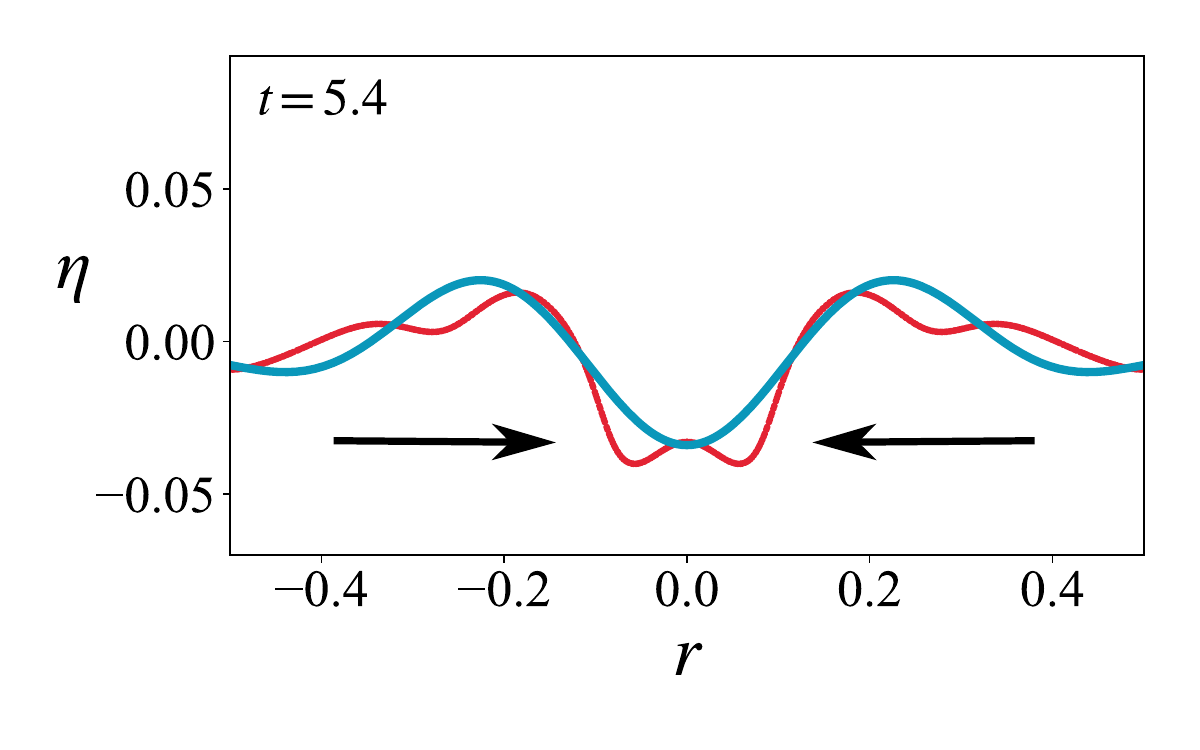}\label{fig18g}}\quad
        \subfloat[]{\includegraphics[scale=0.28]{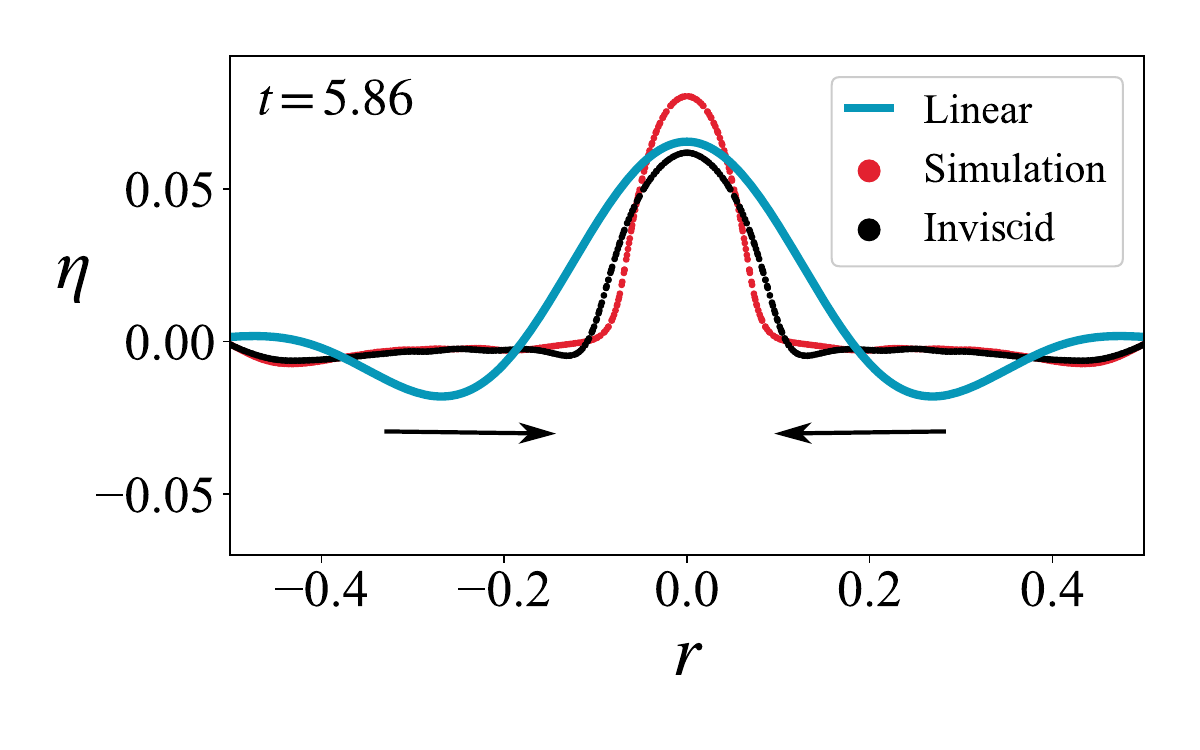}\label{fig18h}}\\
        \caption{Viscous DNS (indicated as `Simulation' with red dots in the legend to panel (a)) with $\varepsilon=0.091$ and $Oh = 1.17\times10^{-4}$ corresponding to case $4$ in table \ref{tab:sim_params}. In contrast to fig. \ref{fig17}, increasing the value of $\varepsilon$ and a corresponding reduction in viscosity, has a dramatic effect in the simulations. We note that viscous linear theory is no longer adequate particularly during the focussing process in panels (f)-(h). In panel (h), we also provide a comparison of the interface at this time-instant, for the inviscid numerical simulation ($Oh=0$) with the same $\varepsilon$. It is seen that the viscous simulation has a crest which at the indicated instant of time, is taller than the one obtained from the inviscid simulation.}
        \label{fig18}
    \end{figure}

Here $\mathcal{L}^{-1}$ is the inverse Laplace operator. We stress that expression~\ref{eqn_visc1} accounts for dissipation in the bulk liquid and boundary layer, as demonstrated by \cite{prosperetti1976viscous} in Cartesian coordinates and by \cite{farsoiya2017axisymmetric} in cylindrical coordinates. Equation~\ref{eqn_visc2} is compared with DNS for two different values of $\varepsilon$ and $Oh$ in Figs.~\ref{fig17} and~\ref{fig18}, where inverse Laplace transforms were performed using the Cohen method by \cite{Cohen} which is a default method in  \cite{mpmath}, a free Python library for arbitrary-precision floating-point arithmetic. See \cite{url} for the code. Fig.~\ref{fig17} benchmarks the theory at a relatively small $\varepsilon = 0.006$, where linear viscous theory is expected to be accurate. Excellent agreement with linear viscous theory is observed in Fig. \ref{fig17}. Conversely, Fig.~\ref{fig18} shows a clear distinction between linear and nonlinear predictions.

To further investigate the impact of viscosity, Fig.~\ref{19_a} presents the interfacial velocity at $r=0$ from DNS for various $Oh$ values \textcolor{black}{while Fig.~\ref{19_b} represents the interface displacement at $r=0$, in the same time window}. The most notable observation is that the peak velocity at $r=0$ during wave focussing occurs in the \textit{viscous} simulation rather than the inviscid one.
This non-monotonous behavior as a function of the Ohnesorge number is a well-known phenomenon in other contexts \citep{duchemin2002jet,ghabache2014physics}, indicating a significant effect of viscosity. In our analysis of converging waves, we attribute the observed non-monotonic behavior to viscous dissipation within the boundary layer at the gas-liquid interface. Even as the Ohnesorge number approaches zero ($Oh = 0^+$), this boundary layer remains significant, similar to the dissipative anomaly seen in fully developed turbulence \citep{prandtl1904, onsager1949statistical, dubrulle2019beyond, eggers2018role} and recently explored in contexts such as sheet retraction \citep{sanjay2022taylor} and drop impact \citep{sanjay2023does, sanjay2024unifying} interfacial flows. Consequently, this non-zero viscous dissipation intensifies the focussing of capillary waves, thereby increasing the velocity at the center ($r = 0$). To validate this hypothesis, in the next section, we next employ the viscous potential flow approach, which accounts for bulk viscous dissipation but neglects dissipation in the gas-liquid boundary layer, to model the converging waves. 

\textcolor{black}{We emphasize that direct numerical simulations for $Oh < 1.17 \times 10^{-4}$ exhibit grid dependency, as indicated by the pink shaded region in figs. ~\ref{anomaly_lin} and ~\ref{anomaly_nonlin}. This dependency arises from insufficient grid resolution to properly resolve the boundary layer in low-viscosity liquids, a challenge analogous to those encountered in wall-bounded turbulence studies \citep{lohse2024ultimate} and classical contact line simulations \citep{snoeijer2013moving}. These fields continue to grapple with resolving multiple scales spanning orders of magnitude. We designate this unresolved region in pink, highlighting an open problem for future multi-scale simulations.}
\textcolor{black}{The $Oh=0$ simulation, represented by symbols in figure \ref{19_a}, demonstrates grid dependency in velocity at $r=0$ for resolutions up to $2048^2$ (maximum adaptive level). This manifests as isolated `spike' points in figure \ref{inviscid_velocity}. In contrast, the nonlinear analytical prediction, depicted by the green curve labeled `N' in figure \ref{19_a}, does not exhibit such spikes.
In the (inviscid) Euler limit, our results align with inviscid non-linear theory (see figs.~\ref{fig9} and \ref{fig10}). However, we stress that this scenario also exhibits grid dependency. The one-fluid approximation used in Basilisk to solve Euler equations creates an over-constrained system by enforcing continuity of tangential velocity at the gas-liquid interface, which is incompatible with Euler equations. Consequently, indefinite grid refinement generates deviations, as evident in panel (a) of fig.~\ref{velocity_grid_convergence}. Lastly, despite setting $Oh = 0$, our simulations retain a non-zero, grid-dependent viscosity. These factors should be considered when interpreting comparisons between inviscid, potential flow theory (where tangential velocity at the interface is discontinuous) and our numerical results obtained from Basilisk.}

\subsection{Viscous potential flow}
To further elucidate viscous effects, we incorporate viscosity into the nonlinear equations using the viscous potential flow model \citep{joseph2006potential}. Unlike the linear case discussed previously, this method does not account for the boundary layer formed at the free surface, since it does not enforce the zero shear stress boundary condition \citep{moore1963boundary}.
As is well-known, in this approach the normal stress boundary condition (eqn. \ref{eq1} c) is modified to incorporate the effect of bulk viscous damping to obtain

\begin{eqnarray}
	&&\left(\frac{\partial\phi}{\partial t}\right)_{z=\eta}+\eta+2\;b\;Oh\; \sqrt{\alpha}\; \left(\frac{\partial^2\phi}{\partial z^2}\right)_{z=\eta}+\frac{1}{2}\left\lbrace\left(\frac{\partial\phi}{\partial r}\right)^2+\left(\frac{\partial\phi}{\partial z}\right)^2\right\rbrace_{z=\eta} \nonumber \\
	&&-\alpha\left(\frac{\partial^2\eta}{\partial r^2}+\frac{1}{r}\frac{\partial\eta}{\partial r}\right) =0
	\label{vp1}
\end{eqnarray}

\noindent We follow the same strategy as the inviscid case and obtain a modified differential equation for $\eta_n$, i.e. the viscous counterpart of eqn. \ref{eq6} leading to

\begin{eqnarray}\label{vp2}
	&&\dfrac{d^2\eta_n}{d t^2}+\omega_n^2\eta_n+2\;b\;Oh\; \sqrt{\alpha}\; k_n^2 \frac{d\eta_n}{dt}+2b\;Oh\;\sqrt{\alpha}\; k_n \sum_{m,p} k_m^2 C_{npm} \frac{d\eta_m}{dt}\eta_p \nonumber \\
	&&+ k_n\sum_{m,p}\left[1+\frac{k_p^2-k_m^2-k_n^2}{2k_mk_n}\right]C_{npm}\left(\dfrac{d^2\eta_m}{d t^2}\right)\eta_p \nonumber\\
	&+&\dfrac{1}{2}k_n\sum_{m,p}\left[1+\dfrac{k_p^2+k_m^2-k_n^2}{2k_mk_p}+\dfrac{k_p^2-k_m^2-k_n^2}{k_mk_n}\right]C_{npm}\left(\dfrac{d\eta_m}{d t}\right)\left(\frac{d\eta_p}{d t}\right)=0
\end{eqnarray}

In fig. \ref{fig20}, we compare the nonlinear analytical inviscid solution (referred to as `Inviscid' in the legend), the viscous potential flow (VPF) solution for $\varepsilon = 0.091$ and the viscous DNS (referred to as `Simulation') for Case $4$ in table \ref{tab:sim_params}. It is seen that the VPF solution, is indistinguishable from the inviscid one in the limit of $Oh = 0^+$, highlighting the importance of resolving the viscous boundary layer in theory.

\begin{figure}
        \centering
        \subfloat[]{\includegraphics[scale=0.25]{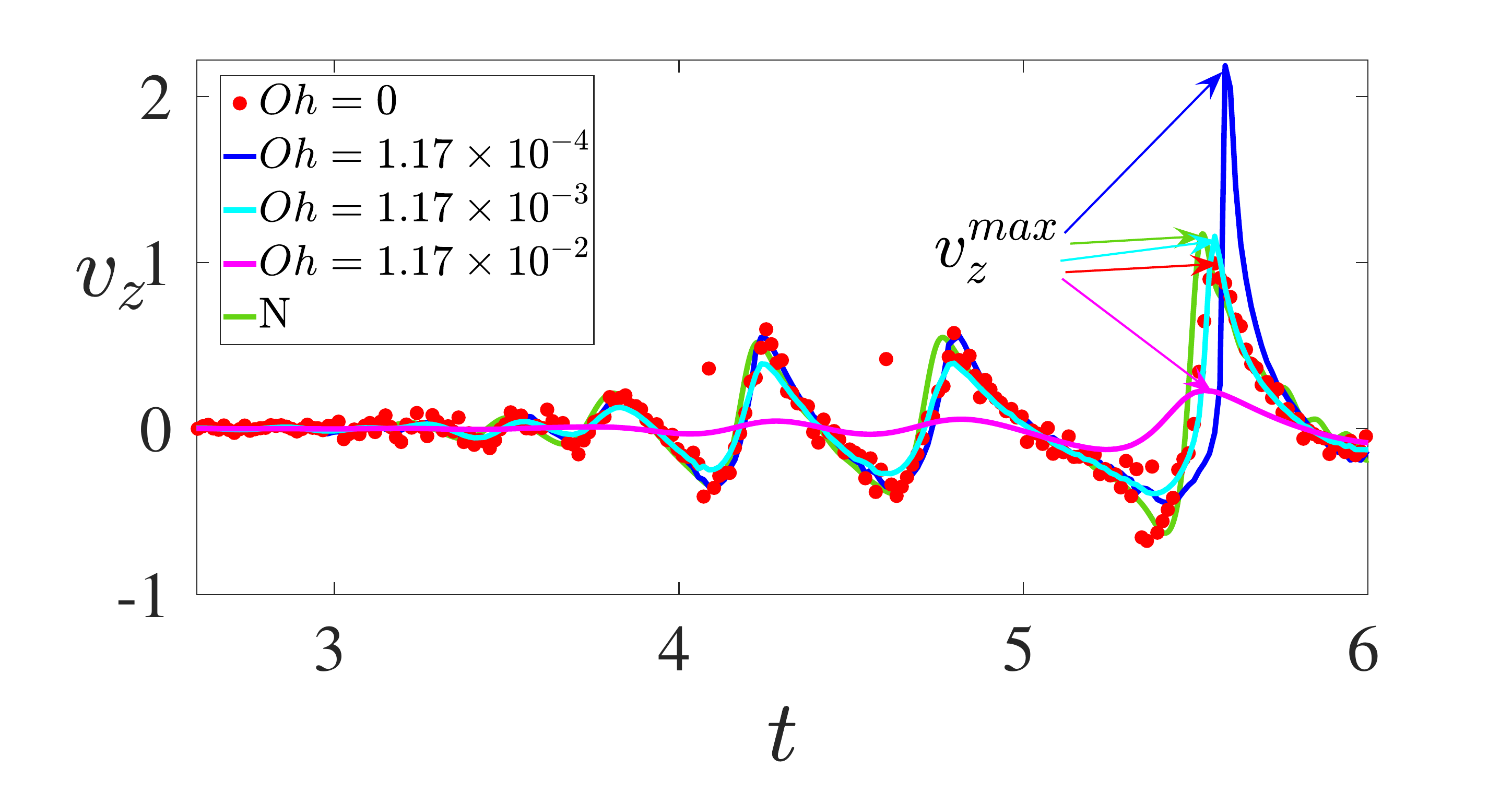}\label{19_a}}\\
        \subfloat[]
        {\includegraphics[scale=0.25]{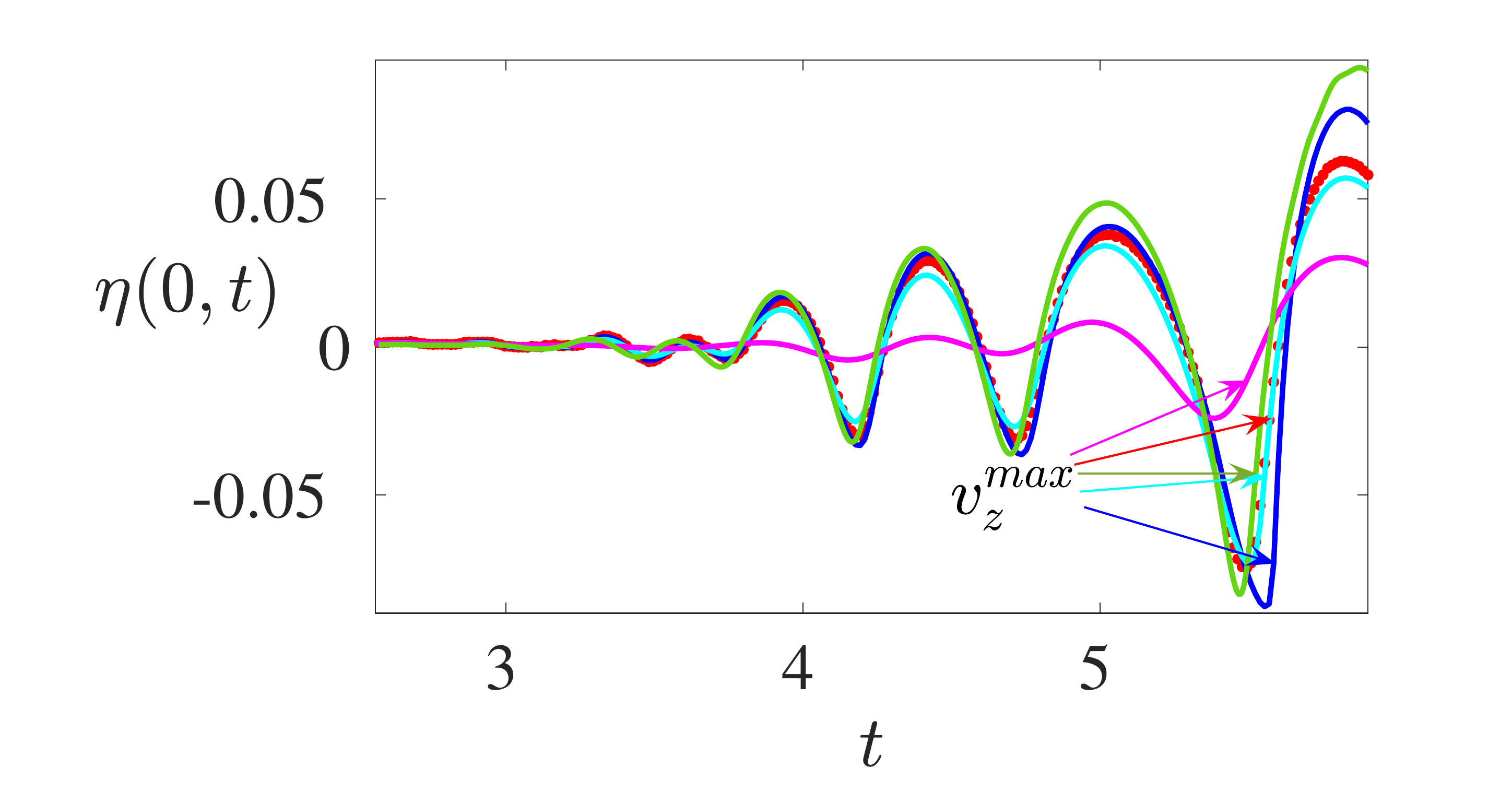}\label{19_b}}
        \caption{(a) Velocity at the interface at $\hat{r}=0$ for different values of $Oh$ and fixed $\varepsilon = 0.091$. Note that the viscous DNS for $Oh = 1.17\times10^{-4}$ (solid deep blue line) produces the largest velocity peak around $t\approx 5.7$. Note in particular that the inviscid signal ($Oh=0$, red symbols) has a peak which is shorter by a factor of half. \textcolor{black}{This difference is because in the $Oh=0$ case, we are not resolving the \textit{numerically} generated boundary-layer at the current grid resolution. As discussed in the text, this introduces a degree of grid dependency in the inviscid simulations which cannot be resolved in the numerical framework of the open-source code Basilisk \citep{popinet-basilisk}. However for $Oh\geq 1.17\times10^{-4}$, we are resolving the boundary layers and the results are grid convergent. (b) The interface height $\eta(0,t)$ with the same color scheme as panel (a).} 
        We refer the reader to Appendix C where the grid convergence results for this (and other) simulations are provided. 
        }
        \label{fig19}
     \end{figure}

\begin{figure}
        \centering
        \subfloat[]{\includegraphics[scale=0.2]{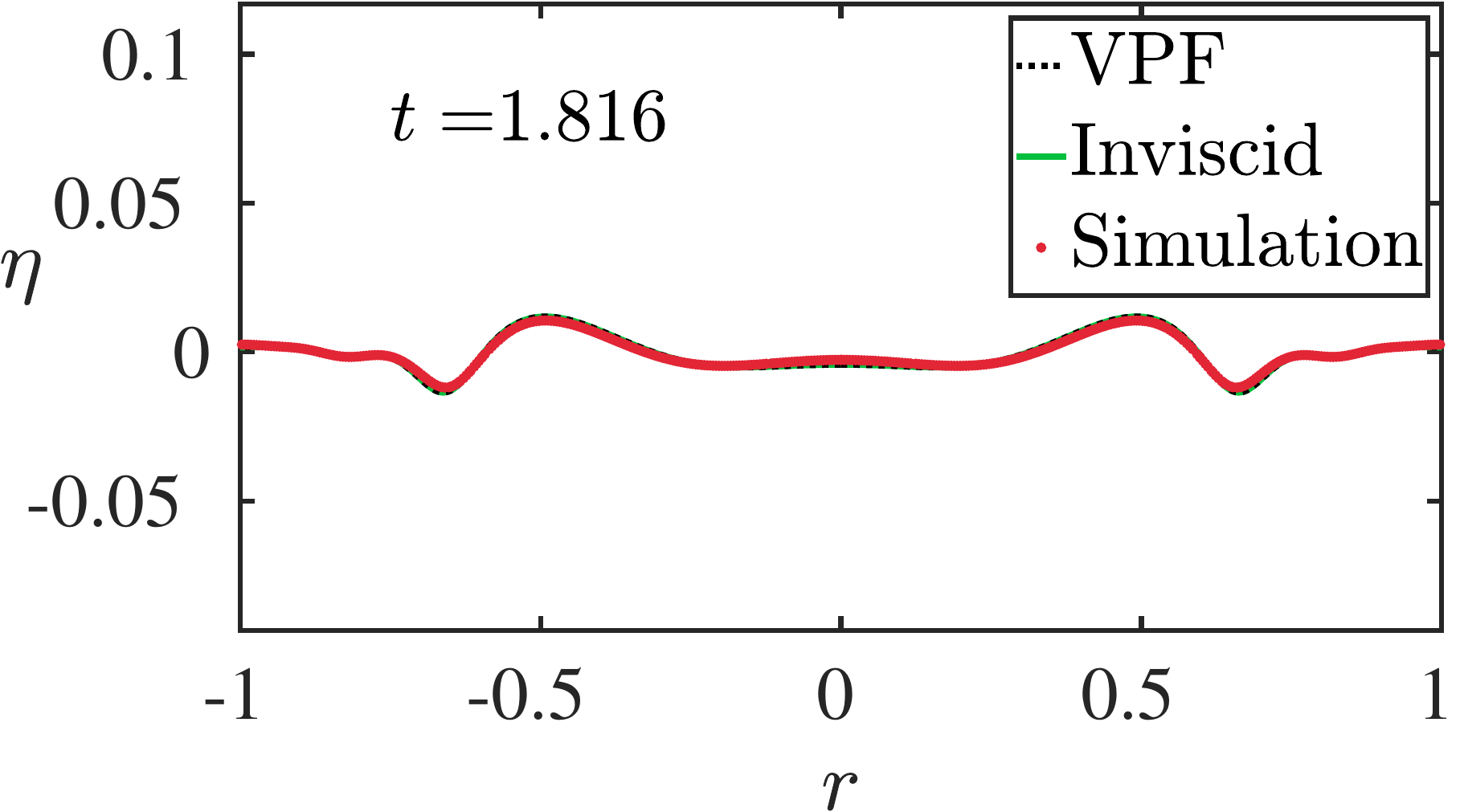}\label{vp_1}}\quad
        \subfloat[]{\includegraphics[scale=0.2]{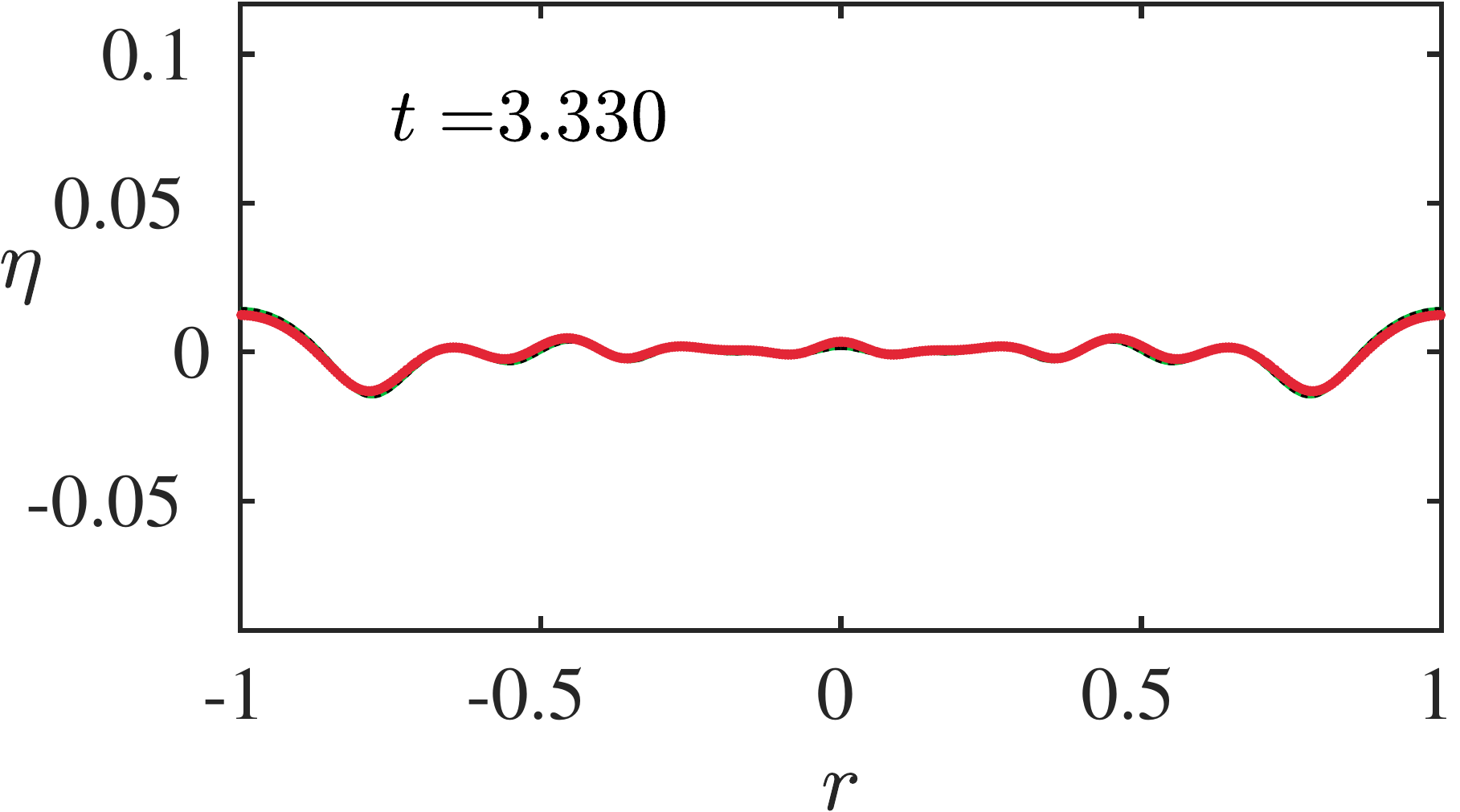}\label{vp_2}}\\
        \subfloat[]{\includegraphics[scale=0.2]{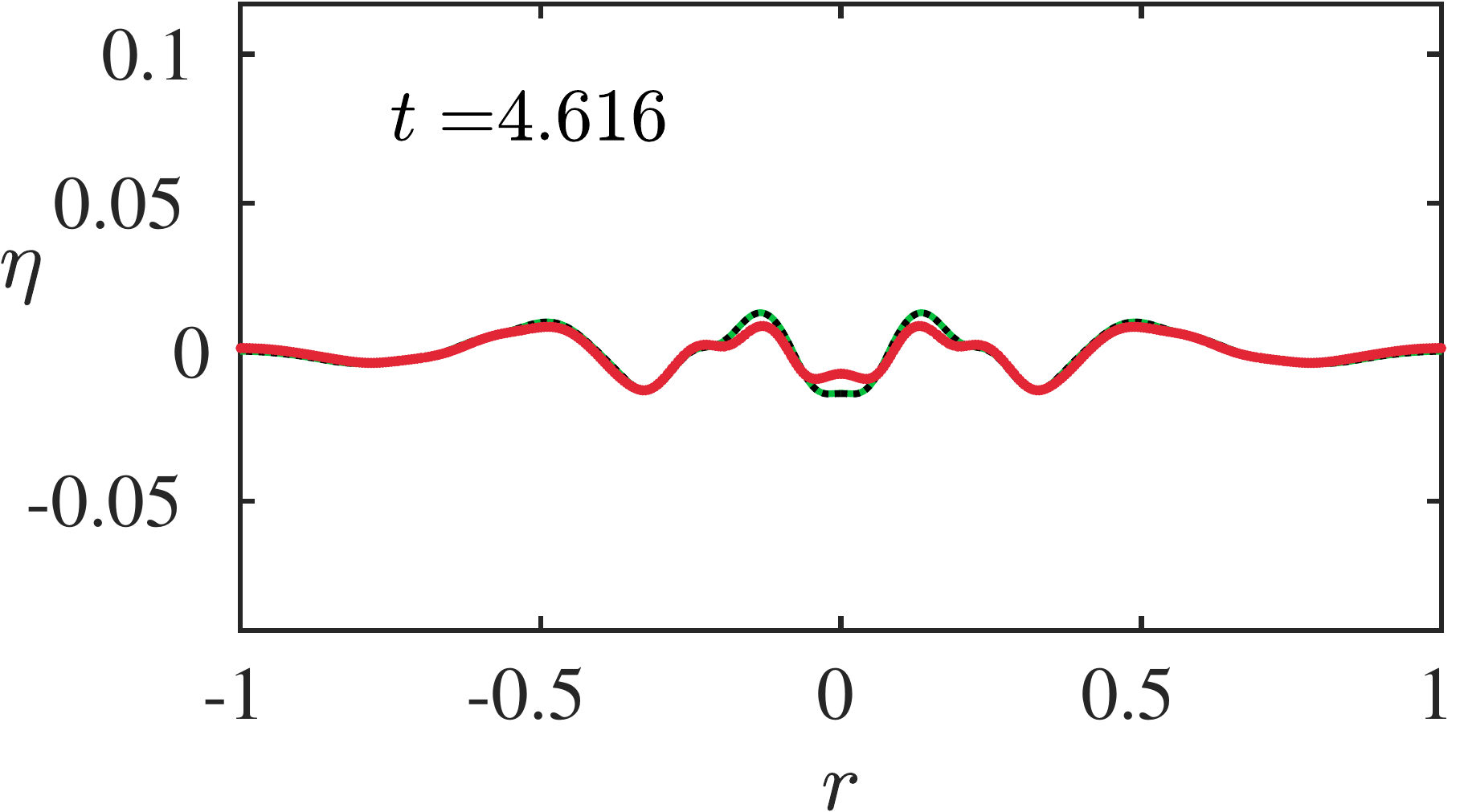}\label{vp_3}}\quad
        \subfloat[]{\includegraphics[scale=0.2]{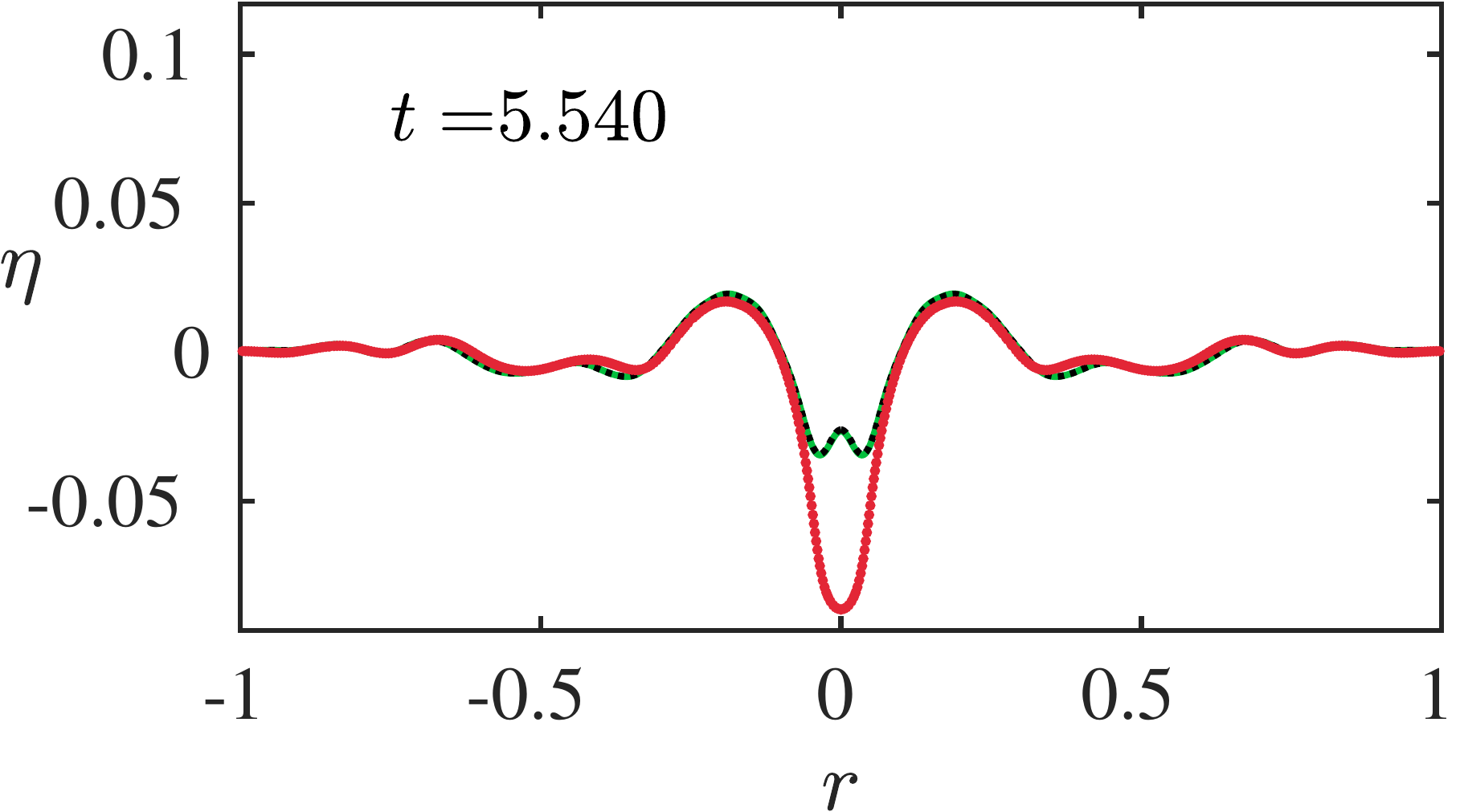}\label{vp_4}}
        \caption{Comparison of the viscous potential solution (VPF, black dotted line), inviscid solution (green solid line) and DNS (red dots) at $\varepsilon=0.091$ and $Oh=1.17\times 10^{-4}$, case $4$ in table \ref{tab:sim_params}. 
        }
        \label{fig20}
    \end{figure}

\begin{figure}
        \centering
        \subfloat[Shallow cavity with $\varepsilon = 0.006$]{\includegraphics[scale=0.23]{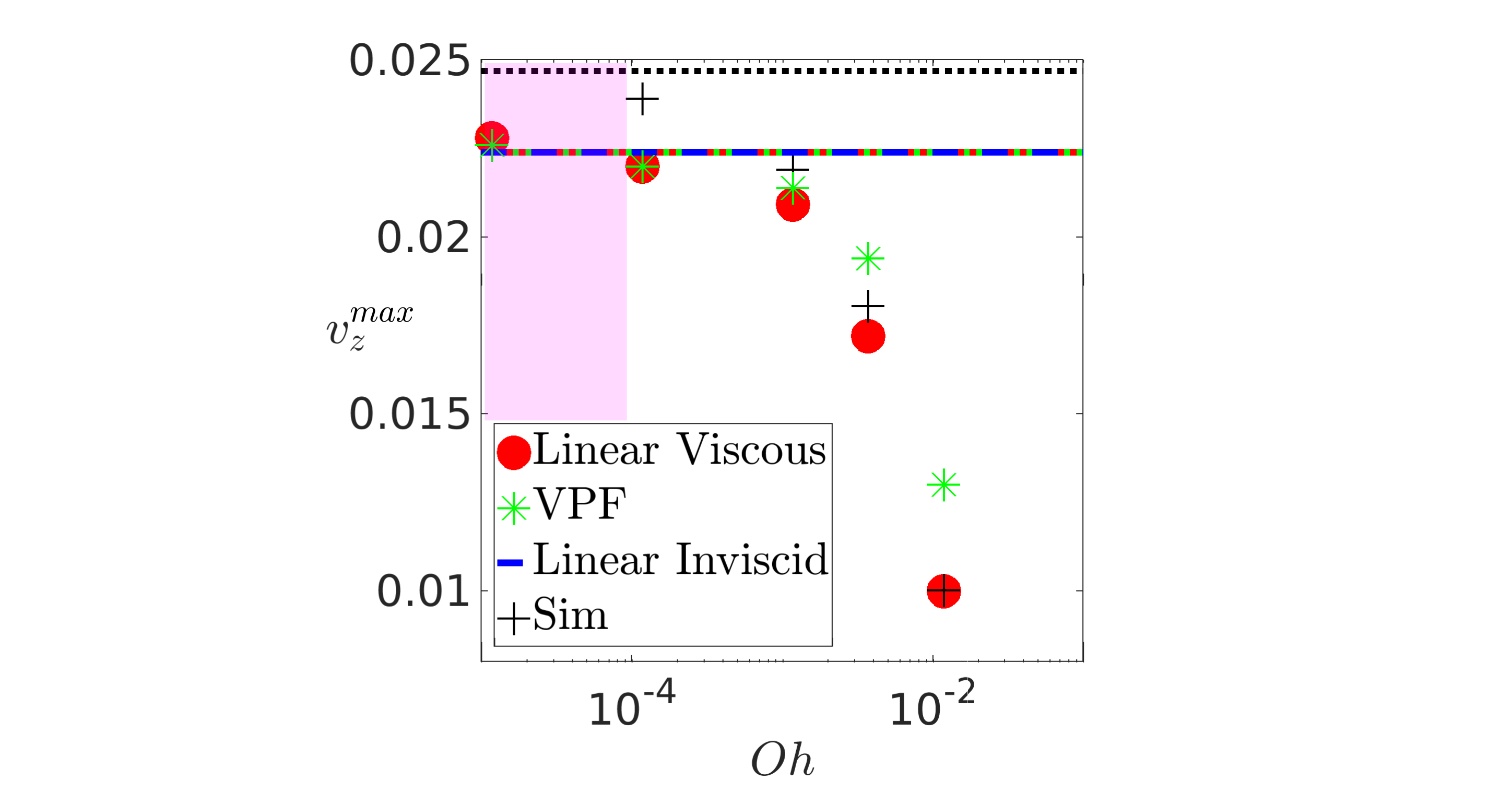}\label{anomaly_lin}}\quad
        \subfloat[Deep cavity with $\varepsilon=0.091$]{\includegraphics[scale=0.23]{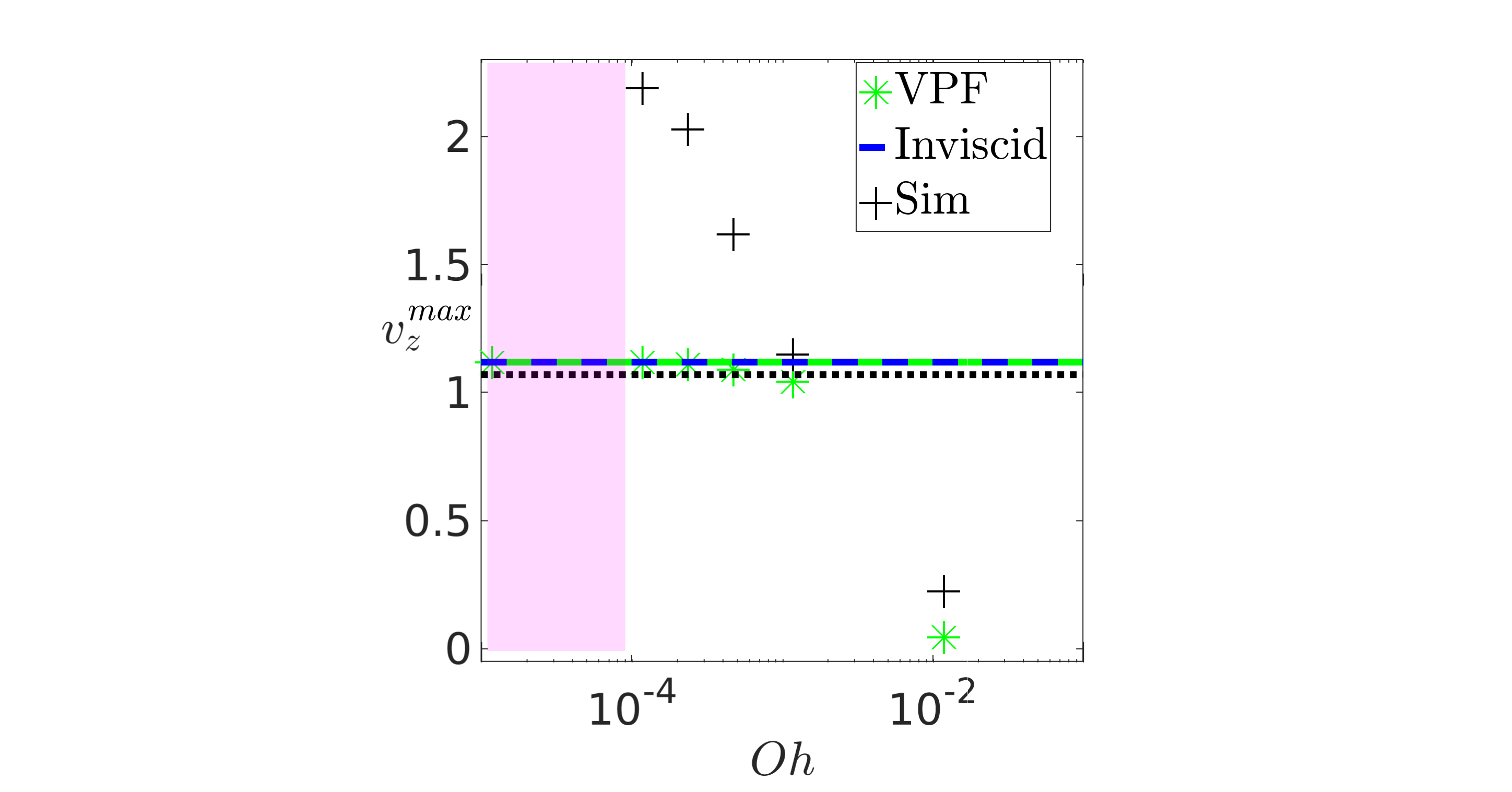}\label{anomaly_nonlin}}
        \caption{Comparison of the maximum velocity at $r=0$ i.e. $v_z^{max}$ (see arrows in fig.~\ref{19_a}) after reflection for different Ohnesorge number for a shallow cavity, cases $7,9,10,11,12$ in table \ref{tab:sim_params} (panel a) and for a deep cavity, case $2,4,5,6,13,14$ (panel b). Panel (a): `+' symbols represent DNS with finite viscosity. \textcolor{black}{Black dotted line represents DNS with zero viscosity}. Red symbols represent the linear viscous solution obtained by numerical inversion of eqn. \ref{eqn_visc1}. Green symbols indicate viscous potential flow (VPF) approximation obtained from solving eqn. \ref{vp2}. At the $Oh=0$ limit, VPF (green dashed line) and linear viscous theory (red dashed line) coincide with the linear inviscid theory (blue dashed line). Panel (b): \textcolor{black}{Symbols have the same meaning as panel (a), the only difference is that we have employed nonlinear inviscid theory (blue dashed line) in this case.} Note that non-monotonicity in the velocity at $r=0$ as a function of $Oh$. The viscous potential flow approximation (VPF) despite being nonlinear is unable to describe this non-monotonicity, presumably because of its inability to resolve the boundary layer at the free-surface. The dotted black line represents the velocity of inviscid DNS which \textcolor{black}{shows grid dependency}. In the current figure, below a certain value of $Oh$ (pink shaded region) grid dependency persists in our simulations, due to the presence of an unresolved thin boundary layer. We do not depict this data here due to the lack of this convergence. For $Oh > 1.17\times10^{-4}$ however, the boundary layers are resolved for simulation points `+' and the data are grid converged. Note that the nonlinear inviscid theory ($Oh=0$, dashed blue line) predicts $v_z^{max}(r=0)$ which is smaller than the prediction by DNS for $Oh\approx1.17\times 10^{-4}$ by a factor $\approx 2$. \textcolor{black}{A similar albeit significantly more intensification at an optimal value of $Oh$ was first noted in the case of bubble bursting in the seminal study by \citet{duchemin2002jet}; see their fig. $12$. } 
        }
        \label{anomaly}
    \end{figure}

To further quantify the comparison between these cases, in fig. \ref{anomaly_lin} shows the \textcolor{black}{maximum velocity} at the axis of symmetry within a shallow cavity during focussing. The linear viscous theory, which accounts for the boundary layer at the free surface, describes the change in $v_z$ with Ohnesorge number slightly better than the VPF model.
Fig.~\ref{anomaly_nonlin} presents results for a deeper cavity where non-linearity plays a significant role, and the non-monotonic behavior observed in fig.~\ref{19_a} as a function of $Oh$ is evident. The VPF model fails to capture this non-monotonic behavior, highlighting the importance of resolving the boundary layer at the gas-liquid interface, as discussed in \S~\ref{subsec:linVisc}.
We propose developing a nonlinear-viscous theory superior to the VPF model to explain the observations in Fig. \ref{fig19} and Fig. \ref{anomaly_nonlin} in future work.

    \section{Conclusion \& outlook}
    \label{sec:4}

    In this study, we have discussed the dynamics of a localised free-surface perturbation in a cylindrical pool of liquid, which generates a train of waves.  These waves, upon reflecting from the container walls, converge back towards the axis of symmetry, leading to progressively increasing free-surface oscillations at the center. Using the potential flow approximation, we derived a set of ordinary differential equations governing the evolution of amplitudes up to second order.

    For shallow cavities, linear theory suffices to explain the wave evolution. However, as the cavity depth increases, the limitations of linear theory become evident, particularly in predicting the focussing effects at $r=0$. Our findings demonstrate that linear dispersive focussing alone is inadequate to describe the intricate dimple shape forming at the axis of symmetry for deeper cavities. A nonlinear theory that accounts \textcolor{black}{for the generation of bound components} is found to be essential for accurately modeling the focussing process. The role of bound components is particularly critical in capturing the \textcolor{black}{interface evolution at the symmetry axis}.

    A notable observation is the significant influence of viscosity on the focussing process. \textcolor{black}{Interestingly}, the maximum velocity at the axis of symmetry is higher for a slightly viscous fluid than for an inviscid one. This non-monotonic behavior with respect to the Ohnesorge number $Oh$ is not captured by either the linear viscous model \citep{farsoiya2017axisymmetric, prosperetti1976viscous} or the nonlinear viscous potential flow (VPF) model \citep{joseph2006potential}. \textcolor{black}{The VPF model's failure, stemming from its neglect of boundary layer effects, underscores the critical role of these layers in the $Oh \to 0^+$ limit. As the VPF model converges to an inviscid solution in this limit, it further emphasizes the boundary layers' importance in velocity enhancement. The singular nature of the $Oh \to 0^+$ limit arises from fundamental disparities between Navier--Stokes and Euler equations. Even as $Oh$ approaches zero, the no-slip condition at the liquid-gas interface necessitates a boundary layer, preserving viscous effects.}

    \textcolor{black}{In conclusion, we emphasize some interesting observations and hypothesis made in \cite{zhang2008satellite} concerning capillary wave focussing, albeit on a spherical bubble unlike the flat surface treated here. Some of these find qualitative support from our theory. In page $9$, first column, first paragraph of \cite{zhang2008satellite}, the authors remark insightfully that the wave convergence process is itself not necessarily nonlinear, as the large amplitude oscillations seen in their fig. $14$ are also predicted by linear theory. However, linear amplification itself may not be enough to trigger pinch-off, emphasizing the \textit{local} importance of nonlinearity at the focal point. Our analysis establishes that this is qualitatively true, cf. fig. \ref{fig18}. In their section C (page $9$), \cite{zhang2008satellite} also emphasize that the phase and amplitude of the wave-train is very important to the convergence process. Our nonlinear analysis establishes the importance of the amplitude of the wave-train while the viscous analysis demonstrates that the VPF model (which does not resolve the interface boundary layer) is unable to capture the non-monotonic dependence of vertical velocity at $r=0$ on $Oh$. Presumably, this non-monotonocity will be predicted from a nonlinear, viscous model, which is still simpler than the full Navier-Stokes equation and this is proposed as future work. Upcoming research thus needs to develop a comprehensive, nonlinear viscous theory that incorporates boundary layer effects and also accounts for the nonlinearity associated with focussing. Additionally, extending this work to non-Newtonian fluids, such as viscoplastic or viscoelastic liquids \citep{sanjay_lohse_jalaal_2021}, could reveal new insights and broaden the applicability of our theoretical framework.}\\\\

	\noindent{\bf Acknowledgements\bf{.}}  RD thanks Prof. B. Sutherland for sharing the study by \cite{smith1976giant}. VS thanks Prof. D. Lohse for stimulating discussions on viscous anomaly.\\

\noindent{\bf Funding\bf{.}} We gratefully acknowledge financial support from DST-SERB (Govt. of India) grants MTR/2019/001240, CRG/2020/003707 and SPR/2021/000536 on topics related to waves, jet formation, cavity collapse and the viscous Cauchy-Poisson problem. The Ph.D. tenure of L.K. is supported by the Prime-Minister’s Research Fellowship (PMRF), Govt. of India and is gratefully acknowledged.\\

\noindent{\bf Declaration of Interests\bf{.}} The authors report no conflict of interest.

	\section*{Appendix A}
	\textcolor{black}{	The expressions for the third-order approximation to $\eta(r, \tilde{t};\tilde{\epsilon})$ by \cite{mack1962periodic} expressed in our notation are}
	\begin{eqnarray}
		&&T_0(r;\tilde{\epsilon}) \equiv \frac{1}{4} \tilde{\epsilon}^2 k_1(J_{01}^2 - J_{11}^2), \nonumber \\
		&&T_1(r;\tilde{\epsilon}) \equiv \tilde{\epsilon} J_{01} + \frac{\tilde{\epsilon}^3 k_1^2 }{4} \Biggl\{ \Biggr. J_{01}^3 - \frac{5}{2} J_{01} J_{11}^2 \nonumber\\
		&&\qquad  + \frac{J_{11}^3}{k_1  r} +
		\frac{1}{8} \sum_{i=1}^\infty \frac{k_{i}\Gamma_i J_{01}J_{0i}}{k_1} - \frac{1}{4} \sum_{i=1}^\infty \frac{k_{i}^2\Gamma_i J_{01}J_{0i}}{k_1^2} - \sum_{i=1}^\infty \frac{1}{8}\frac{k_{i}^2\Gamma_i J_{11}J_{1i}}{k_1^2} \nonumber \\
		&&\qquad  + \frac{1}{4} \sum_{i=1}^\infty \frac{k_{i}\Gamma_i J_{01}J_{0i}}{k_1} + \sum_{i=2}^\infty \frac{(k_i/k_1)J_{0i}}{1 - (k_i/k_1)} \Biggl[\Biggr. - \alpha_n[J_{01}^3] -\alpha_n[J_{01}J_{11}^2] \nonumber\\
		&&\qquad  + \alpha_n\left[\frac{J_{11}^3}{k_1 r}\right] -\frac{1}{4}\sum_{j=1}^\infty \frac{k_j^2 \Gamma_j \alpha_n[J_{01}J_{0j}]}{k_1^2} + \frac{1}{2}\sum_{j=1}^\infty \frac{k_j \Gamma_j \alpha_n[J_{01}J_{0j}]}{k_1} \nonumber\\
		&&\qquad -\frac{1}{8}\sum_{j=1}^\infty \frac{k_j^2 \Gamma_j\alpha_n[J_{11}J_{1j}]}{k_1^2} \Biggl.\Biggr] \Biggl.\Biggr\} \Biggl., \nonumber \\
		&&T_2(r;\tilde{\epsilon}) \equiv \frac{\tilde{\epsilon}^2 k_1}{4} \Biggl( \Biggr. J_{01}^2  -J_{11}^2 - \frac{1}{4}\sum_{i=1}^\infty \frac{k_i \Gamma_i J_{0i}}{k_1} \Biggl.\Biggr), \nonumber \\
		&& T_3(r;\tilde{\epsilon}) \equiv \frac{\tilde{\epsilon}^3 k_1^2}{12} \Biggl( \Biggr. J_{01}^3 - \frac{7}{2}J_{01}J_{11}^2 + \frac{J_{11}^3}{k_1 r} - \frac{1}{8} \sum_{i=1}^\infty \frac{k_{i}\Gamma_i J_{01}J_{0i}}{k_1} \nonumber \\
		&&\qquad - \frac{1}{4} \sum_{i=1}^\infty \frac{k_{i}^2\Gamma_i J_{01}J_{0i}}{k_1^2} + \sum_{i=1}^\infty \frac{1}{8}\frac{k_{i}^2\Gamma_i J_{11}J_{1i}}{k_1^2} + \nonumber \\
		&&\qquad \frac{1}{4} \sum_{i=1}^\infty \frac{k_{i}\Gamma_i J_{01}J_{0i}}{k_1} -\frac{1}{9}\sum_{i=1}^\infty \frac{(k_i/k_1)J_{0i}}{1 - (k_i/9k_1)  } \Biggl[ \Biggr. 5\alpha_n[J_{01}^3] + \frac{7}{2}\alpha_n[J_{01}J_{11}^2] \nonumber \\
		&&\qquad -\alpha_n\left[\frac{J_{11}^3}{k_1 r}\right] - \frac{5}{2}\sum_{j=1}^\infty \frac{k_j \Gamma_j \alpha_n[J_{01}J_{0j}]}{k_1} + \frac{1}{4}\sum_{j=1}^\infty \frac{k_j^2 \Gamma_j \alpha_n[J_{01}J_{0j}]}{k_1^2} \nonumber \\
		&&\qquad + \frac{1}{8}\sum_{j=1}^\infty \frac{k_j^2 \Gamma_j \alpha_n[J_{11}J_{1j}]}{k_1^2}  - \sum_{j=1}^\infty \frac{k_j \Gamma_j \alpha_n[J_{11}J_{1j}]}{k_1} \Biggl.\Biggr]  \Biggl.\Biggr) \nonumber
	\end{eqnarray}

	\textcolor{black}{The $\mathcal{O}(\tilde{\epsilon}^3)$ accurate, non-linear frequency $\omega(\tilde{\epsilon},q=1)$ is given by}
	\begin{equation}
		\begin{split}
			\omega^2(\tilde{\epsilon},q=1) &= k_1  \Biggl\{\Biggr. 1 + \frac{\tilde{\epsilon}^2 k_1^4}{4}\Biggl( \Biggr. -\alpha_1[J_{01}^3] - \alpha_1[J_{01} J_{11}^2] + \alpha_1\left[\frac{J_{11}^3}{k_1 r}\right] - \frac{1}{4} \sum_{i=1}^\infty \frac{k_{i}^2\Gamma_i J_{01} J_{0i}}{k_1^2} \\
			&\qquad + \frac{1}{2}\sum_{i=1}^\infty \frac{k_i \Gamma_i \alpha_1[J_{01} J_{0i}]}{k_1} - \frac{1}{8}\sum_{i=1}^\infty \frac{k_i^2 \Gamma_i \alpha_1[J_{11} J_{1i}]}{k_1}  \Biggl.\Biggr) \Biggl.\Biggr\} \nonumber
		\end{split}
	\end{equation}
	\textcolor{black}{with the functionals $\alpha\left[\cdot\right]$ and $\gamma\left[\cdot\right]$ defined as}
	\begin{eqnarray}
		\alpha_n\left[F(r)\right] &\equiv& \frac{\int_0^1 r F(r) J_{0n} dr}{\frac{1}{2} (J_0(k_{n}))^2},\quad
		\Gamma_i\left[F(r)\right] \equiv \frac{2 \alpha_n[ J_{01}^2 ] + 2 \alpha_n[ J_{11}^2 ]  }{1 - \frac{k_n}{4k_1}} \nonumber
	\end{eqnarray}
	\textcolor{black}{We have used the definitions $J_1(k_n) = 0 (n=1,2,3\ldots)$ and the short-hand notation $J_{ij} = J_i(k_j r)$.}

    \section*{Appendix B}
    We derive the relation between the nonlinear interaction coefficients $C_{mnq}$ and $D_{mnq}$ discussed in eqn. \ref{eq5}. This relation has been provided in \cite{nayfeh1987surface} and \cite{Miles_1976} without proof and the same is presented here. Following \cite{nayfeh1987surface}, we represent eqn. \ref{eq2} a in (semi) basis-independent notation as
	\begin{eqnarray}
		\phi(\bx,z,t) = \sum_{m=1}^{\infty}\phi_m(t)\Psi_m(\bx)\exp(k_mz) \label{a1}
	\end{eqnarray}
    where $\mathbf{x}$ is the horizontal position vector and $\Psi_m$ satisfies the equation $\nabla_H^2\Psi_m  + k_m^2 \Psi = 0$ as a consequence of $\phi$ satisfying the Laplace eqn; note that  $\nabla^2 = \nabla_H^2 + \frac{\partial^2}{\partial z^2}$.  We assume that $\Psi_m(\mathbf{x})$ follow the orthogonality rule $\int\int dS\;\psi_m(\mathbf{x})\psi_q(\mathbf{x}) = \delta_{mq}S$ where $\delta_{mq}$ is the Kronecker delta.
    Using Stokes theorem to relate an area integral (over s) in two dimensions to the line integral, we have for a vector field $\mathbf{F}(\mathbf{x})$
    \begin{eqnarray}
    	\int\int ds\;\bm{\nabla}_H\cdot\mathbf{F} = \int dl\;\left(\mathbf{F}\cdot\mathbf{n}\right). \label{a2}
    \end{eqnarray}
    Choosing $\mathbf{F} = \psi_q\psi_{m}\bm{\nabla}_H\psi_n$, eqn. \ref{a2} leads to
    \begin{eqnarray}
    	\int\int \left[\psi_q\left(\bm{\nabla}_H\psi_m\cdot\bm{\nabla}_H\psi_n\right) + \psi_m\left(\bm{\nabla}_H\psi_q\cdot\bm{\nabla}_H\psi_n\right) + \psi_q\psi_m\nabla_H^2 \psi_n\right]\;ds = 0, \label{a3}
    \end{eqnarray}
    the right hand side following from the no-penetration condition at the wall. Following the same notation as \cite{nayfeh1987surface}, we define,
    \begin{eqnarray}
    	&&\int\int ds\; \psi_m(\bx)\psi_n(\bx)\psi_q(\bx) \equiv S\;C_{mnq},\quad    	\int\int ds\; \left(\bm{\nabla}_H\psi_m(\bx)\cdot\bm{\nabla}_H\psi_n(\bx)\right)\psi_q(\bx) \equiv S\;D_{mnq}. \nonumber \\
    	&& \label{a4}
    \end{eqnarray}
     Note that $D_{nmq} = D_{mnq}$. Using \ref{a4}, eqn. \ref{a3} may be written compactly as
     \begin{eqnarray}
     	D_{mnq}  + D_{qnm} - k_n^2C_{mnq} = 0. \label{a5}
     \end{eqnarray}
     Replacing $m\rightarrow n, n\rightarrow q, q\rightarrow m$ in equation \ref{a5}, we obtain
     \begin{eqnarray}
  		D_{nqm}  + D_{mqn} - k_q^2C_{nqm} = 0. \label{a6}
     \end{eqnarray}
     which may be rewritten as
     \begin{eqnarray}
		D_{qnm}  + D_{qmn} - k_q^2C_{nqm} = 0 \label{a7}
	\end{eqnarray}
     Using \ref{a7} in \ref{a5}, we obtain
     \begin{eqnarray}
     	D_{mnq} = k_n^2C_{mnq} - \left(k_q^2C_{nqm} - D_{qmn}\right). \label{a8}
     \end{eqnarray}
    Replacing once again $m\rightarrow q, n\rightarrow m, q\rightarrow n$ in equation \ref{a5}, we obtain
     \begin{eqnarray}
    	D_{qmn}  + D_{nmq} - k_m^2C_{qmn} = 0. \label{a9}
    \end{eqnarray}
    Combining \ref{a8} and \ref{a9} and the fact that $D_{mnq} = D_{nmq}$, we obtain
    \begin{eqnarray}
    	D_{nmq} = \frac{1}{2}\left(k_n^2 + k_m^2 - k_q^2\right)C_{nmq} \label{a10}
    \end{eqnarray}
After some manipulation, expression \ref{eq5} follows from the above expression.

	\section*{Appendix C}
\begin{figure}
		\centering
		\subfloat[]{\includegraphics[scale=0.13]{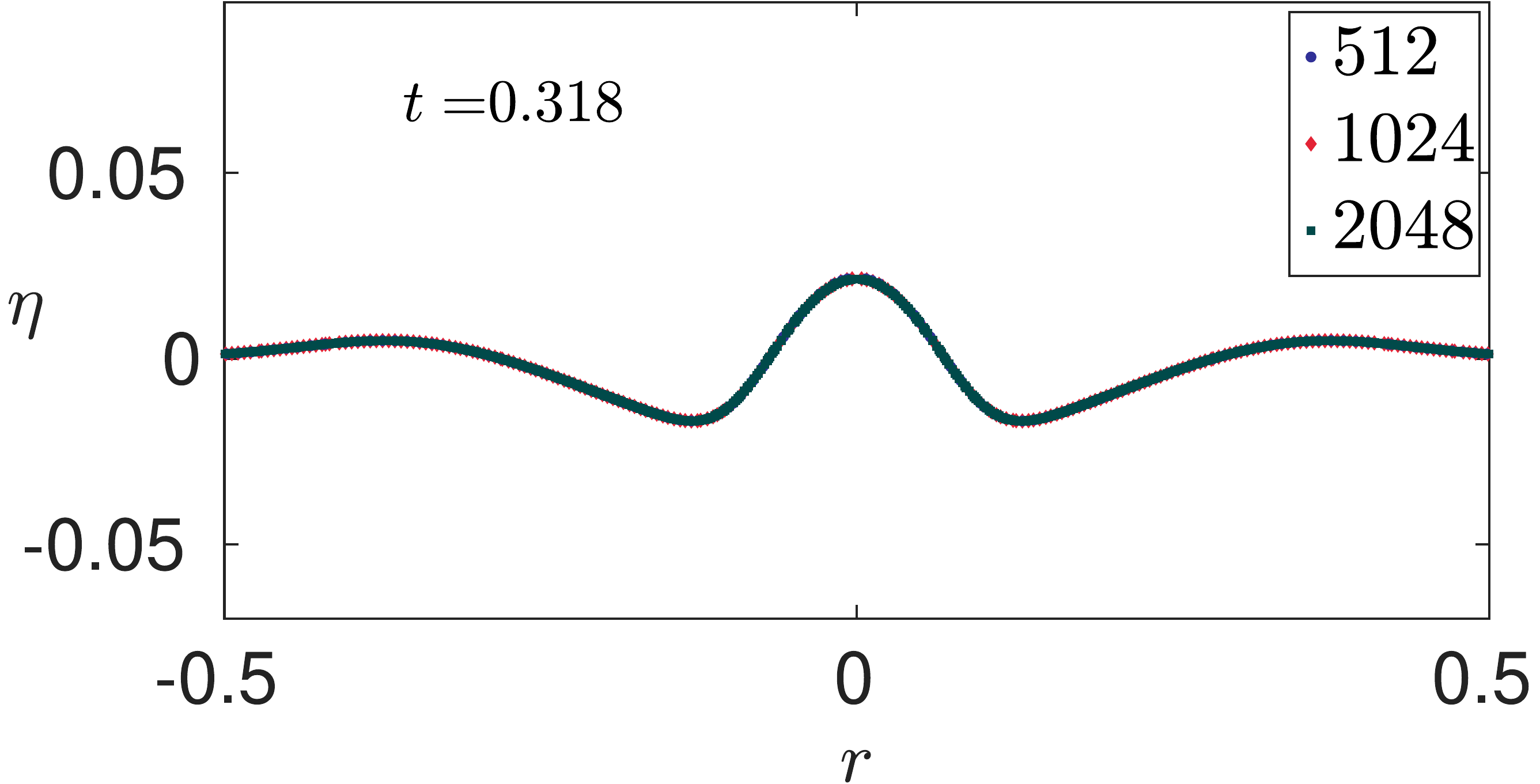}\label{conv_visc_1}}\quad
		\subfloat[]{\includegraphics[scale=0.13]{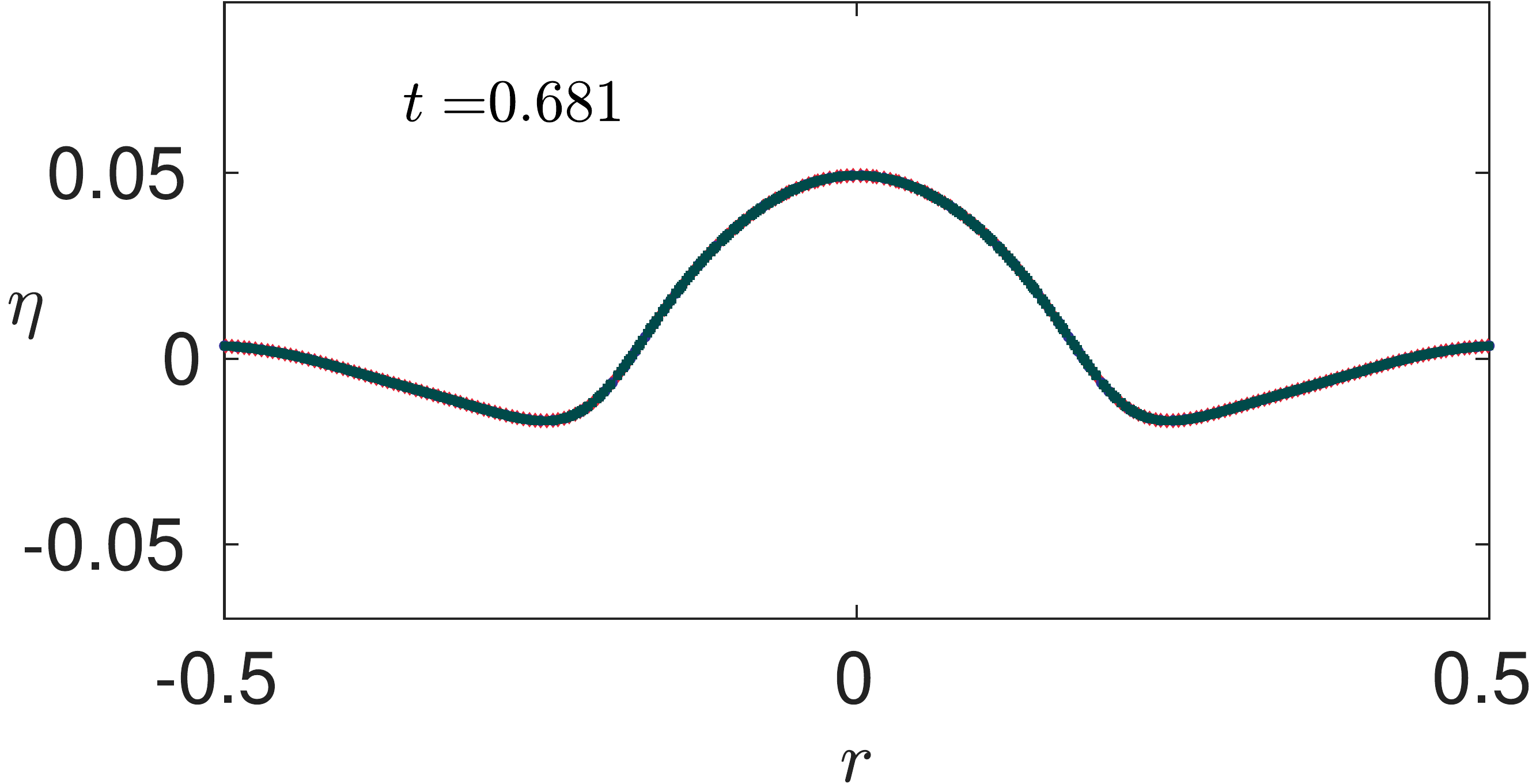}\label{conv_visc_2}}\\
		\subfloat[]{\includegraphics[scale=0.13]{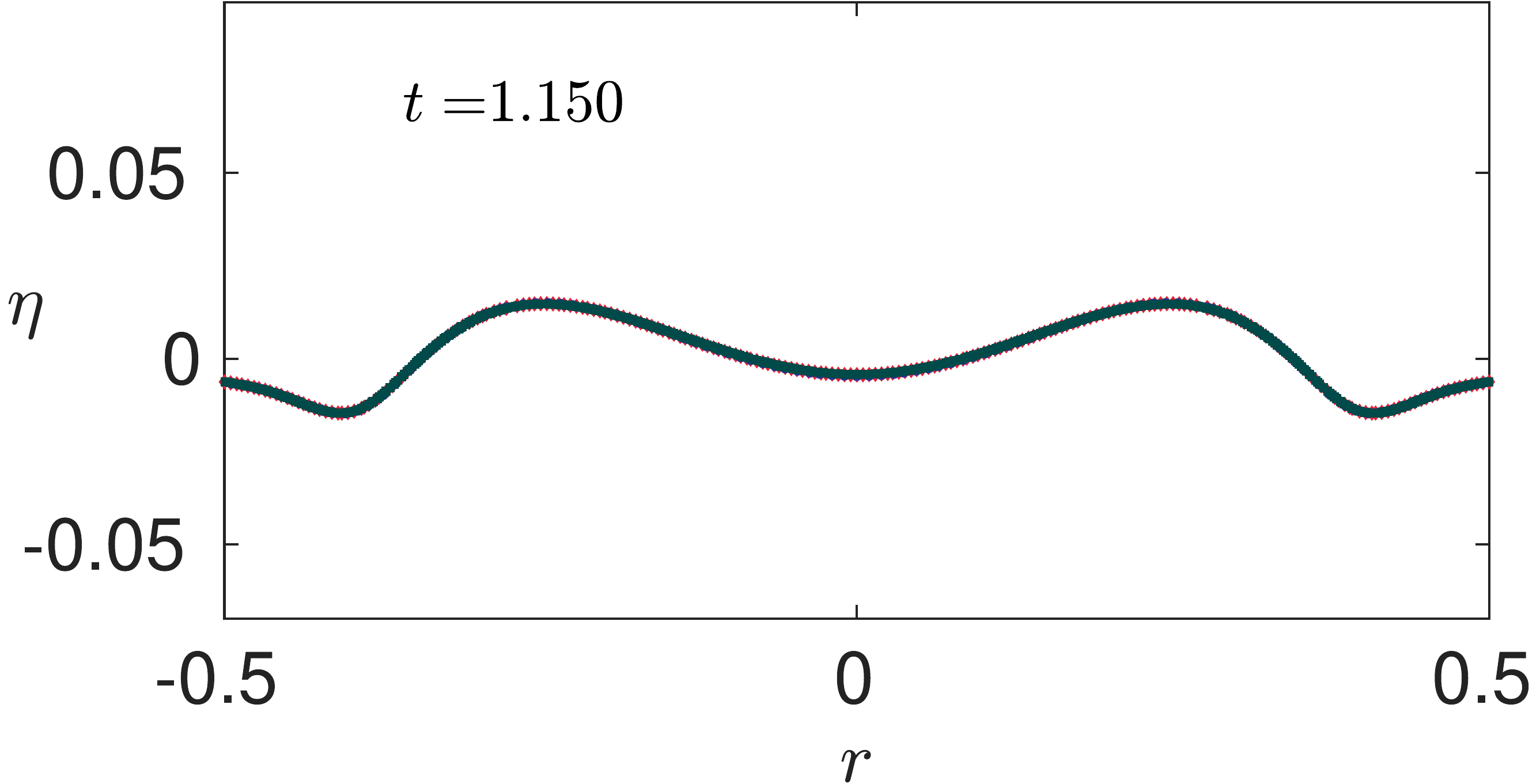}\label{conv_visc_3}}\quad
		\subfloat[]{\includegraphics[scale=0.13]{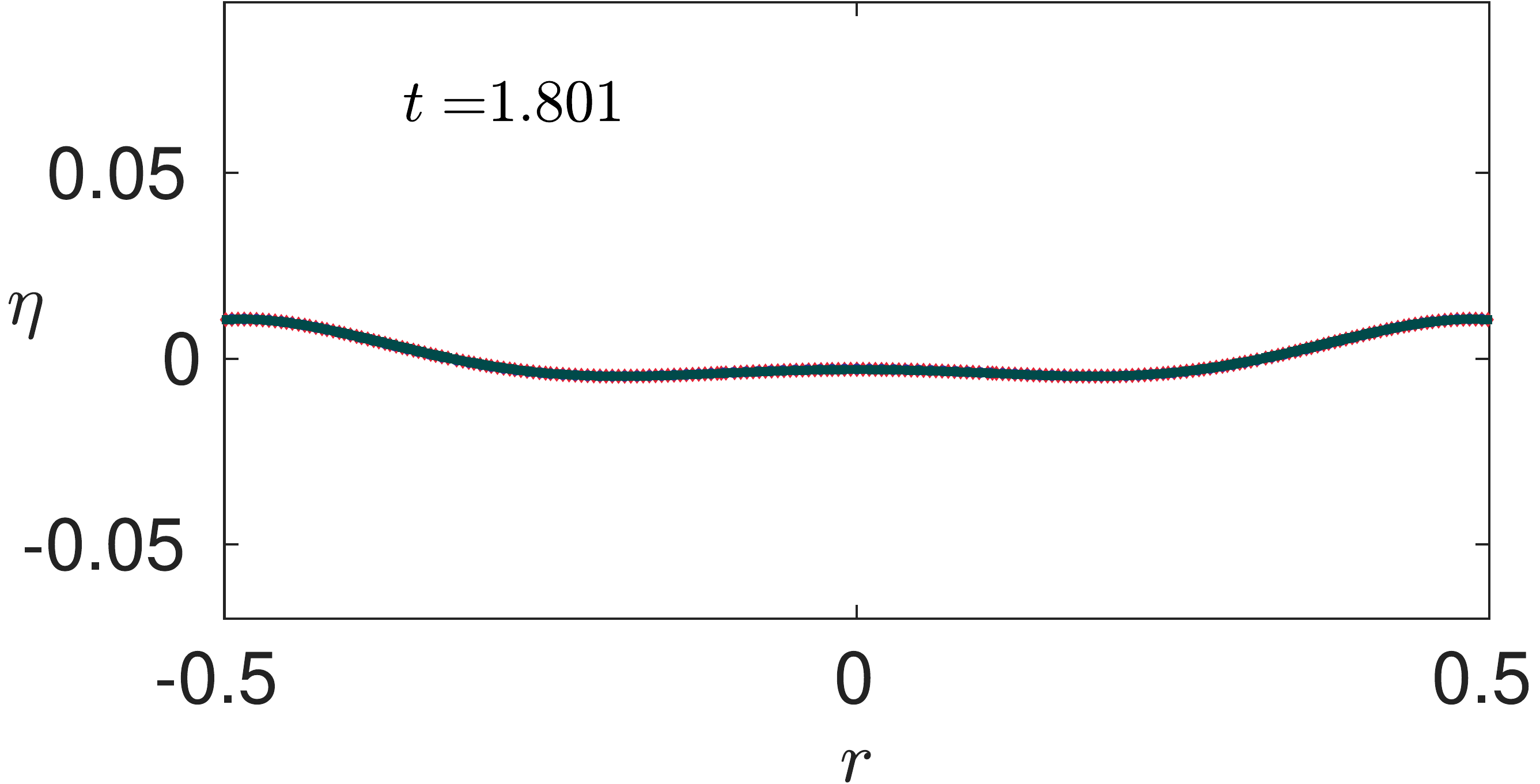}\label{conv_visc_4}}\\
		\subfloat[]{\includegraphics[scale=0.13]{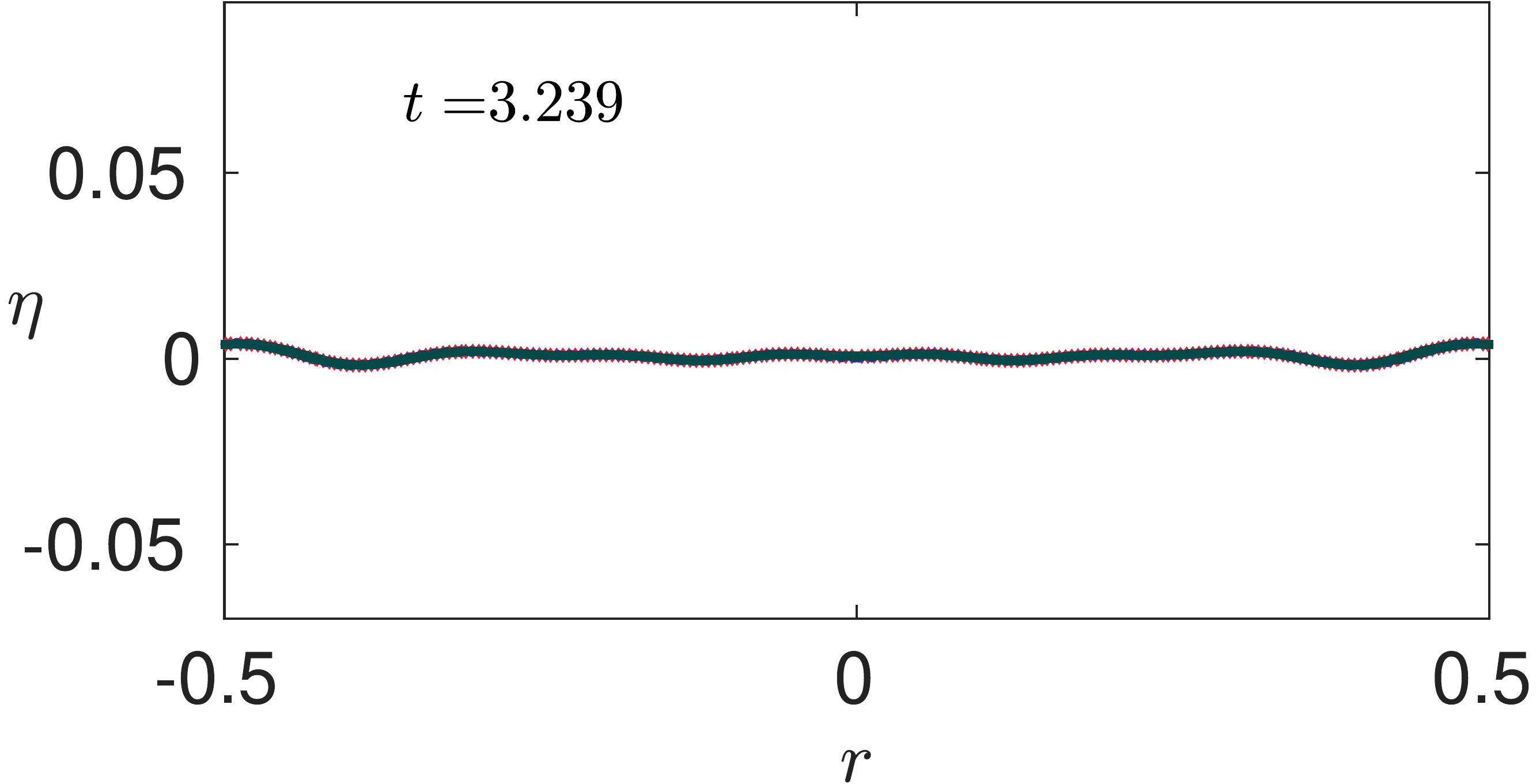}\label{conv_visc_5}}\quad
		\subfloat[]{\includegraphics[scale=0.13]{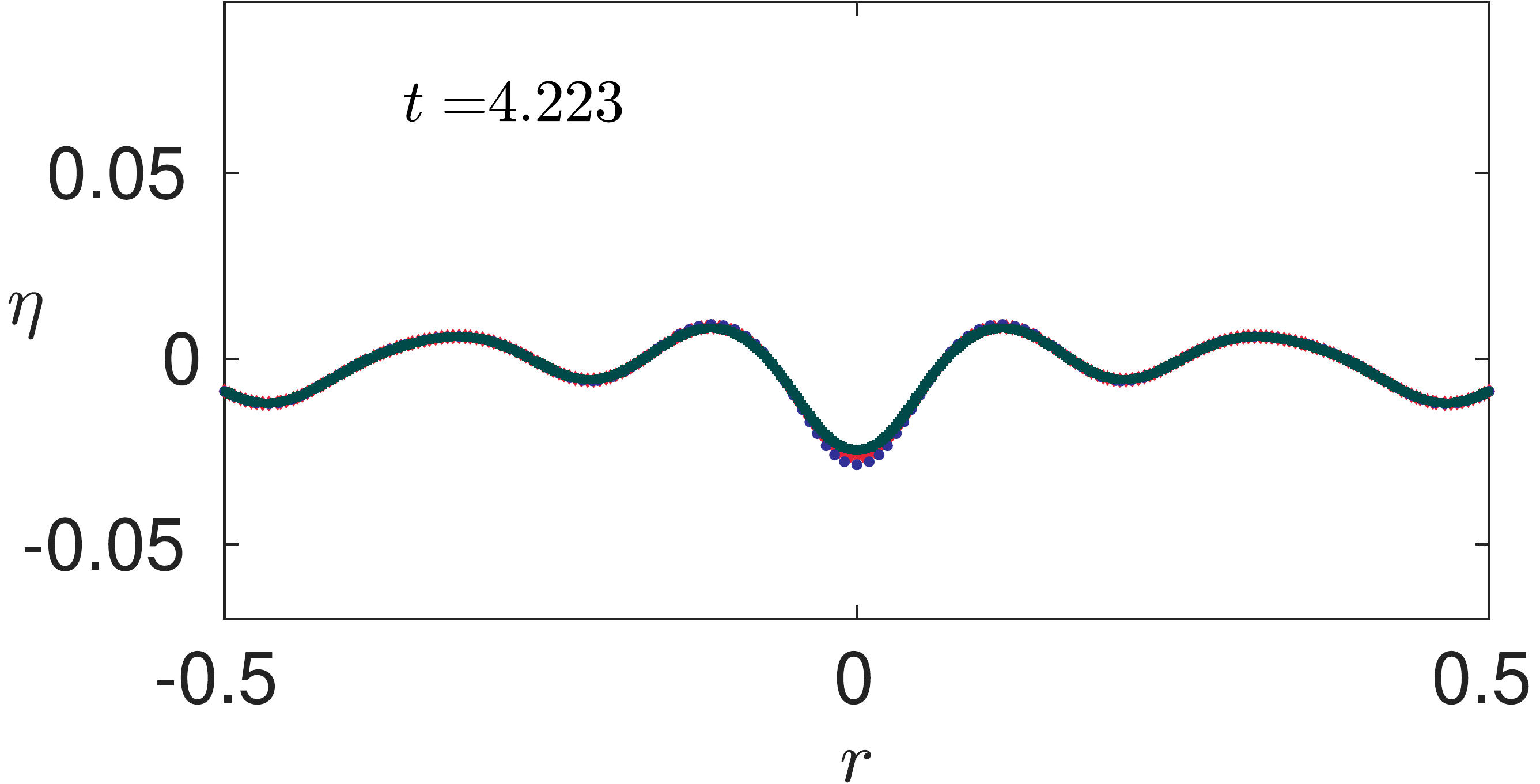}\label{conv_visc_6}}\\
		\subfloat[]{\includegraphics[scale=0.13]{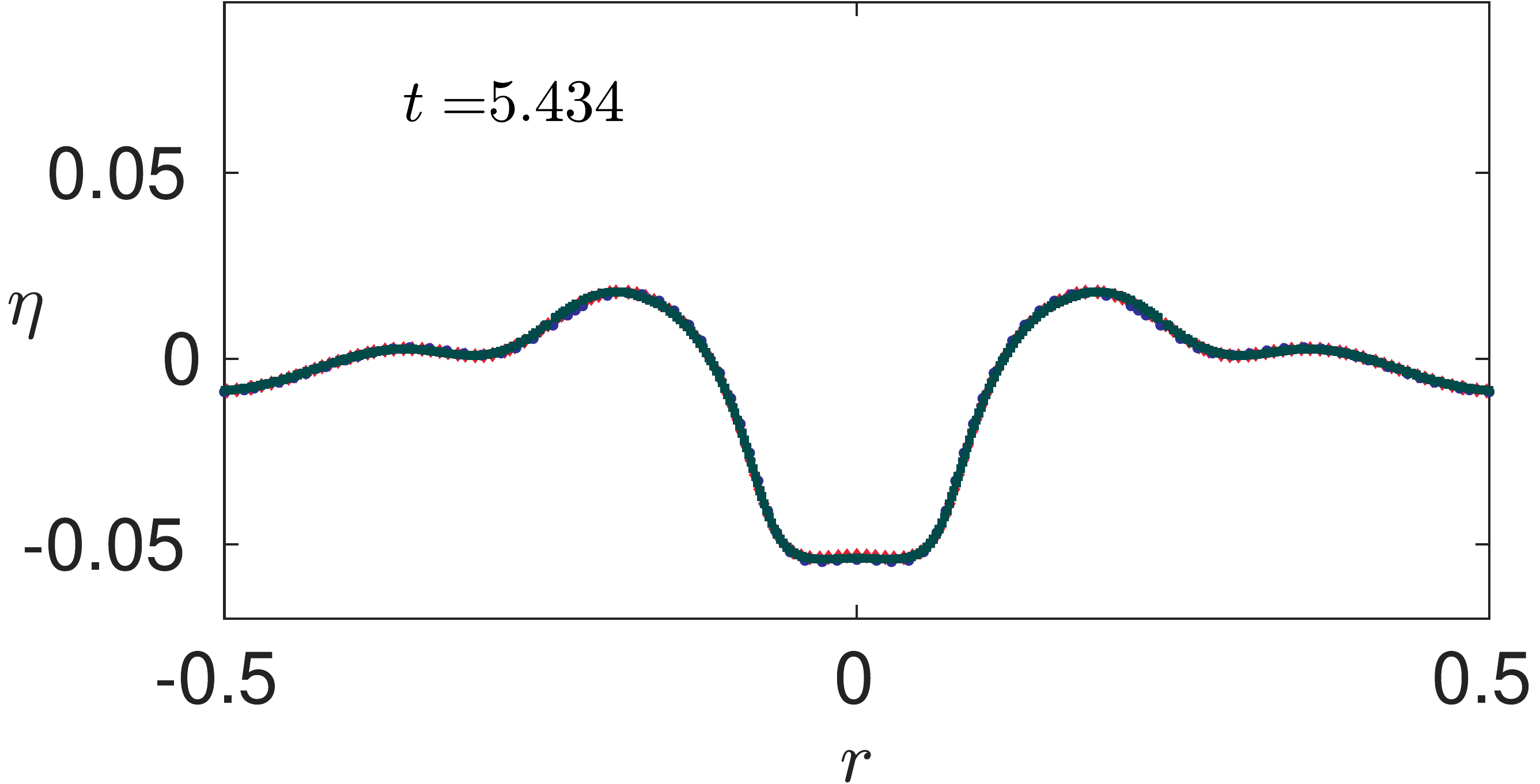}\label{conv_visc_7}}\quad
		\subfloat[]{\includegraphics[scale=0.13]{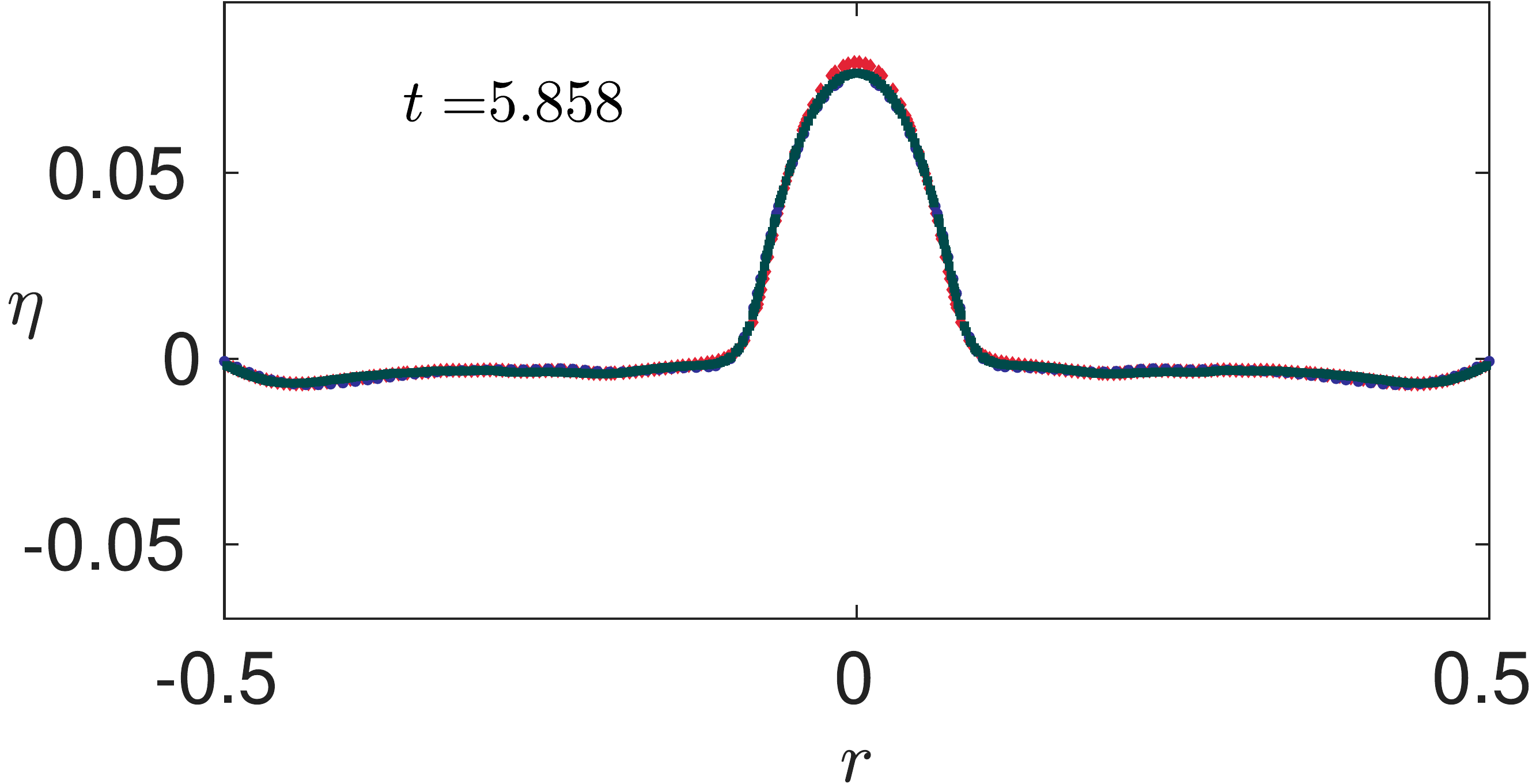}\label{conv_visc_8}}\\
		\caption{Comparison of interface profile for case $4$ in table \ref{tab:sim_params} at three different grid resolutions, $512^2$ (blue dots), $1024^2$ (red dots) and $2048^2$ (green dots)}
		\label{conv_visc}
	\end{figure}

\begin{figure}
		\centering
		\subfloat[]{\includegraphics[scale=0.13]{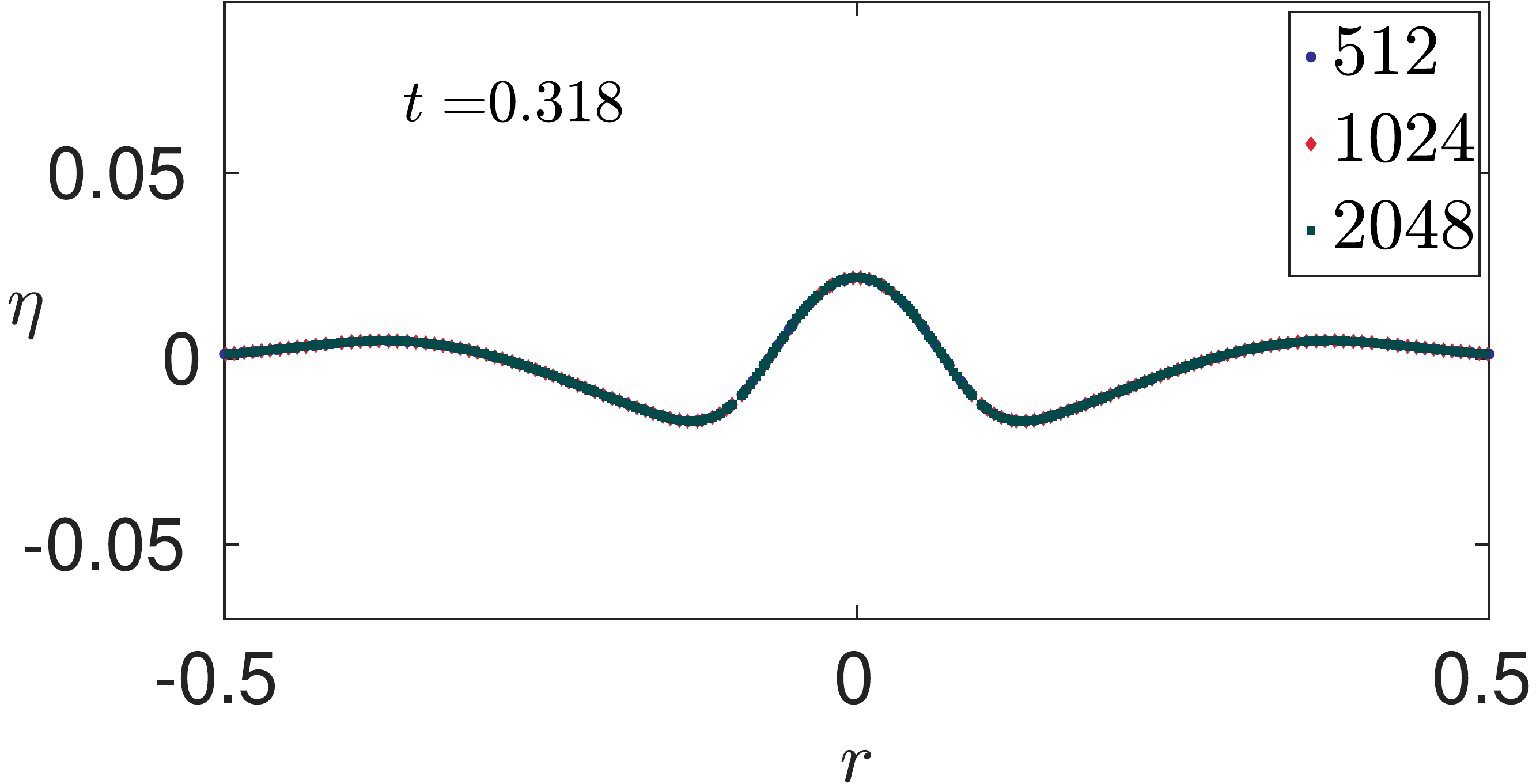}\label{conv_ivisc_1}}\quad
		\subfloat[]{\includegraphics[scale=0.13]{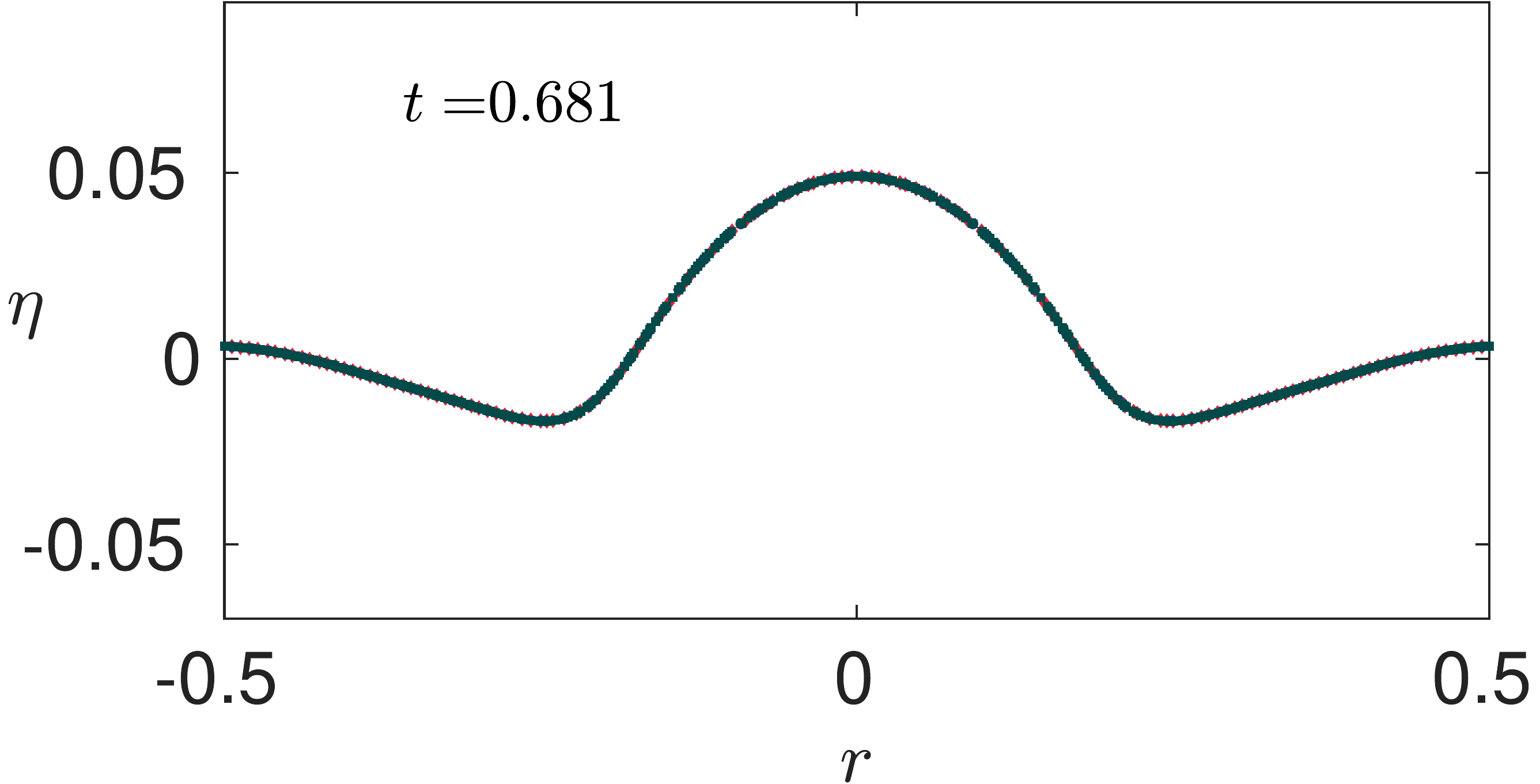}\label{conv_ivisc_2}}\\
		\subfloat[]{\includegraphics[scale=0.13]{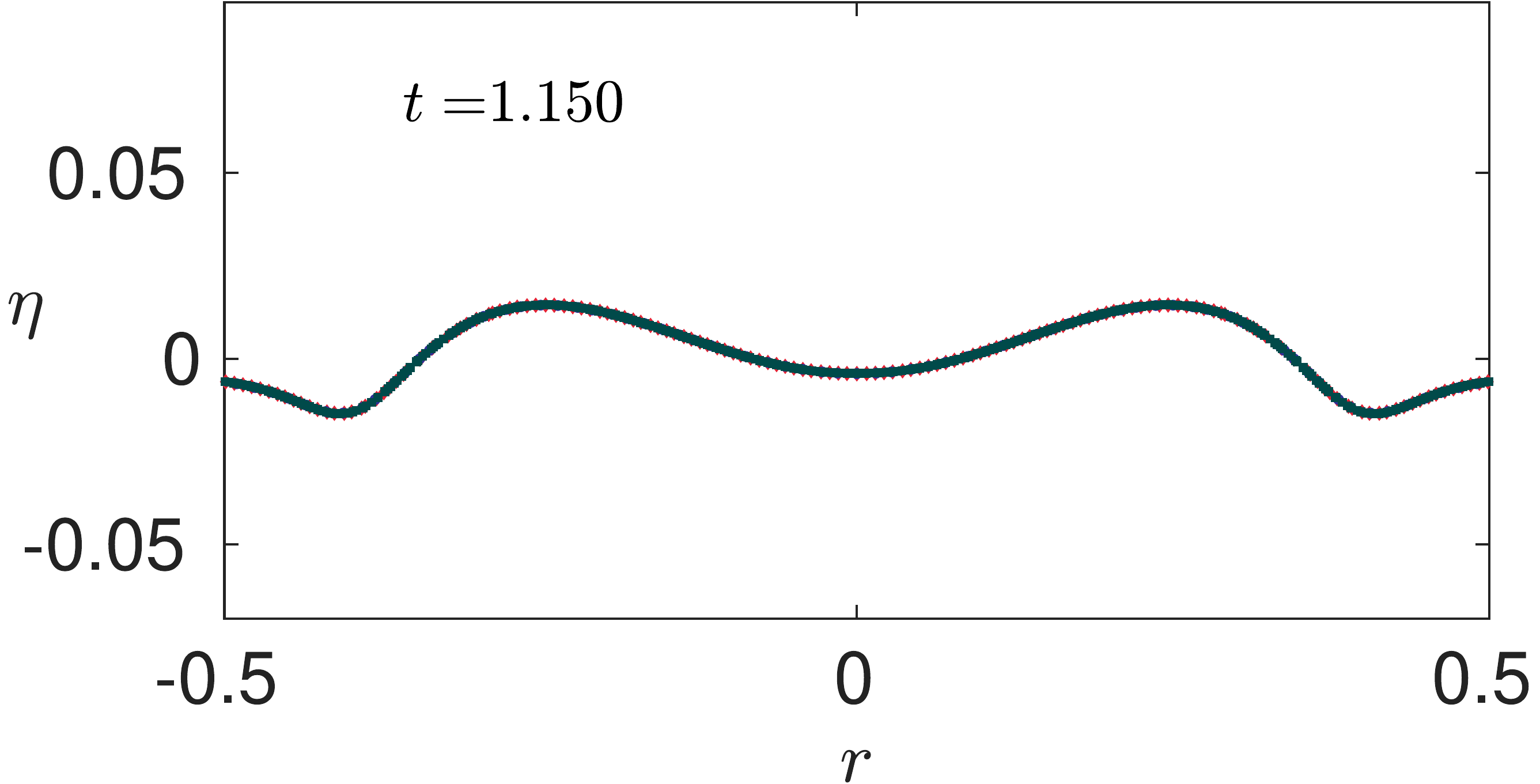}\label{conv_ivisc_3}}\quad
		\subfloat[]{\includegraphics[scale=0.13]{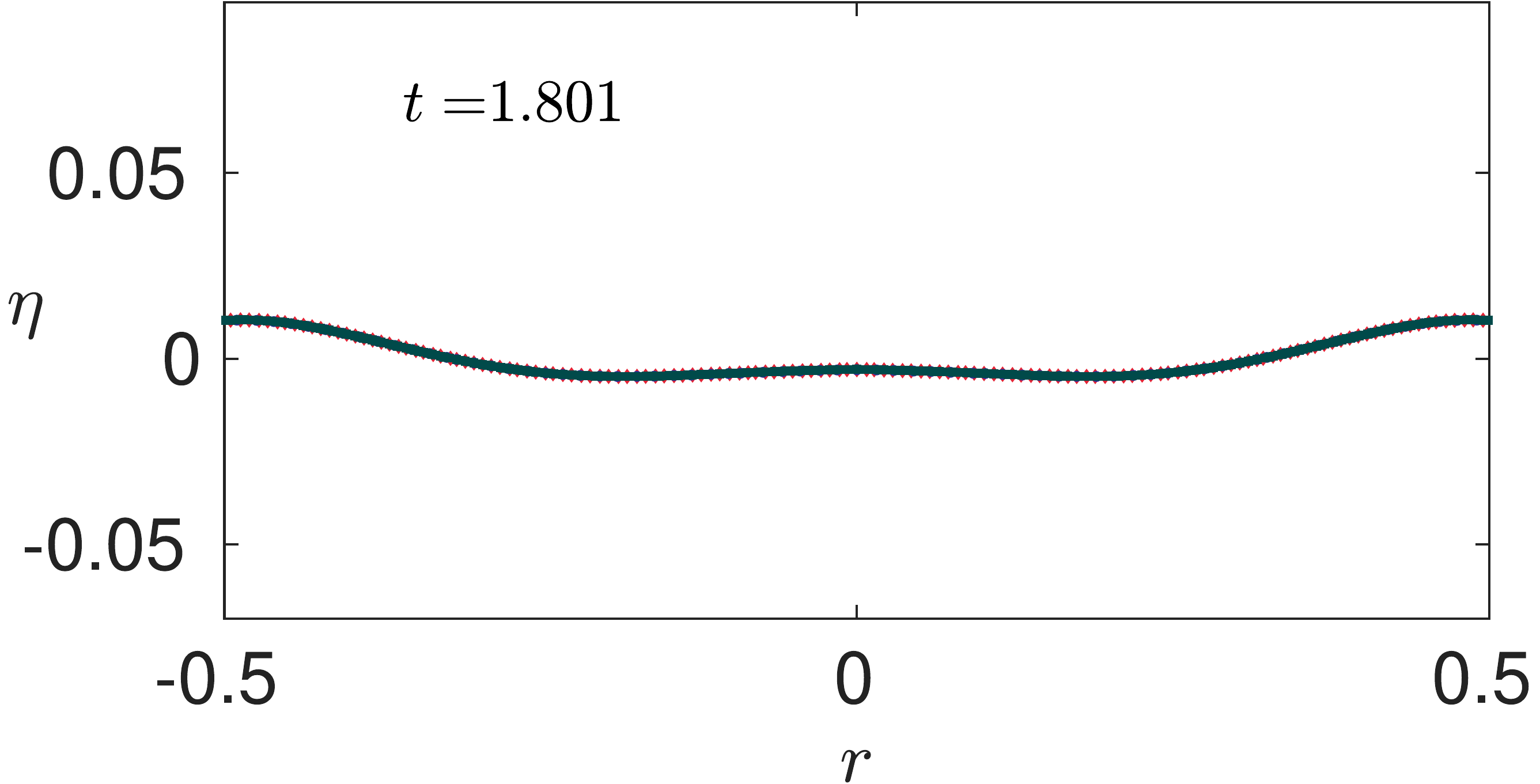}\label{conv_ivisc_4}}\\
		\subfloat[]{\includegraphics[scale=0.13]{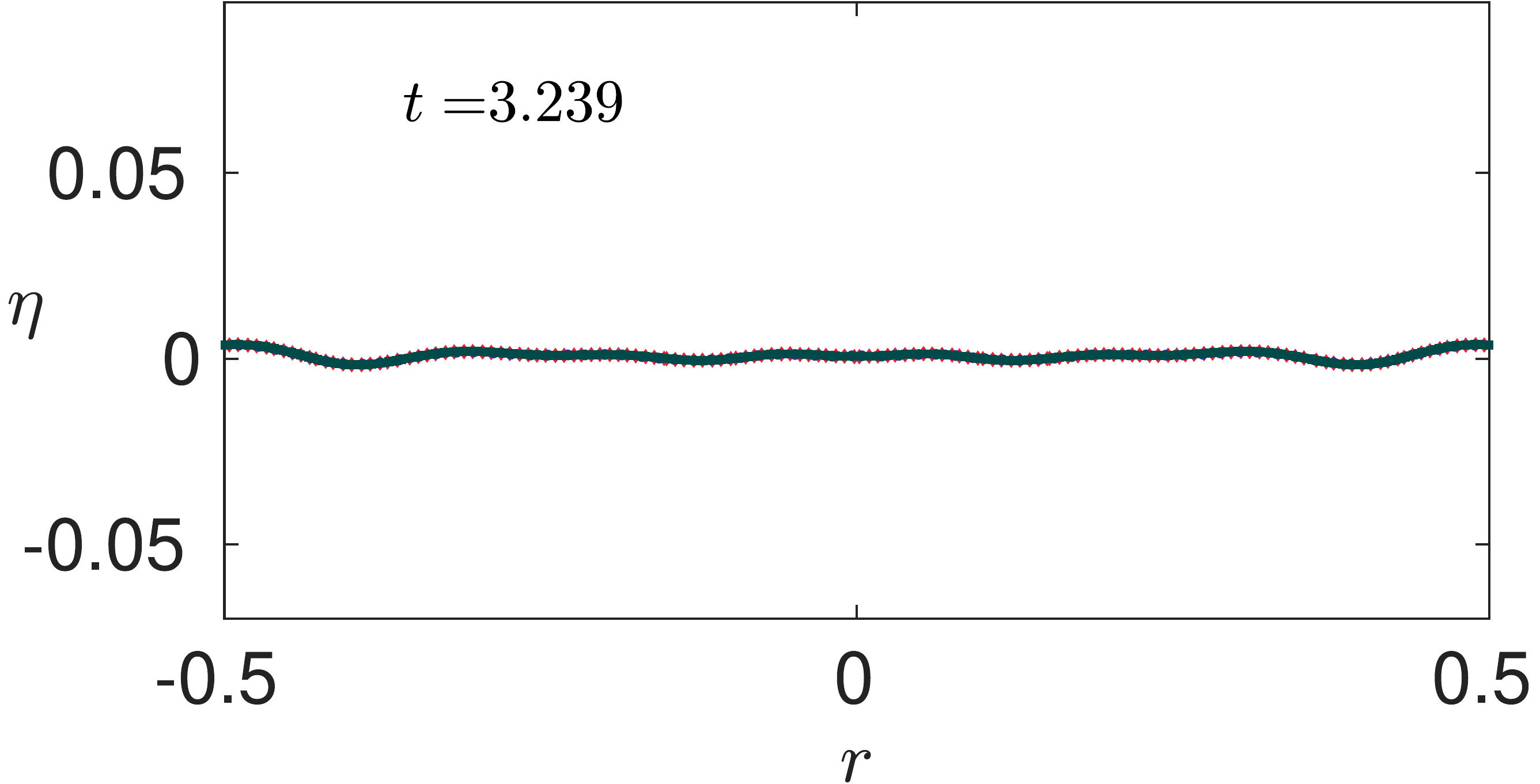}\label{conv_ivisc_5}}\quad
		\subfloat[]{\includegraphics[scale=0.13]{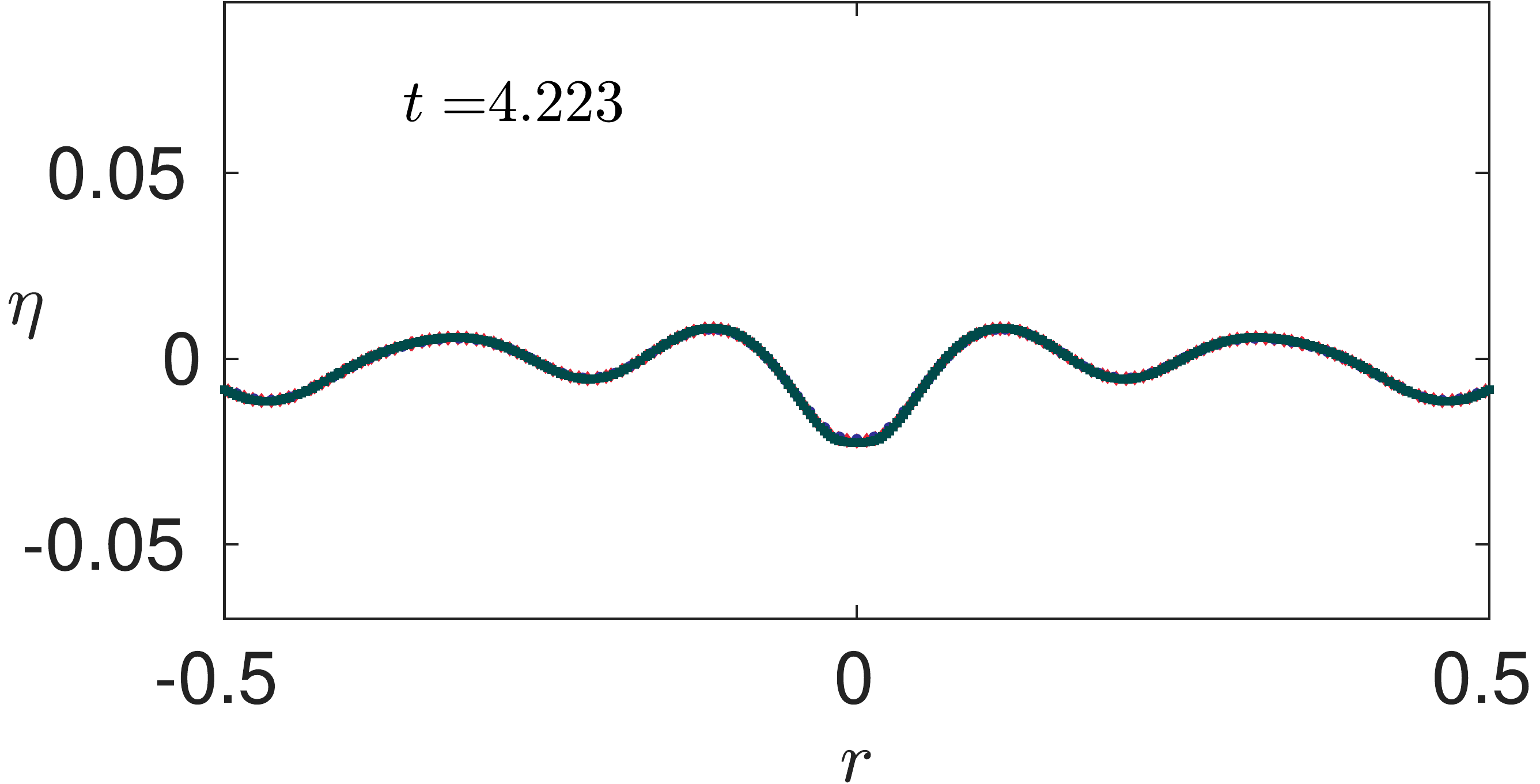}\label{conv_ivisc_6}}\\
		\subfloat[]{\includegraphics[scale=0.13]{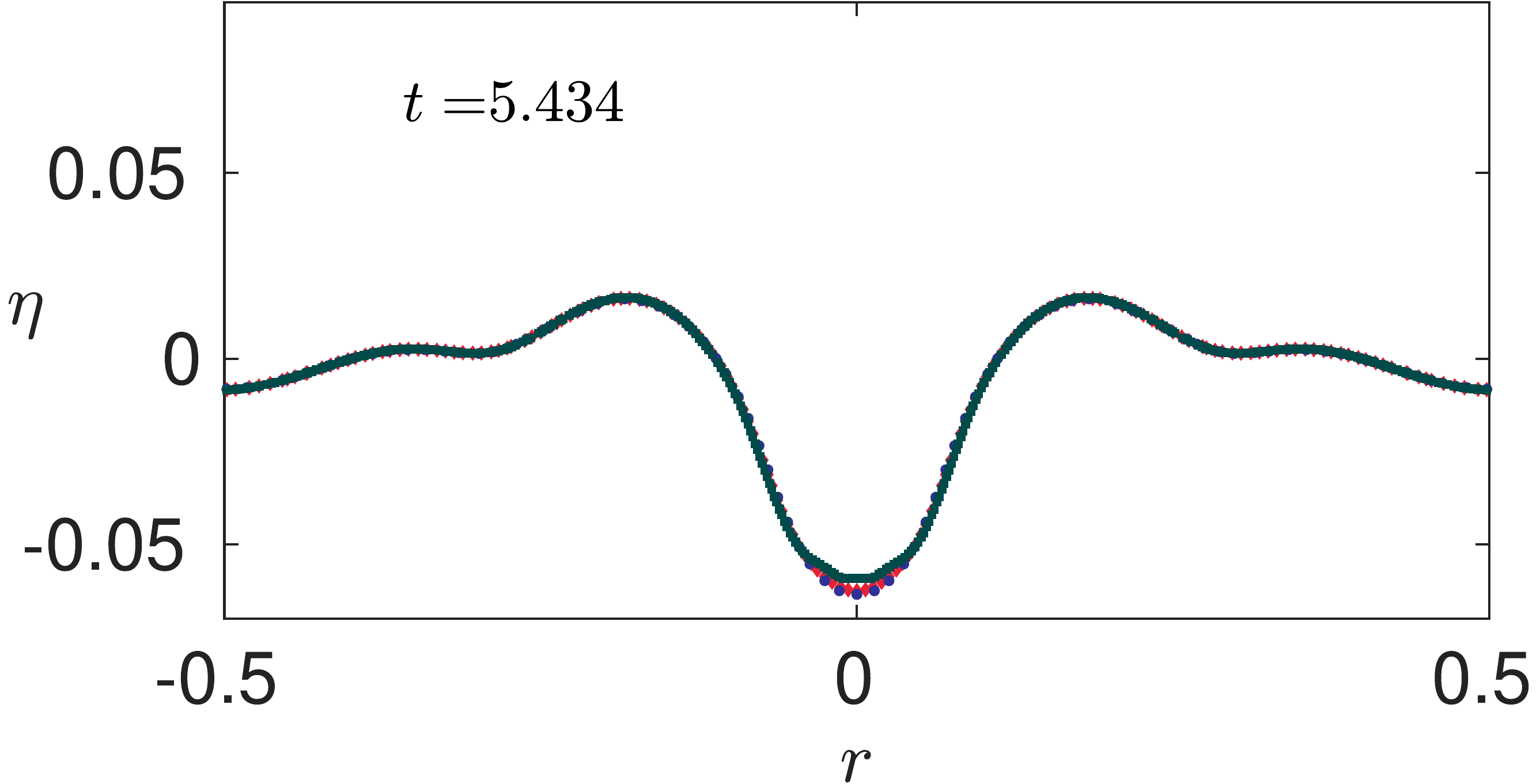}\label{conv_ivisc_7}}\quad
		\subfloat[]{\includegraphics[scale=0.13]{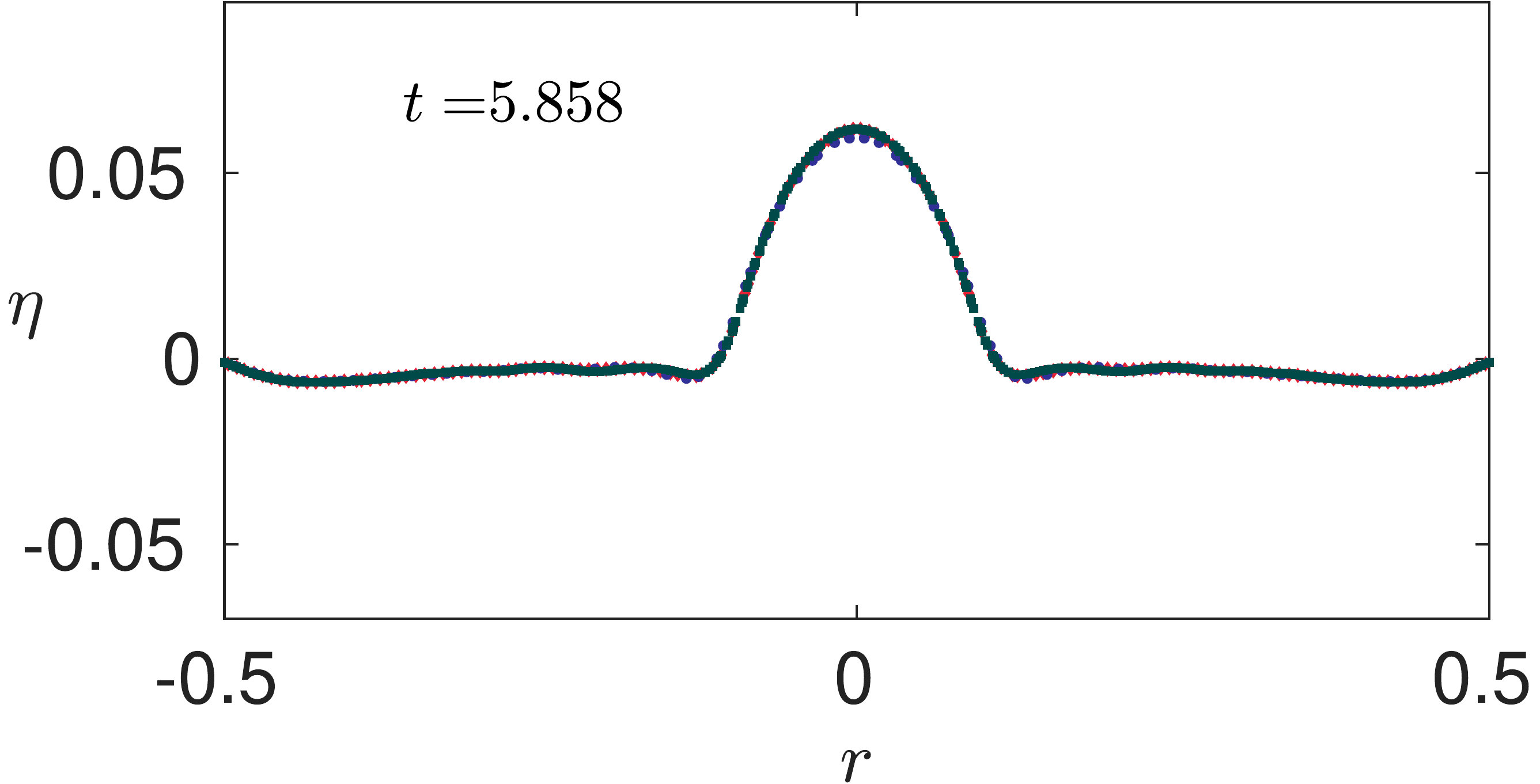}\label{conv_ivisc_8}}\\
		\caption{Comparison of interface profile for case $2$ in table \ref{tab:sim_params} for three different grid resolutions, $512^2$ (blue dots), $1024^2$ (red dots) and $2048^2$ (green dots)}
		\label{conv_ivisc}
	\end{figure}

\begin{figure}
     	\centering
     	\subfloat[$Oh=0$]{\includegraphics[scale=0.2]{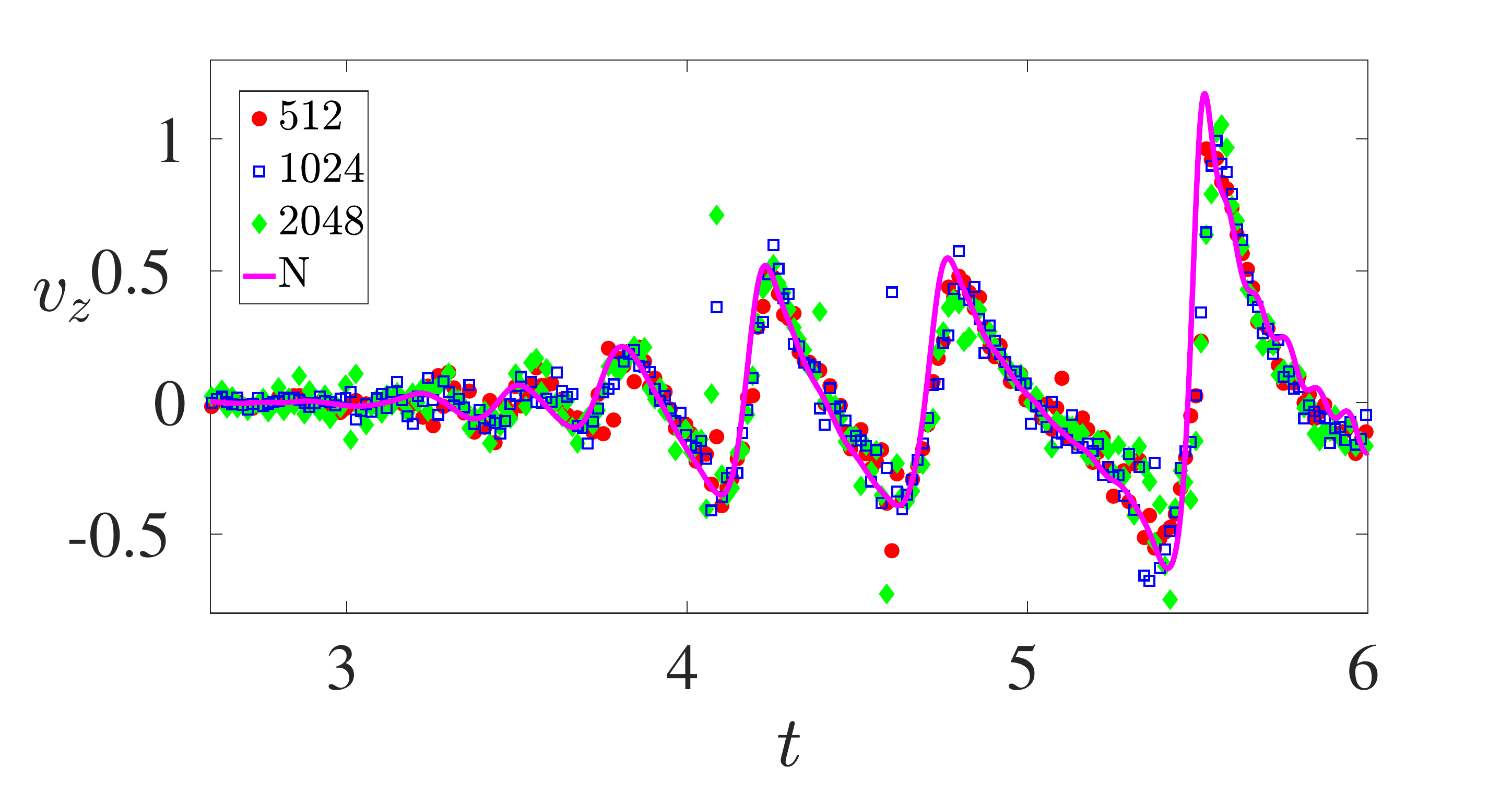}\label{inviscid_velocity}}\quad
     	\subfloat[$Oh=1.17\times 10^{-4}$]{\includegraphics[scale=0.2]{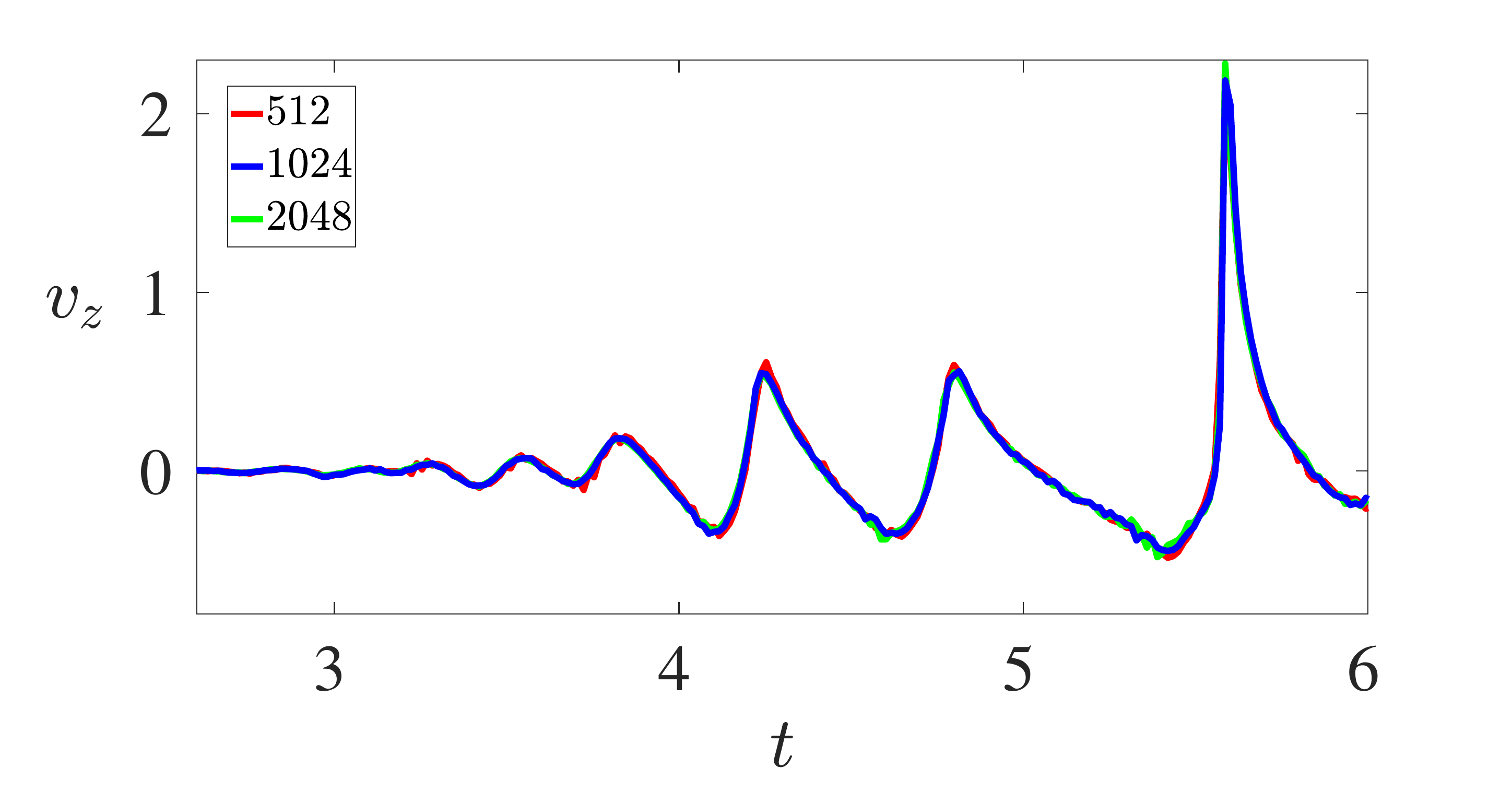}\label{viscous_velocity}}
     	\caption{Comparison of the vertical velocity for case 2 (panel a) and case 4 (panel b) for three different grid resolutions, $512^2$ (red solid line), $1024^2$ (blue solid line) and $2048^2$ (green solid line). \textcolor{black}{In panel (a), we also provide the prediction from the numerical solution to the analytical model from eqn. \ref{eq6} indicated as `N' in the figure.} 
      }
     	\label{velocity_grid_convergence}
     \end{figure}

	Figures~\ref{conv_visc} and \ref{conv_ivisc} illustrate grid convergence results at three resolutions ($512^2$, $1024^2$ and $2048^2$) for cases 4 and 2 respectively from table \ref{tab:sim_params}. Figures~\ref{inviscid_velocity} and ~\ref{viscous_velocity} present grid convergence for the velocity at the symmetry axis.
	
	\textcolor{black}{For the inviscid case ($Oh = 0$, fig.~\ref{inviscid_velocity}), while the overall vertical velocity trend remains consistent, the presence of spikes and the magnitude exhibit grid refinement sensitivity. To provide a robust reference, we include the inviscid, nonlinear analytical solution in figure~\ref{velocity_grid_convergence}, panel (a) denoted by `N'. This demonstrates good agreement with the $Oh = 0$, DNS solution, capturing the main temporal velocity variation features without spurious peaks, thus validating the observed simulation behavior.}
	\textcolor{black}{As $Oh$ approaches zero, our one-fluid approximation made in the solver Basilisk \citep{popinet-basilisk} imposes a no-slip condition at the liquid-gas interface. Resolving this boundary layer requires a minimum grid size of $\Delta \sim KLOh^2$, where $L$ is the characteristic length and $K$ a system-dependent prefactor. This establishes a critical $Oh$ above which results converge well. We empirically determined this as $Oh = 1.17 \times 10^{-4}$ through grid independence testing. Results below this critical value remain unresolved due to insufficient grid resolution, indicated by the pink shaded region in figure~\ref{anomaly}b. Further computational method improvements are needed to resolve cases where $Oh < 1.17 \times 10^{-4}$.}

    
   	\section*{Appendix D}
\begin{figure}
   		\centering
   		\subfloat[$\varepsilon=0.091$ : wave focussing ]{\includegraphics[scale=0.13]{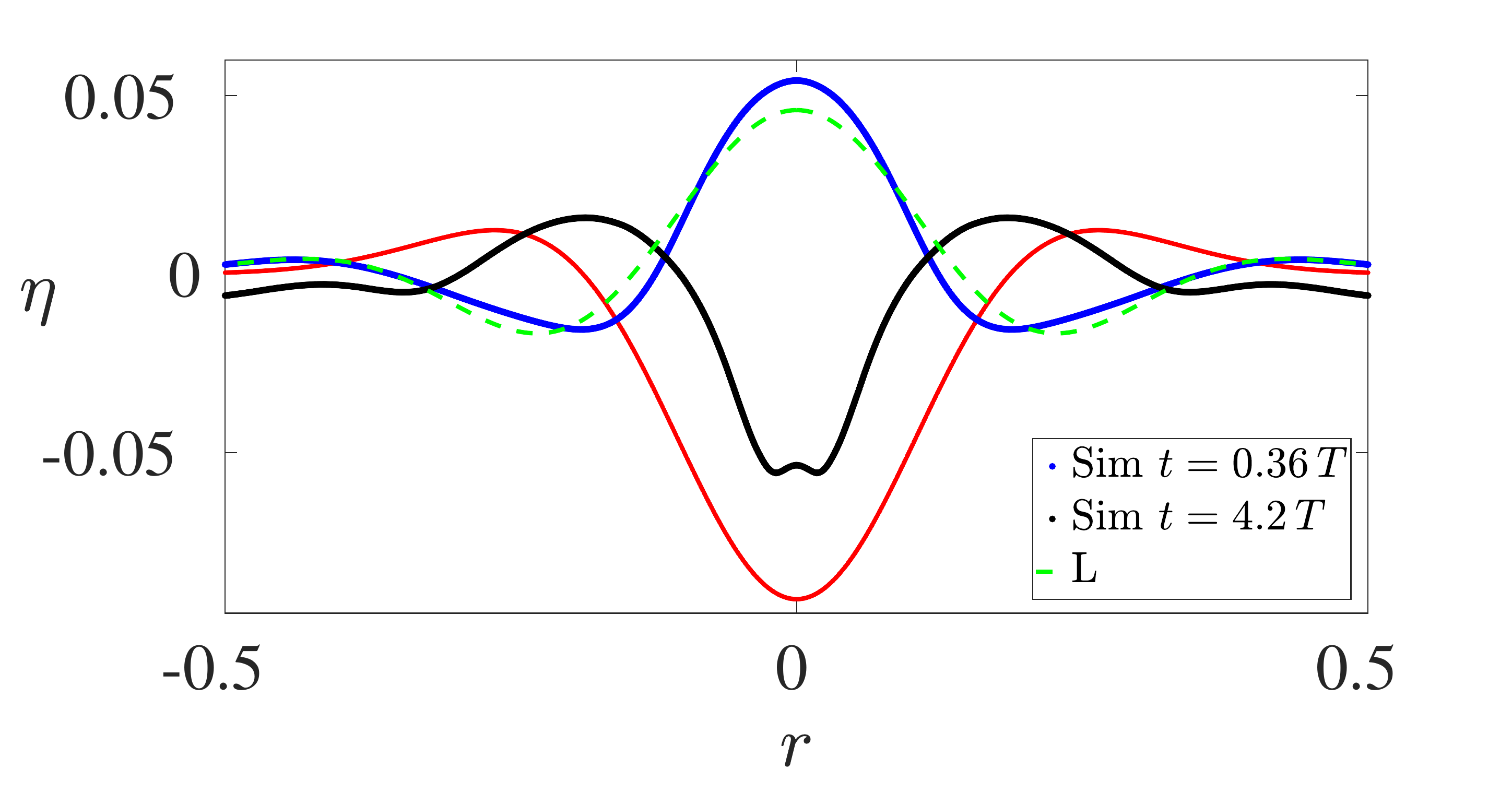}}
   		\subfloat[$\varepsilon=0.242$ : flow focussing]{\includegraphics[scale=0.13]{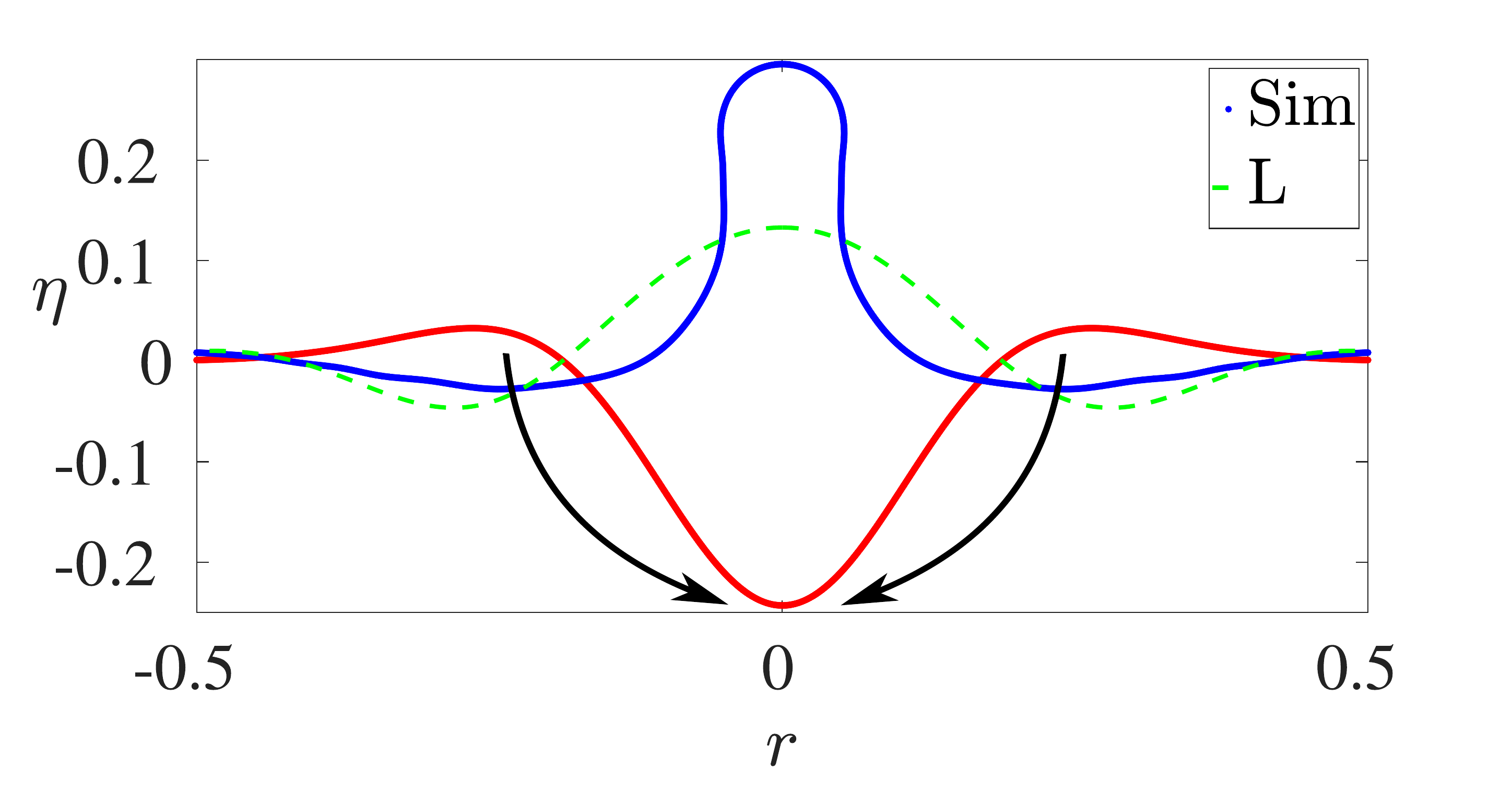}}
   		\caption{\textcolor{black}{Both panels start with the interface deformed as a cavity shown in red with different $\epsilon$. (Left panel) (Red) interface at $t=0$, (green dashed curve) numerical simulation and (blue) linear approximation at $t=0.36T$, when the interface reaches its maximum height and (black curve) at a much later time $t=4.2T$. (Right panel): same color code as left panel, (blue) for $t=0.48$ T. Here $T$ is the dominant mode time-period in the initial spectrum (linear approximation). The arrows on the right panel, indicate the approximate direction of flow resulting from the initial (capillary) pressure gradient. The jet which was studied in \cite{basak2021jetting} from $\eta(r,t=0)=\varepsilon J_0(k_5 r)$ is closely related to the jet in panel (b). Note the lack of any visible wave-train when this jet is produced. In this case, pressure difference arising due to the initial steep interface distortion around $r=0$, triggers a radially inward flow towards the same (indicated with arrows in panel (b) of fig. \ref{fig1}). The radial component of this inflow produces a stagnation zone of high pressure at the base of the cavity and a resultant upward jet. We label this situation as ``flow focussing''. This jet in \cite{basak2021jetting} is associated with a large stagnation pressure at its base, involving conversion of kinetic energy (nonlinear term in Bernoulli equation) to pressure energy. In contrast in panel (a), no significant stagnation pressure zone develops initially (as the initial cavity is comparatively less steep compared to panel (b)). In this case nonlinear effects become manifest much later when the wave-train focusses on to the symmetry axis, producing rapid interfacial oscillations at $r=0$. The apparent importance of nonlinearity around $r=0$ in this case is somewhat akin to linear dispersive focussing of surface waves, where nonlinearity becomes locally important at the focal point.}}
   		\label{fig_flow_wave}
   	\end{figure}

   	Fig. \ref{fig_flow_wave} depicts the qualitative difference in behaviour starting from a cavity with $\varepsilon=0.091$ (left panel). Here no jet is seen initially and the wave-train focusses at $r=0$ after some time. In panel (b) with $\varepsilon=0.242$, a jet is seen at a much earlier time, and no focussing wave-train is apparent. In this study, we focus on the regime indicated in panel (a). The jet in panel (b) is close to the one that was reported in \cite{basak2021jetting}, albeit from a single Bessel function.

    \section*{Appendix E (comparison with \cite{gordillo2019capillary}):}
\begin{figure}
    	\centering
    	\includegraphics[scale=1]{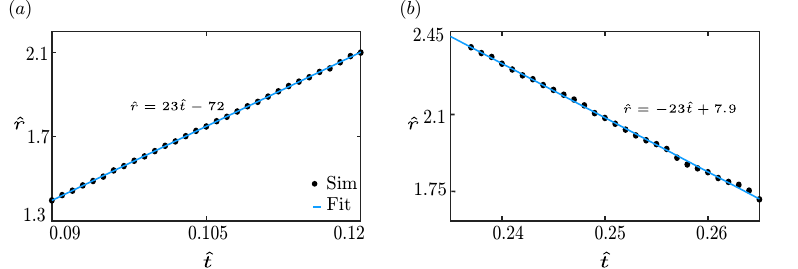}
    	\caption{\textcolor{black}{Measurement of interface elevation from simulations for the largest crest following it, for (a) outward wave propagation and (b) inward propagation. The crests are generated from an initial cavity with $\varepsilon\approx 0.091$ (case $2$ in table \ref{tab:sim_params}). The slope of the linear fit indicates the propagation velocity which is approximately equal to the phase speed of the dominant Bessel function. Similar to the fig. 4b in \cite{gordillo2019capillary}, we observe a good agreement with the linear propagation speed, before and after the reflection.}} \label{i-o velocity}
    \end{figure}

\begin{figure}
    	\centering
    	\includegraphics[scale=0.2]{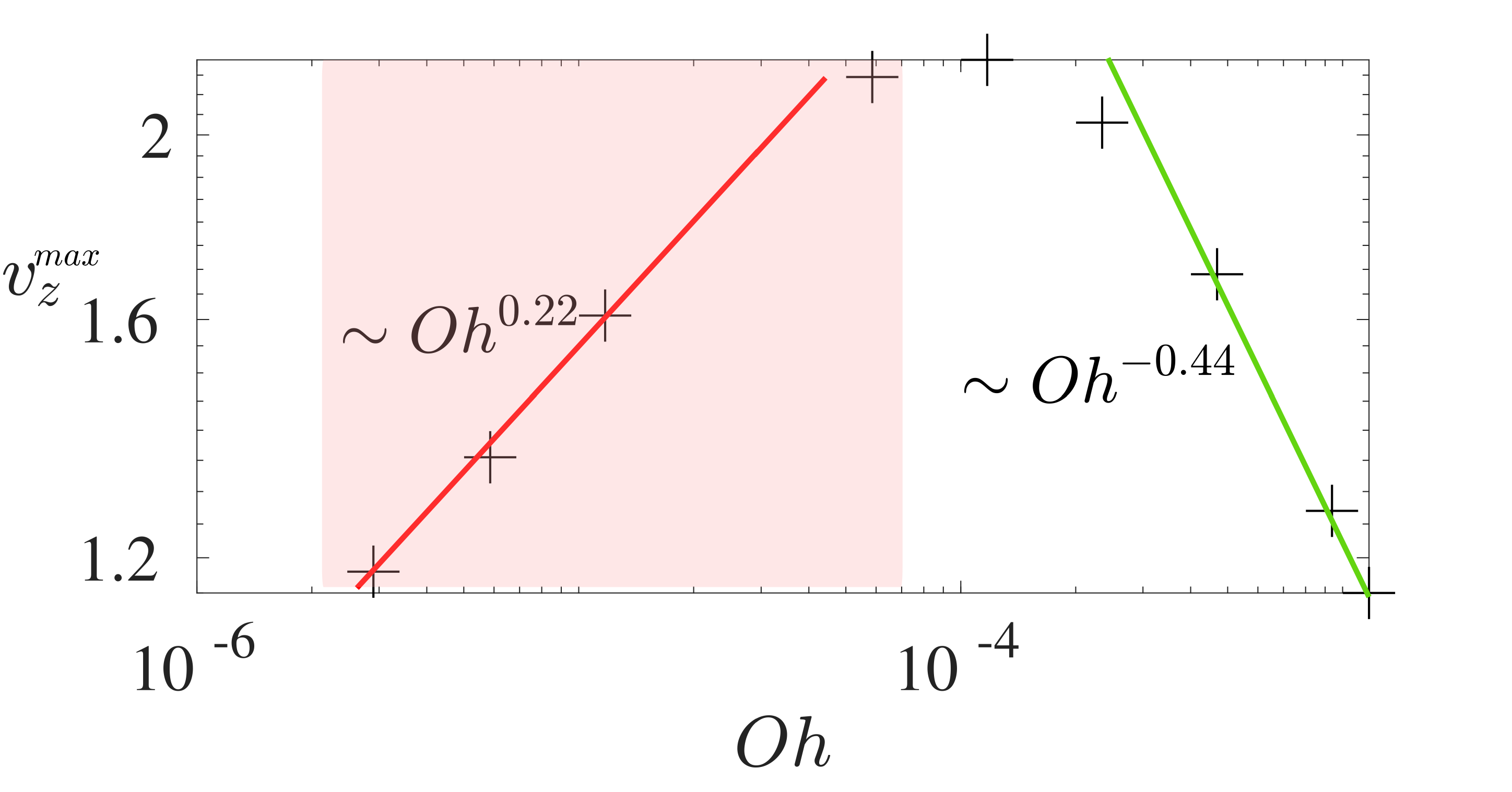}
    	\caption{\textcolor{black}{The maximum vertical velocity at the symmetry axis $v_z^{max}$ for  different $Oh$. This figure is a superset of simulational data provided in fig. \ref{anomaly}b with additional data points and power-law fits. The data for $Oh < 1.17\times 10^{-4}$ is indicated as a hashed region to indicate the grid-sensitivity of this data as discussed earlier in Appendix C.}}\label{weber}
    \end{figure}

\textcolor{black}{Although the wave-train in case of bubble bursting ($Bo<<1$) \citep{gordillo2019capillary} is different from the one we study here, some qualitative comparisons can be obtained between the two. In this section, we estimate the speed of propagation of the dominant crest for $\varepsilon=0.091$ and $Oh=0$. Similar to the fig. 4b in \cite{gordillo2019capillary}, we observe that the waves propagating outwards and inwards propagate with the linear speed. This is validated in fig. \ref{i-o velocity} by tracking the local maxima on the free surface in simulations, before and after reflection. In both cases, the propagation speed agrees with the phase speed of the dominant \textcolor{black}{Bessel function} ($k_4$), see fig. \ref{fig5b}. Fig. \ref{weber} provides an approximate scaling relation (to act as guides only) for the dependence of $v_z$ on Oh. We stress that we do not have theoretical description of these power laws and they are distinct from the $v_z^{max} \sim V_\gamma Oh$ established for bursting bubbles for $Oh > Oh_c$. Unlike the case of bubble bursting seen in fig. $12$ in the seminal study by \cite{duchemin2002jet}, the intensification seen in $v_z^{max}$ for our case for the optimal $Oh_c$ compared to $Oh\rightarrow0$ is only about a factor of two. In the case of bubble bursting, this factor can be as high as $10$ \citep{gordillo2023theory}. Note that $Oh_c$ appears as a fitting parameter in all existing bursting bubble theories and further work is needed to estimate this value. Our analysis shows that a first-principles theory for this may have to include non-linearity and also resolve the boundary layer at the interface.}

	\bibliographystyle{jfm}%
	\bibliography{jfm}
\end{document}